\documentclass[aps,floats,nofootinbib]{revtex4-2}
\usepackage{amsmath,amssymb}
\usepackage{graphicx,epsfig}
\usepackage[greek,english]{babel}
\usepackage{bbold}

\usepackage{mathtools}
\usepackage{makecell,tabularx}
\setcellgapes{1pt}
\makegapedcells

\makeatletter\AtBeginDocument{\let\@elt\relax}\makeatother
 
\DeclareMathOperator{\erf}{erf}

\begin{document}
\bibliographystyle {plain}

\pdfoutput=1
\def\oppropto{\mathop{\propto}} 
\def\opsimeq{\mathop{\simeq}}
\def\opoverderline{\mathop{\overline}}
\def\operarrow{\mathop{\longrightarrow}}
\def\opsim{\mathop{\sim}}

\def\opmin{\mathop{\min}} 
\def\opmax{\mathop{\max}} 
\def\oplim{\mathop{\lim}}

\title{ Conditioning two diffusion processes with respect to their first-encounter properties  } 


\author{Alain Mazzolo}
\affiliation{Universit\'e Paris-Saclay, CEA, Service d'\'Etudes des R\'eacteurs et de Math\'ematiques Appliqu\'ees, 91191, Gif-sur-Yvette, France}

\author{C\'ecile Monthus}
\affiliation{Universit\'e Paris-Saclay, CNRS, CEA, Institut de Physique Th\'eorique, 91191 Gif-sur-Yvette, France}


\begin{abstract}
We consider two independent identical diffusion processes that annihilate upon meeting in order to study their conditioning with respect to their first-encounter properties. For the case of finite horizon $T<+\infty$, the maximum conditioning consists in imposing the probability $P^*(x,y,T ) $ that the two particles are surviving at positions $x$ and $y$ at time $T$, as well as the probability $\gamma^*(z,t) $ of annihilation at position $z$ at the intermediate times $t \in [0,T]$. The adaptation to various conditioning constraints that are less-detailed than these full distributions is analyzed via the optimization of the appropriate relative entropy with respect to the unconditioned processes. For the case of infinite horizon $T =+\infty$, the maximum conditioning consists in imposing the first-encounter probability $\gamma^*(z,t) $ at position $z$ at all finite times $t \in [0,+\infty[$, whose normalization $[1- S^*(\infty )]$ determines the conditioned probability $S^*(\infty ) \in [0,1]$ of forever-survival. This general framework is then applied to the explicit cases where the unconditioned processes are respectively two Brownian motions, two Ornstein-Uhlenbeck processes, or two tanh-drift processes, in order to generate stochastic trajectories satisfying various types of conditioning constraints. Finally, the link with the stochastic control theory is described via the optimization of the dynamical large deviations at Level 2.5 in the presence of the conditioning constraints that one wishes to impose.

\end{abstract}

\maketitle


\section{ Introduction} 

\subsection{ Conditioned Markov processes }

Since the pioneering work of Doob \cite{refDoob}, 
the conditioning of Markov processes has attracted a lot of attention 
both in mathematics \cite{refbookDoob,refbookKarlin,refbookRogers,refBaudoin,Borodin}
and in physics (see the recent review \cite{refMajumdarOrland} and references therein),
with applications in many fields including ecology \cite{refHorne}, finance \cite{refBrody} and
nuclear engineering \cite{refMulatier,refbookPazsit}. The simplest example is the stochastic bridge,
where the Markov process is known to be in the configuration $C_0$ at the initial time $t=0$ and in configuration $C_f$ at the final time $t=t_f$ : 
the conditional probability $B^*(C,t)$ to be in configuration $C$ at some internal time $t \in ]0,T[$ 
can then be computed from the unconditioned propagator $P(C_2, t_2 \vert C_1, t_1)$ via the bridge formula
\begin{eqnarray}
B^*(C,t) =  \frac{P(C_f,t_f \vert C,t) P(C,t \vert C_0,0)}{P(C_f,t_f \vert C_0,0)}
\label{bridge}
\end{eqnarray}
which is normalized over the configuration $C$ as a consequence of the Chapman-Kolmogorov property.
The conditioned dynamics of this stochastic bridge can then be obtained from the backward dynamics of the unconditioned propagator $P(C_f,t_f \vert C,t) $ with respect to its initial variables $(C,t)$
and the forward dynamics of the unconditioned propagator $P(C,t \vert C_0,0) $ with respect to its final variables $(C,t)$. In the field of diffusion processes, the basic example of the Brownian bridge has been extended to many other conditioning constraints, including the Brownian excursion \cite{refMajumdarExcursion,refChung}, the Brownian meander \cite{refMajumdarMeander},
the taboo process \cite{refKnight,refPinsky,refKorzeniowski,refGarbaczewski,refAdorisio,refAlainTaboo},
or non-intersecting Brownian bridges \cite{grela}.
Besides diffusion processes, the stochastic bridges have been studied for many other Markov processes, including discrete-time random walks and L\'evy flights \cite{refGarbaczewski_Levy,bruyne_discrete,Aguilar}, 
continuous-time Markov jump processes \cite{Aguilar},
run-and-tumble trajectories \cite{bruyne_run}, or
processes with resetting \cite{refdeBruyne2022}.

The bridge problem of Eq. \ref{bridge} can be also adapted to analyze the conditioning with respect to some global dynamical constraint as measured by a time-additive observable ${\cal A}$ of the stochastic trajectories: the idea is then to consider the bridge formula for the joint process $(C,A)$ instead of the configuration $C$ alone \cite{refMazzoloJstat,Alain_OU,refdeBruyne2021,c_microcanonical,us_LocalTime}. This  'microcanonical conditioning', where the time-additive observable is constrained
to reach a given value after the finite time window $T$ is the counterpart of 
 the 'canonical conditioning' based on generating functions of additive observables
that has been much studied recently in the field of non-equilibrium Markov processes
\cite{peliti,derrida-lecture,tailleur,sollich_review,lazarescu_companion,lazarescu_generic,jack_review,vivien_thesis,lecomte_chaotic,lecomte_thermo,lecomte_formalism,lecomte_glass,kristina1,kristina2,jack_ensemble,simon1,simon2,simon3,Gunter1,Gunter2,Gunter3,Gunter4,chetrite_canonical,chetrite_conditioned,chetrite_optimal,refSzavits,chetrite_HDR,touchette_circle,touchette_langevin,touchette_occ,touchette_occupation,derrida-conditioned,derrida-ring,bertin-conditioned,garrahan_lecture,Vivo,chemical,touchette-reflected,touchette-reflectedbis,c_lyapunov,previousquantum2.5doob,quantum2.5doob,quantum2.5dooblong,c_ruelle,lapolla,chabane}.
In these studies, as explained in detail in the two complementary papers \cite{chetrite_conditioned,chetrite_optimal}, the Doob conditioning, meant to generate atypical values of the time-additive observable in the large deviation regime with respect to the time-window $T$, produces time-independent generators that can be computed from the spectral properties of the appropriate deformation of the unconditioned Markov generator.

Another important extension of the bridge formula of Eq. \ref{bridge} occurs when one wishes to impose the joint probability $E^*(C_f,t_f)$ of the final configuration $C_f$ and of the final time $t_f$
normalized over $C_f$ and over $t_f$
\begin{eqnarray}
\sum_{C_f} \int_0^{+\infty} dt_f E^*(C_f,t_f) =1
\label{normaEstar}
\end{eqnarray}
while the initial configuration $C_0$ at the initial time $t=0$ is still fixed. The conditioned probability to be in configuration $C$ at time $t$ can then be reconstructed via an average of Eq. \ref{bridge} over the probability $E^*(C_f,t_f)$ that one wishes to impose 
\begin{eqnarray}
P^*(C,t)  =\sum_{C_f} \int_t^{+\infty} dt_f E^*(C_f,t_f)   \frac{P(C_f,t_f \vert C,t) P(C,t \vert C_0,0)}{P(C_f,t_f \vert C_0,0)}
\label{sumoverbridge}
\end{eqnarray}
In particular, this formula has been applied
 to impose an arbitrary final distribution $E^*(C_f,T) $ of $C_f$ at some fixed horizon
$T$ \cite{refBaudoin,refMultiEnds}
or to analyze the conditioning with respect to the first-passage-time properties of one-dimensional diffusions at some level $a$ \cite{refBaudoin,refMultiEnds,us_DoobFirstPassage}, as well as the conditioning of diffusion processes with killing rates \cite{us_DoobKilling}.
 
A natural question is then how the formula of Eq. \ref{sumoverbridge} should be adapted 
when one wishes to impose 
conditioning constraints that are less detailed that the whole joint distribution $E^*(C_f,t_f) $. 
It is then useful to adopt the perspective of the famous paper by E. Schr\"odinger \cite{Schrodinger}
(see the recent detailed commentary \cite{CommentSchrodinger} accompanying its english translation),
where the conditioning constraints are considered as the atypical result of
the Gedankenexperiment concerning a large number $N$ of unconditioned processes.
Via this point of view, the theory of Doob conditioning becomes connected
to the fields of large deviations and of stochastic control, as explained in detail in the commentary \cite{CommentSchrodinger} of the Schr\"odinger paper,
as well as in the two reviews \cite{ControlSchrodinger,MongeSchrodinger} written from the viewpoint of stochastic control. In particular, this interpretation allows to give some precise meaning to conditioning constraints that are less detailed that the whole distribution $E^*(C_f,t_f)$ via the optimization of the appropriate relative entropy.


\subsection{ Goals of the present work }

In the present paper, our goal is to apply the general framework described above
to the conditioning of two diffusion processes with respect to their first-encounter properties. 
 Indeed, the first-encounter problem is very important for many applications
 (see the recent works \cite{grebenkov,grebenkov_d} and references therein) 
and belongs to the broad field of first-passage problems that has attracted a lot of interest
\cite{refbookRedner,Bray,metzler,FirstPassage,Redner_2022}.

More precisely, we will consider that the unconditioned process $[X(t);Y(t)]$
corresponds to two independent identical diffusion processes $X(t)$ and $Y(t)$ 
on the full line $]-\infty,+\infty[$ that annihilate upon meeting. 
As long as they are not annihilated, the two processes satisfy the Ito Stochastic Differential Equations 
\begin{eqnarray}
dX(t) = \mu(X(t)) dt + \sqrt{ 2 D(X(t)) } dW(t)
\nonumber \\
dY(t) =\mu(Y(t)) dt + \sqrt{ 2 D(Y(t)) } d{\tilde W}(t)
\label{ito}
\end{eqnarray}
that involve the same drift $\mu(x)$ and the same diffusion coefficient $D(x)$,
while the two Wiener processes $W(t)$ and ${\tilde W}(t) $ are independent.

As explained above on the examples of Eqs \ref{bridge} and \ref{sumoverbridge},
when one wishes to impose some conditioning constraints,
one should first write
the corresponding conditioned probability in the product form
\begin{eqnarray}
P^*(x,y,t)  =   Q(x,y,t) P(x,y,t \vert x_0,y_0,0) 
\label{conditional}
\end{eqnarray}
where $P(x,y,t \vert x_0,y_0,0) $ represents the unconditioned propagator,
while the remaining function $Q(x,y,t) $ has to be computed in terms of the precise conditioning constraints.
One should then analyze the dynamics of $P^*(x,y,t) $ of Eq. \ref{conditional},
based on the forward Fokker-Planck dynamics satisfied by the unconditioned propagator
$P(x,y,t \vert x_0,y_0,0) $ and
on the backward Fokker-Planck dynamics satisfied the function $Q(x,y,t) $.
In the present setting, the conclusion of this dynamical analysis will be that
the function $Q(x,y,t) $ allows to compute the conditioned drift 
\begin{eqnarray}
\mu^*_X (x,y,t) && =\mu(x) + 2 D(x) \partial_x \ln Q(x,y,t) 
\nonumber \\
\mu^*_Y (x,y,t) && =\mu(y) + 2 D(y) \partial_y \ln Q(x,y,t) 
\label{driftdoob}
\end{eqnarray}
that can be plugged into the Ito system analog to Eq. \ref{ito}
\begin{eqnarray}
dX^*(t) && = \mu^*_X (X^*(t),Y^*(t),t) dt + \sqrt{ 2 D(X^*(t))} dW(t)
\nonumber \\
dY^*(t) && = \mu^*_Y (X^*(t),Y^*(t),t) dt + \sqrt{ 2 D(Y^*(t))} d{\tilde W}(t)
\label{itostar}
\end{eqnarray}
in order to generate stochastic trajectories of the conditioned process $[X^*(t),Y^*(t)]$
with annihilation upon meeting.

In summary, for each type of conditioning constraints that we will consider,
we will write the appropriate function $Q(x,y,t) $ of Eq. \ref{conditional}
in order to compute the corresponding conditioned drifts $[\mu^*_X (x,y,t),\mu^*_Y (x,y,t)] $ via Eq. \ref{driftdoob}.
Some examples of the conditioned drifts that will be derived 
are given in the three following Tables.

\begin{table}[!h]
\setcellgapes{4pt}
\begin{tabular}{|p{5cm}||c|c|}
\hline
Conditioning towards annihilation at position $z^*$ at time $T^*$ & $\mu^*_X (x,y,t)$ & $\mu^*_Y (x,y,t)$  \\
\hline \hline
Brownian motions & $\frac{1}{x-y} + \frac{z^*-x}{T^*-t}$ & $- \frac{1}{x-y} +  \frac{z^*-y}{T^*-t} $  \\
\hline
Ornstein-Uhlenbeck processes  & $- k x + \frac{1}{x-y}  + \frac{ k \left[ z^* -x e^{- k (T^*-t)}\right]}{\sinh \left[ k (T^*-t)\right]}$ & $- k y - \frac{1}{x-y}  + \frac{ k \left[ z^* -y e^{- k (T^*-t)}\right]}{\sinh \left[ k (T^*-t)\right]}$  \\
\hline
Tanh-drift processes & $\frac{1}{x-y} + \frac{z^*-x}{T^*-t}$ & $- \frac{1}{x-y} +  \frac{z^*-y}{T^*-t}$  \\
\hline
\end{tabular}
\caption{Conditioned drifts $[\mu^*_X (x,y,t),\mu^*_Y (x,y,t)] $ towards annihilation at position $z^*$ at time $T^*$. Observe that the conditioned drifts are identical for the Brownian motions and the tanh-drift processes.} 
\label{table1}
\end{table}

\begin{table}[!h]
\setcellgapes{4pt}
\begin{tabular}{|p{4cm}||c|c|}
\hline
Conditioned bridges towards the positions $x_*$ and $y_*$ at time $T$ without meeting & $\mu^*_X (x,y,t)$ & $\mu^*_Y (x,y,t)$  \\
\hline \hline
Brownian bridges & $- \frac{x }{(T-t)}
 +  \frac{(x_*+y_*)  }{2(T-t)}
 +  \frac{(x_*-y_*)  }{2(T-t)}  \coth \left(\frac{(x_*-y_*)(x-y)  }{2(T-t)} \right) $ & $- \frac{y }{(T-t)}
+  \frac{(x_*+y_*)  }{2(T-t)}
  -  \frac{(x_*-y_*)  }{2(T-t)}  \coth \left(\frac{(x_*-y_*)(x-y)  }{2(T-t)} \right) $  \\
\hline
Ornstein-Uhlenbeck bridges  & $\begin{aligned} & - k x 
- k \frac{ x   e^{- k (T-t)}    }{ \sinh \left[ k (T-t) \right] } 
+ k \frac{  (x_*+y_* )    }{2 \sinh \left[ k (T-t) \right] } \\ & + k \frac{  (x_*-y_* )       }{2 \sinh \left[ k (T-t) \right] }
 \coth \left[ k \frac{  (x_*-y_* ) (x-y )      }{2 \sinh \left[ k (T-t) \right] }  \right] \end{aligned}$ & 
 $\begin{aligned} & - k y
- k \frac{y  e^{- k (T-t)}    }{ \sinh \left[ k (T-t) \right] } 
+ k \frac{  (x_*+y_* )    }{2 \sinh \left[ k (T-t) \right] }
\\&  - k \frac{  (x_*-y_* )       }{2 \sinh \left[ k (T-t) \right] }
 \coth \left[ k \frac{  (x_*-y_* ) (x-y )      }{2 \sinh \left[ k (T-t) \right] }  \right]\end{aligned}$  \\
\hline
Tanh-drift bridges & $- \frac{x }{(T-t)}
 +  \frac{(x_*+y_*)  }{2(T-t)}
 +  \frac{(x_*-y_*)  }{2(T-t)}  \coth \left(\frac{(x_*-y_*)(x-y)  }{2(T-t)} \right)$ & $- \frac{y }{(T-t)}
+  \frac{(x_*+y_*)  }{2(T-t)}
  -  \frac{(x_*-y_*)  }{2(T-t)}  \coth \left(\frac{(x_*-y_*)(x-y)  }{2(T-t)} \right)$  \\
\hline
\end{tabular}
\caption{
Conditioned drifts $[\mu^*_X (x,y,t),\mu^*_Y (x,y,t)] $ for two stochastic bridges 
ending at positions $x_*$ and $y_*$ at time $T$ without meeting. Observe that the conditioned drifts are identical for the Brownian bridges and the tanh-drift bridges.} 
\label{table2}
\end{table}

\begin{table}[!h]
\setcellgapes{4pt}
\begin{tabular}{|p{5cm}||c|c|}
\hline
Conditioning towards the forever survival 
without meeting & $\mu^*_X (x,y)$ & $\mu^*_Y (x,y)$  \\
\hline \hline
Brownian motions & $\frac{1}{x-y} $ & $- \frac{1}{x-y}  $  \\
\hline
Ornstein-Uhlenbeck processes  & $- k x + \frac{1}{x-y} $ & $- k y - \frac{1}{x-y} $  \\
\hline
Tanh-drift processes & $\frac{\alpha}{ \tanh [\alpha(x-y) ]}$ & $- \frac{\alpha}{ \tanh [\alpha(x-y) ]}$  \\
\hline
\end{tabular}
\caption{Time-independent conditioned drifts $[\mu^*_X (x,y),\mu^*_Y (x,y)] $
 for two forever-surviving processes without meeting.} 
\label{table3}
\end{table}


\subsection{ Organization of the paper }

The paper is organized as follows. The section \ref{sec_unconditioned} describes the properties of the unconditioned process $[X(t);Y(t)]$, where $X(t)$ and $Y(t)$ are two independent identical diffusion processes that annihilate upon meeting. Section \ref{sec_finitehorizon} is devoted to the construction of the conditioned process $[X^*(t);Y^*(t)]$ with respect to the finite horizon $T<+\infty$, when one imposes the full surviving distribution $P^*(x,y,T)$ at time $T$ and the full annihilation distribution $\gamma^*(z,t) $ for the intermediate times $t \in [0,T]$. The adaptation to various conditioning constraints that are less-detailed than the full distributions $\left[ P^*(.,.,T) ; \gamma^*(.,.)  \right] $ is discussed in section \ref{sec_lessdetailed}. The limit of the infinite horizon $T=+\infty$ is considered in section \ref{sec_infinitehorizon}, as a function of the conditioned forever-survival probability $ S^*(\infty ) \in [0,1]$ that one wishes to impose. This general framework is then applied to three cases with diffusion coefficient $D(x)=1/2$, while the unconditioned drift $\mu(x)$ is either vanishing $\mu(x)=0$ in section \ref{sec_Brown}, corresponds to the Ornstein-Uhlenbeck linear drift $\mu(x)=- k x$ towards the origin $x=0$ in section \ref{sec_OU}, or is given by the drift 
$\mu(x)=\alpha \tanh (\alpha x)  $ pushing away from the origin $x=0$ in section \ref{sec_tanh}. In these three sections, explicit examples are studied in order
to generate stochastic trajectories satisfying various types of conditioning constraints and Monte Carlo simulations illustrate our findings.
Our conclusions are summarized in section \ref{sec_conclusion}. The appendix \ref{app_control} describes the link with the stochastic control theory
via the dynamical large deviations at Level 2.5.


\section{ Unconditioned process : two diffusions that annihilate upon meeting }

\label{sec_unconditioned}

In this section, we recall some useful properties for two independent identical diffusions that annihilate upon meeting.

\subsection{ Properties of the propagator $p(x_2,t_2 \vert x_1,t_1)$ for a single diffusion process $X(t)$}

\label{subsec_1particle}

The Fokker-Planck generator associated to the Ito Stochastic Differential Equation of Eq. \ref{ito}
for the single diffusion process $X(t)$
\begin{eqnarray}
{\cal F}_x = \mu(x) \partial_x + D(x) \partial^2_{x} 
\label{generator}
\end{eqnarray}
governs the backward dynamics of the propagator $p(x_2,t_2 \vert x_1,t_1)$
 with respect to its initial variables $(x_1,t_1)$ 
\begin{eqnarray}
-\partial_{t_1} p(x_2,t_2 \vert x_1,t_1) = {\cal F}_{x_1} p(x_2,t_2 \vert x_1,t_1) 
= \mu(x_1) \partial_{x_1} p(x_2,t_2 \vert x_1,t_1) + D(x_1) \partial^2_{x_1} p(x_2,t_2 \vert x_1,t_1)
\label{backward}
\end{eqnarray}
while the adjoint operator of the generator of Eq. \ref{generator}
\begin{eqnarray}
{\cal F}^{\dagger}_x = -  \partial_x \mu(x)+ \partial^2_{x}  D(x)
\label{adjoint}
\end{eqnarray}
governs the forward dynamics of the propagator $p(x_2,t_2 \vert x_1,t_1)$
 with respect to the its final variables $(x_2,t_2)$
\begin{eqnarray}
\partial_{t_2} p(x_2,t_2 \vert x_1,t_1) ={\cal F}^{\dagger}_{x_2}  p(x_2,t_2 \vert x_1,t_1)
= -  \partial_{x_2} \left[  \mu(x_2) p(x_2,t_2 \vert x_1,t_1) \right]+ \partial^2_{x_2} 
\left[ D(x_2) p(x_2,t_2 \vert x_1,t_1) \right]
\label{forward}
\end{eqnarray}


\subsection{ Propagator $P(x_2,y_2,t_2 \vert x_1,y_1,t_1) $ for two independent diffusion processes that annihilate upon meeting }

We are interested in the case where the two diffusion processes $X(t)$ and $Y(t)$ 
generated by Ito Stochastic Differential Equations of Eqs \ref{ito} 
annihilate upon meeting.
Let us consider the initial conditions  $Y(t_1)=y_1<x_1=X(t_1)$ at the time $t_1$.
The probability $P(x_2,y_2,t_2 \vert x_1,y_1,t_1) $ that these two processes are still surviving 
at position $x_2$ and $y_2$ at time $t_2$ 
is given for $y_2<x_2$ by the following Karlin-McGregor $2 \times 2$ determinant 
\cite{1959,Karlin,Karlin_theorem}
\begin{eqnarray}
P(x_2,y_2,t_2 \vert x_1,y_1,t_1) 
&& =  \begin{vmatrix} 
p(x_2,t_2 \vert x_1,t_1) &  p(y_2,t_2 \vert x_1,t_1)  \\
p(x_2,t_2 \vert y_1,t_1) &  p(y_2,t_2 \vert y_1,t_1) 
 \end{vmatrix} 
\nonumber \\
&&  = p(x_2,t_2 \vert x_1,t_1)  p(y_2,t_2 \vert y_1,t_1) 
- p(x_2,t_2 \vert y_1,t_1)  p(y_2,t_2 \vert x_1,t_1)  
\label{pdet}
\end{eqnarray}
involving the 1-particle propagators $p(.,.\vert.,.)$ discussed in the previous subsection \ref{subsec_1particle}.
The Karlin-McGregor determinant \cite{1959,Karlin,Karlin_theorem}
concerns of course the much more general problem
 of an arbitrary number $N$ of non-crossing independent processes,
and plays in particular an essential role in the theory of random matrices, 
as explained in detail in the recent 
PhD Thesis \cite{Tristan} containing an extensive bibliography.

Here we will only consider the simple case of $N=2$ processes,
 where the physical interpretation of the determinant of Eq. \ref{pdet} is straightforward :

(i) the first term 
\begin{eqnarray}
P^{indep}(x_2,y_2,t_2 \vert x_1,y_1,t_1) = p(x_2,t_2 \vert x_1,t_1)  p(y_2,t_2 \vert y_1,t_1)
\label{pindep}
\end{eqnarray}
 represents the propagator when
the two processes are independent and do not annihilate upon meeting.

(ii) the second term 
\begin{eqnarray}
P^{exchange}(x_2,y_2,t_2 \vert x_1,y_1,t_1) = p(x_2,t_2 \vert y_1,t_1)  p(y_2,t_2 \vert x_1,t_1)
\label{pexchange}
\end{eqnarray}
where the two initial conditions $(x_1,y_1)$
are interchanged with respect to Eq. \ref{pindep},
is meant to kill all contributions of Eq. \ref{pindep} where the two paths are crossing each other, as explained in greater detail in the original work of Karlin and McGregor \cite{1959}.

Another useful way to understand the two-particles propagator of Eq. \ref{pdet}
 is to consider $[X(t);Y(t)]$ as a two-dimensional process
living in the half-plane $x > y$ : the point $(x_1'=y_1;y_1'=x_1)$ 
then corresponds to the image of the initial condition $(x_1,y_1)$ with respect 
to the diagonal $x=y$ that represents the absorbing boundary condition.

The propagator $P(x_2,y_2,t_2 \vert x_1,y_1,t_1)  $ of Eq. \ref{pdet}
inherits from Eq. \ref{backward}
the backward dynamics with respect to its initial variables $(x_1,y_1,t_1)$
\begin{eqnarray}
&& -\partial_{t_1} P(x_2,y_2,t_2 \vert x_1,y_1,t_1)   = \left[ {\cal F}_{x_1} +{\cal F}_{y_1} \right]  P(x_2,y_2,t_2 \vert x_1,y_1,t_1) 
\nonumber \\
&& = \left[ \mu(x_1)   + D(x_1) \partial_{x_1} \right] \partial_{x_1} P(x_2,y_2,t_2 \vert x_1,y_1,t_1)
+  \left[ \mu(y_1)   + D(y_1) \partial_{y_1} \right] \partial_{y_1} P(x_2,y_2,t_2 \vert x_1,y_1,t_1)
\label{backward2}
\end{eqnarray}
and inherits from Eq. \ref{forward}
the forward dynamics with respect to its final variables $(x_2,y_2,t_2)$
\begin{eqnarray}
\partial_{t_2} P(x_2,y_2,t_2 \vert x_1,y_1,t_1) 
&& =\left[ {\cal F}^{\dagger}_{x_2}  +{\cal F}^{\dagger}_{y_2} \right] P(x_2,y_2,t_2 \vert x_1,y_1,t_1)
\nonumber \\
&& = -  \partial_{x_2} J_X (x_2,y_2,t_2 \vert x_1,y_1,t_1)
-  \partial_{y_2} J_Y (x_2,y_2,t_2 \vert x_1,y_1,t_1)
\label{forward2}
\end{eqnarray}
where the interpretation as a continuity equation involves the two components of the current 
\begin{eqnarray}
J_X (x_2,y_2,t_2 \vert x_1,y_1,t_1) && =  \mu(x_2) P(x_2,y_2,t_2 \vert x_1,y_1,t_1) 
- \partial_{x_2} \left[ D(x_2) P(x_2,y_2,t_2 \vert x_1,y_1,t_1)  \right]
\nonumber \\
J_Y (x_2,y_2,t_2 \vert x_1,y_1,t_1) && =  \mu(y_2) P(x_2,y_2,t_2 \vert x_1,y_1,t_1) 
- \partial_{y_2} \left[ D(y_2) P(x_2,y_2,t_2 \vert x_1,y_1,t_1)  \right]
\label{current}
\end{eqnarray}


\subsection{ Survival probability $S(t_2 \vert x_1,y_1,t_1) $ and 
 annihilation probability $\gamma(z_2,t_2 \vert x_1,y_1,t_1) $  at position $z_2$ at time $t_2$}

The total survival probability at time $t_2$ 
can be computed via the integration of the propagator $P(x_2,y_2,t_2 \vert x_1,y_1,t_1)  $ 
over all the possible positions $y_2<x_2$ using the Heaviside theta function $ \theta(x_2-y_2)$
\begin{eqnarray}
S(t_2 \vert x_1,y_1,t_1) = \int_{-\infty}^{+\infty}  dy_2   \int_{-\infty}^{+\infty}  dx_2
\theta(x_2-y_2)P(x_2,y_2,t_2 \vert x_1,y_1,t_1)
\label{survival}
\end{eqnarray}
Its time-decay allows to compute 
the probability $\gamma(t_2 \vert x_1,t_1)$ of annihilation at time $t_2$
\begin{eqnarray}
\gamma(t_2 \vert x_1,y_1,t_1) =
- \partial_{t_2} S(t_2 \vert x_1,y_1,t_1) 
   = \int_{-\infty}^{+\infty}  dy_2   \int_{-\infty}^{+\infty}  dx_2
\theta(x_2-y_2) \left[ - \partial_{t_2} P(x_2,y_2,t_2 \vert x_1,y_1,t_1) \right]
\label{gammatdef}
\end{eqnarray}
Using the forward dynamics of Eq. \ref{forward}
and integrations by parts, Eq. \ref{gammatdef} can be rewritten in terms of the current of Eq. \ref{current}
\begin{eqnarray}
\gamma(t_2 \vert x_1,y_1,t_1)
&& =  \int_{-\infty}^{+\infty}  dy_2   \int_{-\infty}^{+\infty}  dx_2
\theta(x_2-y_2) \left[  \partial_{x_2} J_X (x_2,y_2,t_2 \vert x_1,y_1,t_1)
+ \partial_{y_2} J_Y (x_2,y_2,t_2 \vert x_1,y_1,t_1) \right]
\nonumber \\
&& =  \int_{-\infty}^{+\infty}  dy_2   \int_{y_2}^{+\infty}  dx_2
   \partial_{x_2} J_X (x_2,y_2,t_2 \vert x_1,y_1,t_1)
+ \int_{-\infty}^{+\infty}  dx_2 \int_{-\infty}^{x_2}  dy_2  
 \partial_{y_2} J_Y (x_2,y_2,t_2 \vert x_1,y_1,t_1)
 \nonumber \\
&& =  \int_{-\infty}^{+\infty}  dy_2  
 \left[ J_X (x_2,y_2,t_2 \vert x_1,y_1,t_1) \right]_{x_2=y_2}^{x_2=+\infty}
+ \int_{-\infty}^{+\infty}  dx_2  \left[ J_Y (x_2,y_2,t_2 \vert x_1,y_1,t_1) \right]_{y_2=-\infty}^{y_2=x_2}
 \nonumber \\
&& = - \int_{-\infty}^{+\infty}  dy_2  
 J_X (y_2,y_2,t_2 \vert x_1,y_1,t_1) 
+ \int_{-\infty}^{+\infty}  dx_2   J_Y (x_2,x_2,t_2 \vert x_1,y_1,t_1) 
 \nonumber \\
&& \equiv  \int_{-\infty}^{+\infty}  dz_2  \gamma(z_2,t_2 \vert x_1,y_1,t_1) 
\label{survivalderi}
\end{eqnarray}
where the probability $\gamma(z_2,t_2 \vert x_1,y_1,t_1) $ 
of annihilation at position $z_2$ at time $t_2$
\begin{eqnarray}
\gamma(z_2,t_2 \vert x_1,y_1,t_1) 
\equiv - J_X (z_2,z_2,t_2 \vert x_1,y_1,t_1) +  J_Y (z_2,z_2,t_2 \vert x_1,y_1,t_1) 
\label{gammaztdef}
\end{eqnarray}
corresponds to the current entering the absorbing diagonal $x_2=y_2$. 
Using the explicit currents of Eq. \ref{current} and the vanishing of the propagator
 at coinciding points $P(z_2,z_2,t_2 \vert x_1,y_1,t_1) =0 $,
Eq. \ref{gammaztdef} can be rewritten only in terms of the spatial derivative of the propagator 
in the normal direction with respect to the absorbing diagonal $x_2=y_2$
\begin{eqnarray}
\gamma(z_2,t_2 \vert x_1,y_1,t_1) 
= D(z_2) \left[ 
\left(    \partial_{x_2} -\partial_{y_2} \right)
\left[  P(x_2,y_2,t_2 \vert x_1,y_1,t_1)  \right]
\right] \vert_{x_2=z_2;y_2=z_2}
\label{gammazt}
\end{eqnarray}
The three functions $S(t_2 \vert x_1,y_1,t_1) $, $\gamma(t_2 \vert x_1,y_1,t_1) $ and $\gamma(z_2,t_2 \vert x_1,y_1,t_1)  $
inherit from the propagator $ P(x_2,y_2,t_2 \vert x_1,y_1,t_1)   $
 the backward dynamics of Eq. \ref{backward2} with respect to the initial variables $(x_1,y_1,t_1)$
\begin{eqnarray}
-\partial_{t_1} S(t_2  \vert x_1,y_1,t_1) && = \left[ {\cal F}_{x_1} +{\cal F}_{y_1} \right]   S(t_2  \vert x_1,y_1,t_1)
\nonumber \\
-\partial_{t_1} \gamma(t_2 \vert x_1,y_1,t_1) && =\left[ {\cal F}_{x_1} +{\cal F}_{y_1} \right]   \gamma(t_2 \vert x_1,y_1,t_1)
\nonumber \\
-\partial_{t_1} \gamma(z_2,t_2 \vert x_1,y_1,t_1) && =\left[ {\cal F}_{x_1} +{\cal F}_{y_1} \right]   \gamma(z_2,t_2 \vert x_1,y_1,t_1)
\label{gammabackward}
\end{eqnarray}

The normalization of Eq. \ref{gammatdef}
over the possible finite times $t_2 \in [t_1,+\infty[$ 
\begin{eqnarray}
\int_{t_1}^{+\infty} dt_2 \gamma(t_2 \vert x_1,y_1,t_1) =
- \left[ S(t_2 \vert x_1,y_1,t_1) \right]_{t_2=t_1}^{t_2=+\infty} =1 - S(\infty \vert x_1,y_1,t_1) = 1 - S(\infty \vert x_1,y_1)
\label{gammatnorma}
\end{eqnarray}
involves the probability $S(\infty \vert x_1,y_1) \in [0,1]$ of forever-survival for the two particles starting at $(x_1,y_1)$.


\subsection{ Probability $P(x_2=z_2+\epsilon,y_2=z_2-\epsilon ,t_2\vert x_1,y_1,t_1) $ to be near the absorbing diagonal $x_2=y_2$}

For later purposes, it is also useful to evaluate 
the probability to be near the absorbing diagonal at position $(x_2=z_2+\epsilon,y_2=z_2-\epsilon)$ 
via the Taylor expansion at first order in $\epsilon$ around $P(z_2,z_2,t_2 \vert x_1,y_1,t_1)=0 $ \begin{eqnarray}
&& P(x_2=z_2+\epsilon,y_2=z_2-\epsilon,t_2\vert x_1,y_1,t_1)  
\nonumber \\
&& =P(z_2,z_2,t_2 \vert x_1,y_1,t_1) 
 + \epsilon  \left[ (\partial_{x_2} - \partial_{y_2})P(x_2,y_2,t_2 \vert x_1,y_1,t_1)  \right] \vert_{x_2=z_2;y_2=z_2}
 + O(\epsilon^2)
\nonumber \\
&&   = \epsilon \frac{1}{ D(z_2) } \gamma(z_2,t_2 \vert x_1,y_1,t_1)  + O(\epsilon^2)
\label{Absgamma}
\end{eqnarray}
in terms of the annihilation distribution $\gamma(z_2,t_2 \vert x_1,y_1,t_1) $ of Eq. \ref{gammazt}.


\section{ Conditioned process $[X^*(t);Y^*(t)]$ with respect to the finite horizon $T<+\infty$ }

\label{sec_finitehorizon}

\subsection{ Full conditioning constraints $\left[ P^*(.,.,T) ; \gamma^*(.,.)  \right]  $ associated to the finite horizon $T<+\infty$ }

For the unconditioned diffusion process $[X(t);Y(t)]$ starting at the position $[X(0)=x_0;Y(0)=y_0]$ at time $t=0$ :

(i) the probability for the two particles to be
surviving at time $T$ at the positions $x$ and $y$ is given by the unconditioned propagator $P(x,y,T \vert x_0, y_0,0)$
of Eq. \ref{pdet},
with the corresponding unconditioned survival probability at time $T$ of Eq. \ref{survival}
\begin{eqnarray}
S(T \vert x_0,y_0,0) = \int_{-\infty}^{+\infty}  dx   \int_{-\infty}^{+\infty}  dy
\theta(x-y)P(x,y,T \vert x_0,y_0,0)
\label{survivalT}
\end{eqnarray}

 (ii) the probability to have been annihilated at the position $z_a$ at the time $T_a$ 
 is given by the unconditioned annihilation probability $\gamma(z_a,T_a \vert x_0,y_0,0) $  of Eq. \ref{gammazt}
 where the normalization is complementary to the unconditioned survival probability of Eq. \ref{survivalT}
\begin{eqnarray}
 \int_{0}^T dT_a \int_{-\infty}^{+\infty} dz_a \gamma(z_a,T_a \vert x_0,y_0,0)   = 1- S(T \vert x_0,y_0,0) 
\label{deadT}
\end{eqnarray}

In this section, we wish to construct 
the conditioned diffusion process $[X^*(t);Y^*(t)]$ by imposing instead 
the following other constraints: 

(i) another probability $P^*(x,y,T )$ for the two particles to be
surviving at the positions $x$ and $y$ at time $T$,
with the corresponding conditioned survival probability 
$S^*(T )$ at time $T$
\begin{eqnarray}
   S^*(T)  = \int_{-\infty}^{+\infty}  dx   \int_{-\infty}^{+\infty}  dy \theta(x-y)P^*(x,y,T )
\label{survivalTstar}
\end{eqnarray}

 (ii) another probability $\gamma^*(z_a,T_a ) $
 to have been annihilated at position $z_a$ at the time $T_a$, 
  whose normalization is complementary to Eq. \ref{survivalTstar}
\begin{eqnarray}
 \int_{0}^T dT_a \int_{-\infty}^{+\infty} dz_a \gamma^*(z_a,T_a )   = 1- S^*(T ) 
\label{deadTstar}
\end{eqnarray}
 The conditioned survival probability $S^*(t)$ at any intermediate time $t \in [0,T]$ can be computed via
\begin{eqnarray}
S^*(t ) = 1-  \int_{0}^t dT_a  \int_{-\infty}^{+\infty} dz_a \gamma^*(z_a,T_a)    
\label{survivalstarinter}
\end{eqnarray}


\subsection{ Conditioned probability $P^*(x,y,t ) $ at any intermediate time $t \in ]0,T[$ }

At any intermediate time $t \in ]0,T[$, the conditioned probability $P^*(x,y,t ) $ 
to be still surviving at the positions $x$ and $y$ at time $t$
involves the two contributions
\begin{eqnarray}
P^*(x,y,t) && =  \int_t^{T} dT_a \int_{-\infty}^{+\infty} dz_a \gamma^*( z_a,T_a) 
 \left[\oplim_{\epsilon \to 0} \frac{P(z_a+\epsilon,z_a-\epsilon,T_a \vert x,y,t) P(x,y,t \vert x_0,y_0,0)}
 {P(z_a+\epsilon,z_a-\epsilon,T_a \vert x_0,y_0,0)} \right]
 \nonumber \\ &&
 + \int_{-\infty}^{+\infty}  dx_T   \int_{-\infty}^{+\infty}  dy_T
\theta(x_T-y_T)  P^*(x_T,y_T,T )  \frac{ P(x_T,y_T,T \vert x,y,t) P(x,y,t \vert x_0,y_0,0)}{P(x_T,y_T,T \vert x_0,y_0,0) }
\label{conditionaldef}
\end{eqnarray}
Its normalization over the two positions $x$ and $y$
can be computed using the Chapman-Kolmogorov property of the unconditioned process
and Eq. \ref{survivalstarinter}
\begin{eqnarray}
\int_{-\infty}^{+\infty}  dx   \int_{-\infty}^{+\infty}  dy \theta(x-y) P^*(x,y,t)  && = 
 \int_t^{T} dT_a \int_{-\infty}^{+\infty} dz_a \gamma^*( z_a,T_a) 
 + \int_{-\infty}^{+\infty}  dx_T   \int_{-\infty}^{+\infty}  dy_T
\theta(x_T-y_T)  P^*(x_T,y_T,T ) 
 \nonumber \\  && = \left[ S^*(t) - S^*(T) \right] +S^*(T ) = S^*(t )
\label{conditionalnorma}
\end{eqnarray}
i.e. one obtains the conditioned survival probability $S^*(t ) $ that one wishes to impose.

The initial condition at $t=0$ is the same as for the unconditioned process
using Eqs \ref{survivalTstar} and \ref{deadTstar}
 \begin{eqnarray}
P^*(x,y,t=0)  && =  \int_0^{T} dT_a \int_{-\infty}^{+\infty} dz_a \gamma^*( z_a,T_a) \delta(x-x_0)\delta(y-y_0)
 \nonumber \\ &&
 + \int_{-\infty}^{+\infty}  dx_T   \int_{-\infty}^{+\infty}  dy_T
\theta(x_T-y_T)  P^*(x_T,y_T,T ) \delta(x-x_0)\delta(y-y_0)
 \nonumber \\ && =\delta(x-x_0)\delta(y-y_0)
\label{initialstar}
\end{eqnarray}

In the first contribution of Eq. \ref{conditionaldef}, 
the property of Eq. \ref{Absgamma} allows to rewrite the limit involving $\epsilon \to 0$ in terms of 
the annihilation distributions $ \gamma(z_a,T_a \vert x,y,t)$ and $\gamma(z_a,T_a\vert x_0,y_0,0) $
\begin{eqnarray}
 \oplim_{\epsilon \to 0}  
  \frac{P(z_a+\epsilon,z_a-\epsilon,T_a \vert x,y,t) }
 {P(z_a+\epsilon,z_a-\epsilon,T_a \vert x_0,y_0,0)}
= \frac{\gamma(z_a,T_a \vert x,y,t)}{\gamma(z_a,T_a\vert x_0,y_0,0)}
\label{ratio}
\end{eqnarray}
In summary, the conditioned probability of Eq. \ref{conditionaldef} 
can be written in the product form of Eq. \ref{conditional}
involving the unconditioned propagator $P(x,y,t \vert x_0,y_0,0) $,
while the function 
\begin{eqnarray}
Q_T(x,y,t)  && \equiv 
 \int_t^{T} dT_a \int_{-\infty}^{+\infty} dz_a 
 \frac{\gamma^*( z_a,T_a) }{\gamma(z_a,T_a\vert x_0,y_0,0)}\gamma(z_a,T_a \vert x,y,t)
 \nonumber \\ &&
 + \int_{-\infty}^{+\infty}  dx_T   \int_{-\infty}^{+\infty}  dy_T
\theta(x_T-y_T)    \frac{ P^*(x_T,y_T,T ) }{P(x_T,y_T,T \vert x_0,y_0,0) }P(x_T,y_T,T \vert x,y,t)
 \label{Qdef}
\end{eqnarray}
inherits the backward Fokker-Planck dynamics of Eq. \ref{gammabackward} concerning $\gamma(z_a,T_a \vert x,y,t) $ 
and of Eq. \ref{backward2} concerning $P(x_T,y_T,T \vert x,y,t) $ 
with respect to their initial variables $(x,y,t)$
\begin{eqnarray}
- \partial_t  Q_T(x,y,t) && =\left[ {\cal F}_x +{\cal F}_y \right] Q_T(x,y,t)
\nonumber \\
&& = \left[      \mu(x) \partial_x   +   \mu(y) \partial_y  + D(x)\partial^2_x + D(y)\partial^2_y\right] Q_T(x,y,t)
\label{Qbackward}
\end{eqnarray}
since the derivative with respect to the time $t$ appearing as the lower boundary of the integral of the first
contribution of Eq. \ref{Qdef} gives zero as a consequence 
of the vanishing of the annihilation propability $\gamma(z_a,t \vert x,y,t)=0 $ 
at coinciding times for $x>y$
\begin{eqnarray}
-   \int_{-\infty}^{+\infty} dz_a 
 \frac{\gamma^*( z_a,t) }{\gamma(z_a,t\vert x_0,y_0,0)}\gamma(z_a,t \vert x,y,t) =0
\label{Qderiborneintegrale}
\end{eqnarray}


\subsection{ Forward dynamics of the conditioned process $[X^*(t);Y^*(t)]$  }

Using the forward dynamics of Eq. \ref{forward2} satisfied by the unconditioned 
propagator $P(x,y,t \vert x_0,y_0,0) $
\begin{eqnarray}
\partial_t P(x,y,t \vert x_0,y_0,0) && 
= \left[ {\cal F}^{\dagger}_x  +{\cal F}^{\dagger}_y \right] P(x_2,y_2,t_2 \vert x_1,y_1,t_1)
\nonumber \\
&& =  -  \partial_x \left[  \mu(x) P(x,y,t \vert x_0,y_0,0)  \right]
-  \partial_y \left[  \mu(y) P(x,y,t \vert x_0,y_0,0)  \right]
\nonumber \\
&& + \partial^2_x \left[ D(x) P(x,y,t \vert x_0,y_0,0)  \right]
+  \partial^2_y \left[ D(y) P(x,y,t \vert x_0,y_0,0)  \right]
\label{forwardbis}
\end{eqnarray}
and the backward dynamics of Eq. \ref{Qbackward} satisfied by $Q_T(x,y,t)$,
one obtains that the time derivative of the conditioned probability of Eq. \ref{conditional}
reads
\begin{eqnarray}
  \partial_t P^*(x,y,t\vert x_0,y_00) ) &&  =
  P(x,y,t \vert x_0,y_0,0) \left[   \partial_t  Q_T(x,y,t) \right]  
    +Q_T(x,y,t)  \left[   \partial_t  P(x,y,t \vert x_0,y_0,0)\right]  
\nonumber \\
&& =
 -    \partial_x \left[  \mu(x) Q_T(x,y,t) P(x,y,t \vert x_0,y_0,0)  \right]
 - \partial_y \left[  \mu(y)  Q_T(x,y,t) P(x,y,t \vert x_0,y_0,0)  \right] 
  \nonumber \\
&&   
+ \partial_x \bigg(  \partial_x \left[ D(x) Q_T(x,y,t)   P(x,y,t \vert x_0,y_0,0)  \right]
- 2 D(x) P(x,y,t \vert x_0,y_0,0)  \partial_x Q_T(x,y,t) \bigg)
\nonumber \\
&&  + \partial_y \bigg(  \partial_y \left[ D(y) Q_T(x,y,t)   P(x,y,t \vert x_0,y_0,0)  \right]
- 2 D(y) P(x,y,t \vert x_0,y_0,0)  \partial_y Q_T(x,y,t) \bigg)  
\label{conditionalderi}
\end{eqnarray}
Using Eq. \ref{conditional} to replace the unconditioned propagator
\begin{eqnarray}
P(x,y,t \vert x_0,y_0,0)  =  \frac{ P^*(x,y,t) } {Q_T(x,y,t) }
\label{eliminate}
\end{eqnarray}
into Eq. \ref{conditionalderi}, one obtains that the conditioned propagator
satisfies the following forward Fokker-Planck dynamics with respect to $(x,y,t)$
\begin{eqnarray}
  \partial_t P^*(x,y,t) && =
-  \partial_x \left[  \mu^*_X (x,y,t) P^*(x,y,t)  \right]
-  \partial_y \left[  \mu^*_Y (x,y,t)  P^*(x,y,t)  \right]
\nonumber \\
&&  + \partial^2_x \left[ D(x) P^*(x,y,t)  \right]
+  \partial^2_y \left[ D(y) P^*(x,y,t)  \right]
\label{conditionaldyn}
\end{eqnarray}
where the only difference with respect to the unconditional dynamics of Eq. \ref{forwardbis}
is in the drift whose two components $\mu^*_X (x,y,t) $ and $\mu^*_Y (x,y,t) $ are given in terms of the function $Q_T(x,y,t) $ 
by Eq. \ref{driftdoob}.
The corresponding Ito Stochastic Differential Equations of Eq. \ref{itostar}
can then be used to generate stochastic trajectories of the conditioned process $[X^*(t),Y^*(t)]$
that annihilate upon meeting.


\section{ Conditioning less detailed than the full distributions $\left[ P^*(.,.,T) ; \gamma^*(.,.)  \right] $ at $T$ }

\label{sec_lessdetailed}

In the previous section \ref{sec_finitehorizon}, 
we have described the construction of the conditioned process  $[X^*(t),Y^*(t)]$
when the conditioning constraints correspond to the full distributions $\left[ P^*(.,.,T) ; \gamma^*(.,.)  \right] $ associated to the finite horizon $T$. 
In the present section, we describe how the cases where the conditioning constraints are less detailed
can be analyzed via the notion of relative entropy cost.

\subsection{ Relative entropy cost of the full conditioning constraints
$\left[ P^*(.,.,T) ; \gamma^*(.,.)  \right]$
imposed at the time horizon $T$}

\label{subsec_sanov}

\subsubsection{  Sanov theorem for $N$ independent unconditioned processes observed at the finite time horizon $T$}

Let us consider a large number $N$ of independent realizations  $[X_n(t),Y_n(t)]$ of the unconditioned process
 labelled by $n=1,2,..,N$ starting all at the same initial condition $X_n(0)=x_0$ and $X_n(0)=x_0$
 at time $t=0$

For each realization $n=1,..,N$, 
the final state at the finite time horizon $T$ is characterized by :

(i) either the two positions $X_n(T)=x_n$ and $Y_n(T)=y_n$ if the two processes are still surviving at $T$;

(ii) or the position $z_n$ and the time $t_n \in [0,T]$ of their annihilation.

The global normalization of these events involve the unconditioned propagator $P(x_n,y_n,t \vert x_0,y_0,0) $ 
and the unconditioned annihilation probability $\gamma(z_n,t_n \vert x_0,y_0,0) $
\begin{eqnarray}
1 =\int_{-\infty}^{+\infty}  dy_n   \int_{-\infty}^{+\infty}  dx_n \theta(x_n-y_n)  P(x_n,y_n,t \vert x_0,y_0,0)
 +\int_{-\infty}^{+\infty}  dz_n  \int_0^T dt_n \gamma(z_n,t_n \vert x_0,y_0,0) 
\label{globalnormaT}
\end{eqnarray}

When one considers the $N$ independent processes $X_n(t) $ with $n=1,..,N$
at the finite time horizon $T$, the empirical histogram ${\hat P}(x,y,T) $ of the surviving 
positions at time $T$ 
\begin{eqnarray}
 {\hat P}(x,y,T) \equiv \frac{1}{N} \sum_{n=1}^N \delta( x_n-x) \delta( y_n-y) 
 \label{empiPTN}
\end{eqnarray}
and the empirical  histogram ${\hat \gamma}(z,t) $
of the annihilation events 
\begin{eqnarray}
{\hat \gamma}(z,t) \equiv \frac{1}{N} \sum_{n=1}^N \delta(z_n -z )  \delta(t_n -t )  
\label{empigammaN}
\end{eqnarray}
satisfy the global normalization analog to Eq. \ref{globalnormaT}
\begin{eqnarray}
1 = \int_{-\infty}^{+\infty}  dy   \int_{-\infty}^{+\infty}  dx \theta(x-y)   {\hat P}(x,y,T)
 +\int_{-\infty}^{+\infty}  dz  \int_0^T dt  {\hat \gamma}(z,t)
\label{globalnormaTempi}
\end{eqnarray}

In the field of large deviations (see the reviews \cite{oono,ellis,review_touchette} and references therein),
the empirical histogram of independent identically distributed variables 
 is governed by the Sanov theorem. Its application to the present case yields the following conclusion :
the joint probability to observe the empirical surviving density  ${\hat P}(x,y,T) $  
and the empirical annihilation distribution ${\hat \gamma}(z,t) $ 
satisfy the large deviation form for large $N$
\begin{eqnarray}
{\cal P}^{Sanov}_T \left[ {\hat P }(.,.,T) ; {\hat \gamma}(.,.)  \right]  
&& \opsimeq_{N \to +\infty} 
\delta \left(  \int_{-\infty}^{+\infty}  dy   \int_{-\infty}^{+\infty}  dx \theta(x-y)   {\hat P}(x,y,T)
 +\int_{-\infty}^{+\infty}  dz  \int_0^T dt  {\hat \gamma}(z,t)
-1 \right) 
\nonumber \\
&& \times e^{- N {\cal I}^{Sanov}_T \left[ {\hat P }(.,.,T) ; {\hat \gamma}(.,.)  \right] }
\label{LevelSanov}
\end{eqnarray}
where the delta function imposes the normalization constraint of Eq. \ref{globalnormaTempi},
while the Sanov rate function 
\begin{eqnarray}
{\cal I}^{Sanov}_T \left[ {\hat P }(.,.,T) ; {\hat \gamma}(.,.)  \right]
&& = 
  \int_{-\infty}^{+\infty}  dy   \int_{-\infty}^{+\infty}  dx \theta(x-y)   {\hat P}(x,y,T)
    \ln \left( \frac{ {\hat P}(x,y,T)}{ P(x,y,T \vert x_0, y_0,0)}  \right)
\nonumber \\
&&    +\int_{-\infty}^{+\infty}  dz  \int_0^T dt  {\hat \gamma}(z,t)
      \ln \left(  \frac{  {\hat \gamma}(z,t)  }{\gamma(z,t \vert x_0,y_0,0) }  \right)
\label{RateSanov}
\end{eqnarray} 
corresponds to the relative entropy
of the empirical distributions $\left[ {\hat P }(.,.,T) ; {\hat \gamma}(.,.)  \right] $ with respect to the "true" 
distributions $\left[ P(.,.,T \vert x_0, y_0,0) ;  \gamma(. ,.\vert x_0,y_0,0) \right] $ that are the only ones that survive in the thermodynamic limit $N \to +\infty$.


\subsubsection{ Application  : relative entropy cost of the conditioning constraints
$\left[ P^*(.,.,T) ; \gamma^*(.,.)  \right]$
imposed at the time horizon $T$}

As mentioned in the Introduction,
the above framework involving $N$ independent unconditioned processes
provide an interesting alternative perspective on the 
conditioning constraints imposed at the finite horizon $T<+\infty$ :
one can interpret the imposed distributions $P^*(x,y,T ) $ and $\gamma^*(z,t ) $  
 as the empirical results $\left[ {\hat P }(.,.,T) ; {\hat \gamma}(.,.)  \right]$ 
obtained in an experiment concerning
$N$ independent unconditioned processes, as initially proposed by 
E. Schr\"odinger in his famous paper \cite{Schrodinger}
(see the recent detailed commentary \cite{CommentSchrodinger} accompanying its english translation).

Let us stress the important consequences of this perspective :

(i) the Sanov rate function of Eq. \ref{RateSanov} evaluated for the imposed conditions
$\left[ P^*(.,.,T) ; \gamma^*(.,.)  \right]$ at the horizon $T$
\begin{eqnarray}
{\cal I}^{Sanov}_T \left[ P^*(.,.,T) ; \gamma^*(.,.)  \right]&& = 
 \int_{-\infty}^{+\infty}  dy   \int_{-\infty}^{+\infty}  dx \theta(x-y)   P^*(x,y,T)
    \ln \left( \frac{ P^*(x,y,T)}{ P(x,y,T \vert x_0, y_0,0)}  \right)
\nonumber \\
&&    +\int_{-\infty}^{+\infty}  dz  \int_0^T dt   \gamma^*(z,t)
      \ln \left(  \frac{  \gamma^*(z,t)  }{\gamma(z,t \vert x_0,y_0,0) }  \right)
\label{RateSanovstar}
\end{eqnarray} 
measures how rare it is for large $N$ to see the distributions $\left[ P^*(.,.,T) ; \gamma^*(.)  \right]$ 
different from the 'true' distributions $\left[ P(.,.,T \vert x_0, y_0,0) ;  \gamma(. ,.\vert x_0,y_0,0) \right] $.

(ii) the conditioned process $[X^*(t);Y^*(t)]$
for the intermediate times $t \in ]0,T[$ described in the previous section
can then be interpreted as the most probable empirical dynamics 
that one can infer. This interpretation is corroborated by
the analysis of the relative entropy cost of the empirical dynamics 
during the whole time-window $t \in [0,T]$, as described in Appendix \ref{app_control}.

(iii)  the Sanov rate function ${\cal I}^{Sanov}_T \left[ P^*(.,.,T) ; \gamma^*(.,.)  \right] $
 of Eq. \ref{RateSanovstar} allows
to give some precise meaning to conditioning constraints 
that are less detailed that the whole distributions $\left[ P^*(.,.,T) ; \gamma^*(.,.)  \right] $ considered 
up to now :
the idea is that one needs to optimize the Sanov rate function
in the presence of the less detailed conditioning constraints that one wishes to impose.
Let us now describe the simplest examples in the remainder of this section.


\subsection{ Conditioning towards the surviving distribution 
$P^*(x,y,T) $ at the horizon $T$ alone}

\label{subsec_spatial}

If one wishes to impose only the probability $P^*(x,y,T) $ at time $T$, together
with its corresponding survival probability
\begin{eqnarray}
 S^*(T) \equiv \int_{-\infty}^{+\infty} dy  \int_{-\infty}^{+\infty}  dx \theta(x-y)   P^*(x,y,T) 
  \label{survivalspace}
\end{eqnarray}
one needs to optimize the Sanov rate function ${\cal I}^{Sanov}_T \left[ P^*(.,.,T) ; \gamma^*(.,.)  \right] $ of Eq. \ref{RateSanovstar}
over the annihilation probability $ \gamma^*(.,.) $ satisfying the normalization constraint 
\begin{eqnarray}
\int_{-\infty}^{+\infty} dz \int_0^T dt  \gamma^*(z,t) = 1-S^*(T)
\label{normaKstar}
\end{eqnarray}
It is convenient to introduce the following Lagrangian involving the Lagrange multiplier $\alpha$ to impose this constraint
\begin{eqnarray}
&& {\cal L}^{space} \left[  \gamma^*(.,.)  \right] 
 = {\cal I}^{Sanov}_T \left[ P^*(.,.,T) ; \gamma^*(.,.)  \right]
  + \alpha \left( \int_{-\infty}^{+\infty} dz \int_0^T dt  \gamma^*(z,t) - \left[ 1-  S^*(T) \right] \right) 
   \nonumber \\
 &&
 =   \int_{-\infty}^{+\infty}  dy   \int_{-\infty}^{+\infty}  dx \theta(x-y)   P^*(x,y,T)
    \ln \left( \frac{ P^*(x,y,T)}{ P(x,y,T \vert x_0, y_0,0)}  \right)
\nonumber \\ &&
    +\int_{-\infty}^{+\infty}  dz  \int_0^T dt   \gamma^*(z,t)
      \ln \left(  \frac{  \gamma^*(z,t)  }{\gamma(z,t \vert x_0,y_0,0) }  \right)
 + \alpha \left( \int_{-\infty}^{+\infty} dz \int_0^T dt  \gamma^*(z,t) - \left[ 1-  S^*(T) \right] \right)
 \label{lagrangianspace}
\end{eqnarray}
The optimization of this Lagrangian 
over the distribution $\gamma^*(z,t) $ 
\begin{eqnarray}
0 = \frac{ \partial {\cal L}_T^{space} \left[  \gamma^*(.,.)  \right] }{ \partial  \gamma^*(z,t)} = 
\ln \left(  \frac{\gamma^*(z,t)  }{ \gamma(z,t \vert x_0,y_0,0) } \right) +1+\alpha 
 \label{lagrangianspacederi}
\end{eqnarray}
leads to the optimal solution
\begin{eqnarray}
\gamma^{*opt} (z,t)  =e^{-1-\alpha} \gamma(z,t \vert x_0,y_0,0)
 \label{gammaopt}
\end{eqnarray}
that should satisfy the normalization constraint of Eq. \ref{normaKstar}
\begin{eqnarray}
1- S^*(T) = \int_{-\infty}^{+\infty} dz \int_0^T dt  \gamma^{*opt}(z,t) =
e^{-1-\alpha} \int_{-\infty}^{+\infty} dz \int_0^T dt \gamma(z,t \vert x_0,y_0,0)
 = e^{-1-\alpha} \left[ 1-S(T\vert x_0,y_0,0) \right]
  \label{gammaoptnorm}
\end{eqnarray}
Plugging this value of the Lagrange multiplier $\alpha$ into Eq. \ref{gammaopt}
leads to the final 
optimal solution
\begin{eqnarray}
\gamma^{*opt} (z,t)  = \left( \frac{1- S^*(T) }{1- S(T \vert x_0,y_0,0)} \right)   \gamma(z,t \vert x_0,y_0,0)
 \label{gammaoptfinal}
\end{eqnarray}
The contribution to the Lagrangian of Eq. \ref{lagrangianspace} of this optimal solution 
\begin{eqnarray}
&& \int_{-\infty}^{+\infty} dz \int_0^T dt  \gamma^{*opt}(z,t)    
     \ln \left(  \frac{\gamma^{*opt}(z,t)  }{ \gamma(z,t \vert x_0,y_0,0) } \right)  
 \nonumber \\ &&
  = \int_{-\infty}^{+\infty} dz \int_0^T dt 
\left( \frac{1- S^*(T) }{1- S(T \vert x_0,y_0,0)} \right)  \gamma(z,t \vert x_0,y_0,0)
  \ln \left( \frac{1- S^*(T) }{1- S(T \vert x_0,y_0,0)} \right)
  \nonumber \\ &&
  =  \left( 1- S^*(T)  \right) 
  \ln \left( \frac{1- S^*(T) }{1- S(T \vert x_0,y_0,0)} \right)
 \label{lagrangianspaceopt}
\end{eqnarray}
involves only the conditioned and unconditioned survival propabilities.
The relative entropy cost of the imposed probability $P^*(x,y,T) $
and of its corresponding survival probability $S^*(T) $ of Eq. \ref{survivalspace} then reads
\begin{eqnarray}
&& {\cal I}_T^{space} \left[ P^*(.,.,T) ; S^*(T)  \right] 
 \equiv  {\cal I}^{Sanov}_T \left[ P^*(.,.,T) ; \gamma^{*opt}(.,.)  \right]
 \nonumber \\
&&  =   
 \int_{-\infty}^{+\infty}  dy   \int_{-\infty}^{+\infty}  dx \theta(x-y)   P^*(x,y,T)
    \ln \left( \frac{ P^*(x,y,T)}{ P(x,y,T \vert x_0, y_0,0)}  \right)
  +  \left( 1- S^*(T)  \right) 
  \ln \left( \frac{1- S^*(T) }{1- S(T \vert x_0,y_0,0)} \right)
 \label{ratespacealone}
\end{eqnarray}

In conclusion, if one wishes to impose only the probability $P^*(y,T) $ at time $T$,
one should use the optimal solution $\gamma^{*opt} (z,t) $ of Eq. \ref{gammaoptfinal},
so that the function $Q_T(x,y,t) $ of Eq. \ref{Qdef} becomes
\begin{eqnarray}
&& Q_T^{[ P^*(.,.,T) ; S^*(T) ] }(x,y,t)   = 
 \left( \frac{1- S^*(T) }{1- S(T \vert x_0,y_0,0)} \right)
 \int_t^{T} dT_a \int_{-\infty}^{+\infty} dz_a 
   \gamma(z_a,T_a \vert x,y,t)
 \nonumber \\ &&
 + \int_{-\infty}^{+\infty}  dx_T   \int_{-\infty}^{+\infty}  dy_T
\theta(x_T-y_T)    \frac{ P^*(x_T,y_T,T ) }{P(x_T,y_T,T \vert x_0,y_0,0) }P(x_T,y_T,T \vert x,y,t)
\nonumber \\
&& = 
( 1- S^*(T) ) \left( \frac{ 1-S(T \vert x,y,t) }{1- S(T \vert x_0,y_0,0)} \right)
 + \int_{-\infty}^{+\infty}  dx_T   \int_{-\infty}^{+\infty}  dy_T
\theta(x_T-y_T) P^*(x_T,y_T,T )   \frac{P(x_T,y_T,T \vert x,y,t)  }{P(x_T,y_T,T \vert x_0,y_0,0) }
 \label{Qspacealone}
\end{eqnarray}


\subsection{ Conditioning towards the annihilation distribution $ \gamma^*(z,t) $ for $t \in [0,T]$ alone}

\label{subsec_annihilation}

If one wishes to impose only the annihilation distribution $ \gamma^*(z,t) $ for $t \in [0,T]$, together
with its normalization 
\begin{eqnarray}
\int_{-\infty}^{+\infty} dx \int_0^T dt  \gamma^*(z,t) = 1-S^*(T) 
  \label{survivaltime}
\end{eqnarray}
one needs to optimize the Sanov rate function ${\cal I}^{Sanov}_T \left[ P^*(.,.,T) ; \gamma^*(.,.)  \right] $
of Eq. \ref{RateSanovstar}
over the possible spatial surviving distribution $P^*(x,y,T)  $ normalized to $S^*(T) $.
It is thus convenient to introduce the following Lagrangian involving the Lagrange multiplier $\beta$
\begin{eqnarray}
&& {\cal L}^{annihilation} \left[   P^*(.,.,T) \right]  
 = {\cal I}^{Sanov}_T \left[ P^*(.,.,T) ; \gamma^*(.,.)  \right]
   + \beta \left( \int_{-\infty}^{+\infty} dy   P^*(y,T)  -  S^*(T) \right) 
   \nonumber \\
 &&
 = 
 \int_{-\infty}^{+\infty}  dy   \int_{-\infty}^{+\infty}  dx \theta(x-y)   P^*(x,y,T)
    \ln \left( \frac{ P^*(x,y,T)}{ P(x,y,T \vert x_0, y_0,0)}  \right)
    +\int_{-\infty}^{+\infty}  dz  \int_0^T dt   \gamma^*(z,t)
      \ln \left(  \frac{  \gamma^*(z,t)  }{\gamma(z,t \vert x_0,y_0,0) }  \right)
   \nonumber \\
 &&   + \beta \left( \int_{-\infty}^{+\infty} dy  \int_{-\infty}^{+\infty}  dx \theta(x-y)   P^*(x,y,T)  -  S^*(T) \right) 
 \label{lagrangianannihilation}
\end{eqnarray}
The optimization is therefore very similar to the previous subsection
and leads to the optimal solution
\begin{eqnarray}
P^{*opt}(x,y,T)  = \left( \frac{ S^*(T) }{ S(T \vert x_0,y_0,0)} \right)  P(x,y,T \vert x_0,y_0,0) 
 \label{pyoptfinal}
\end{eqnarray}
with the corresponding contribution to the Lagrangian of Eq. \ref{lagrangianannihilation} 
\begin{eqnarray}  
 \int_{-\infty}^{+\infty}  dy   \int_{-\infty}^{+\infty}  dx \theta(x-y)   P^{*opt}(x,y,T)
    \ln \left( \frac{ P^{*opt}(x,y,T)}{ P(x,y,T \vert x_0, y_0,0)}  \right)
  =  S^*(T) 
   \ln \left( \frac{ S^*(T) }{ S(T \vert x_0,y_0,0)} \right)
 \label{lagrangiantimeopt}
\end{eqnarray}
The relative entropy cost of the imposed annihilation distribution $\gamma^*(z,t) $ 
and of the corresponding survival probability $S^*(T) $ of Eq. \ref{survivaltime}
is thus given by
\begin{eqnarray}
 {\cal I}_T^{annihilation} \left[  \gamma^*(.,.)  ; S^*(T) \right]
 && =  {\cal I}^{Sanov}_T \left[ P^{*opt}(.,T) ; \gamma^*(.,.)  \right]
 \nonumber \\
&&  =   S^*(T)    \ln \left( \frac{ S^*(T) }{ S(T \vert x_0,y_0,0)} \right)
 +\int_{-\infty}^{+\infty}  dz  \int_0^T dt   \gamma^*(z,t)
      \ln \left(  \frac{  \gamma^*(z,t)  }{\gamma(z,t \vert x_0,y_0,0) }  \right)
   \label{rateannihilationalone}
\end{eqnarray}

In conclusion, if one wishes to impose only the annihilation probability 
$\gamma^*(z,t) $ for $t \in [0,T]$,
one should use the optimal solution $P^{*opt}(y,T) $ of Eq. \ref{pyoptfinal},
so that the function $Q_T(x,y,t) $ of Eq. \ref{Qdef} becomes
\begin{eqnarray}
Q_T^{[  \gamma^*(.,.)  ; S^*(T) ]}(x,y,t)  && = 
 \int_t^{T} dT_a \int_{-\infty}^{+\infty} dz_a 
 \frac{\gamma^*( z_a,T_a) }{\gamma(z_a,T_a\vert x_0,y_0,0)}\gamma(z_a,T_a \vert x,y,t)
 \nonumber \\ &&
 + \left( \frac{ S^*(T) }{ S(T \vert x_0,y_0,0)} \right)  \int_{-\infty}^{+\infty}  dx_T   \int_{-\infty}^{+\infty}  dy_T
\theta(x_T-y_T)  
P(x_T,y_T,T \vert x,y,t)
\nonumber \\
&& =  \int_t^{T} dT_a \int_{-\infty}^{+\infty} dz_a \gamma^*( z_a,T_a)
 \frac{ \gamma(z_a,T_a \vert x,y,t)}{\gamma(z_a,T_a\vert x_0,y_0,0)}
 +S^*(T) \left( \frac{S(T \vert x,y,t)  }{ S(T \vert x_0,y_0,0)} \right)  
 \label{Qannihilationalone}
\end{eqnarray}


\subsection{ Conditioning towards the surviving probability $ S^*(T) $ at time $T$ alone}

If one wishes to impose only the value $S^*(T)$ of the conditioned survival probability at time $T$,
one can use the analysis of the two previous subsections \ref{subsec_spatial} and \ref{subsec_annihilation} to obtain the following results.
The relative entropy cost of imposing 
the surviving probability $ S^*(T) $ at time $T$ alone reduces to
\begin{eqnarray}
 {\cal I}^{surviving}_T \left[  S^*(T) \right]
 && =        S^*(T)    \ln \left( \frac{ S^*(T) }{ S(T \vert x_0,y_0,0)} \right)
   +  \left( 1- S^*(T)  \right)   \ln \left( \frac{1- S^*(T) }{1- S(T \vert x_0,y_0,0)} \right)
   \label{ratesurvivalalone}
\end{eqnarray}
In addition, the optimal solutions $\gamma^{*opt} (z,t) $ of Eq. \ref{gammaoptfinal}
and $P^{*opt}(x,y,T) $ of Eq. \ref{pyoptfinal} yields
that the corresponding function $Q_T(x,y,t)$ reads using 
the contributions computed in Eqs \ref{Qspacealone}
and \ref{Qannihilationalone}
\begin{eqnarray}
Q_T^{[  S^*(T)]}(x,y,t)   = 
( 1- S^*(T) ) \left( \frac{ 1-S(T \vert x,y,t) }{1- S(T \vert x_0,y_0,0)} \right)
 +S^*(T) \left( \frac{S(T \vert x,y,t)  }{ S(T \vert x_0,y_0,0)} \right)  
 \label{Qsurvivingalone}
\end{eqnarray}


\subsection{ Conditioning towards the time-annihilation distribution $ \gamma^*(t) = \int_{-\infty}^{+\infty} \gamma^*(z,t) $ for $t \in [0,T]$ alone}

If one wishes to impose only the time-annihilation distribution $ \gamma^*(t) = \int_{-\infty}^{+\infty} \gamma^*(z,t) $ for $t \in [0,T]$, together with its normalization from Eq. \ref{survivaltime}
\begin{eqnarray}
 \int_0^T dt  \gamma^*(t) = 1-S^*(T) 
  \label{survivaltimeannihilation}
\end{eqnarray}
one needs to optimize the rate function ${\cal I}_T^{annihilation} \left[  \gamma^*(.,.)  ; S^*(T) \right] $
of Eq. \ref{rateannihilationalone}
over the possible spatial-dependence in $z$
 of the annihilation distribution  $\gamma^*(z,t)  $,
with the normalization constraint for each $t$
\begin{eqnarray}
\int_{-\infty}^{+\infty} dz \gamma^*(z,t) = \gamma^*(t)
  \label{normaannihilationanty}
\end{eqnarray}
Let us introduce the following Lagrangian involving the Lagrange multiplier $\chi(t)$ for $t \in [0,T]$
\begin{eqnarray}
&& {\cal L}_T^{time} \left[   \gamma^*(.,.) \right]  
 = {\cal I}_T^{annihilation} \left[  \gamma^*(.,.)  ; S^*(T) \right]
    + \int_0^T dt \chi(t)  \left(\int_{-\infty}^{+\infty} dz \gamma^*(z,t) - \gamma^*(t)  \right) 
   \nonumber \\
 &&
= \int_{-\infty}^{+\infty} dz \int_0^T dt  \gamma^*(z,t)     
    \ln \left(  \frac{\gamma^*(z,t)  }{ \gamma(z,t \vert x_0,y_0,0) } \right)    
  +  S^*(T)    \ln \left( \frac{ S^*(T) }{ S(T \vert x_0,y_0,0)} \right)
  \nonumber \\
 &&    + \int_0^T dt \chi(t)  \left(\int_{-\infty}^{+\infty} dz \gamma^*(z,t) - \gamma^*(t)  \right) 
\label{lagrangiantime}
\end{eqnarray}
The optimization is again very similar to the previous subsections
and leads to the optimal solution
\begin{eqnarray}
\gamma^{*opt}(z,t)  = \frac{\gamma^*(t)}{\gamma(t \vert x_0,y_0,0) } \gamma(z,t \vert x_0,y_0,0)
 \label{kstarxoptfinal}
\end{eqnarray}
The corresponding contribution to the Lagrangian of Eq. \ref{lagrangiantime} reads
\begin{eqnarray}
&& \int_{-\infty}^{+\infty} dz \int_0^T dt  \gamma^*(z,t)    
     \ln \left(  \frac{\gamma^*(z,t)  }{ \gamma(z,t \vert x_0,y_0,0) } \right)    
 = \int_{-\infty}^{+\infty} dz \int_0^T dt \frac{\gamma^*(t)}{\gamma(t\vert x_0,y_0,0) } \gamma(z,t \vert x_0,y_0,0) \ln \left(  \frac{\gamma^*(t)  }{ \gamma(t\vert x_0,y_0,0) } \right)  
\nonumber \\
&& = \int_0^T dt \gamma^*(t) \ln \left(  \frac{\gamma^*(t)  }{ \gamma(t\vert x_0,y_0,0) } \right)  
\label{lagrangiantimeoptcontri}
\end{eqnarray}
The relative entropy cost of the time-annihilation distribution $\gamma^*(t) $ 
and of the corresponding survival probability $S^*(T) $ of Eq. \ref{survivaltimeannihilation}
is thus given by
\begin{eqnarray}
 {\cal I}_T^{time} \left[  \gamma^*(.)  ; S^*(T) \right]
 && =  \int_0^T dt \gamma^*(t) \ln \left(  \frac{\gamma^*(t)  }{ \gamma(t\vert x_0,y_0,0) } \right)  
  +  S^*(T)    \ln \left( \frac{ S^*(T) }{ S(T \vert x_0,y_0,0)} \right)
   \label{ratetimeannihilationalone}
\end{eqnarray}

In conclusion, if one wishes to impose only the time-annihilation distribution $\gamma^*(t) $ for $t \in [0,T]$,
one should use the optimal solution $\gamma^{*opt} (z,t) $ of Eq. \ref{kstarxoptfinal},
so that the function $Q_T(x,t) $ of Eq. \ref{Qannihilationalone} becomes
\begin{eqnarray}
Q_T^{[\gamma^*(.)  ; S^*(T) ]}(x,y,t)  &&  
=  \int_t^{T} dT_a \frac{\gamma^*(T_a)}{\gamma(T_a \vert x_0,y_0,0) }
\int_{-\infty}^{+\infty} dz_a
 \gamma(z_a,T_a \vert x,y,t)
 +S^*(T) \left( \frac{S(T \vert x,y,t)  }{ S(T \vert x_0,y_0,0)} \right)  
 \nonumber \\
 && =  \int_t^{T} dT_a \gamma^*(T_a) \frac{ \gamma(T_a \vert x,y,t)}{\gamma(T_a \vert x_0,y_0,0) }
 +S^*(T) \left( \frac{S(T \vert x,y,t)  }{ S(T \vert x_0,y_0,0)} \right)  
 \label{Qtimealone}
\end{eqnarray}


\section{ Conditioned process $[X^*(t);Y^*(t)]$ with respect to the infinite horizon $T=+\infty$ }

\label{sec_infinitehorizon}

In this section, we discuss the limit of the infinite horizon $T \to +\infty$ 
for the conditioned processes $[X^*(t);Y^*(t)]$ constructed in the two previous sections.
It is convenient to distinguish three cases according to the values of
the conditioned forever-survival probability $ S^*(\infty )$ that one wishes to impose.


\subsection{ Cases $ S^*(\infty )=0$ where the conditioning is towards full annihilation before 
the infinite horizon $T=+\infty$ }

When the conditioning corresponds to full annihilation $ S^*(T )=0$ before 
the horizon time $T$ in Eqs \ref{survivalTstar} and \ref{deadTstar}
\begin{eqnarray}
P^*(x,y,T )   && = 0 \ \ {\rm for } \ \  y <x 
 \nonumber \\
 \int_{0}^T dT_a \int_{-\infty}^{+\infty} dz_a \gamma^*(z_a,T_a )    && =1
\label{normatstar0}
\end{eqnarray}
then the function $Q_T(x,t)$ contains only the first contribution of Eq. \ref{Qdef}
involving an integral over the time $T_a \in ]t,T[$ and the position $z_a \in ]-\infty,+\infty[$ 
\begin{eqnarray}
Q_T^{[\gamma^*(.,.);S^*(T )=0]}(x,y,t) = 
 \int_t^{T} dT_a \int_{-\infty}^{+\infty} dz_a 
 \frac{\gamma^*( z_a,T_a) }{\gamma(z_a,T_a\vert x_0,y_0,0)}\gamma(z_a,T_a \vert x,y,t) 
 \label{Qtime}
\end{eqnarray}

The limit of the infinite horizon $T \to +\infty$ can then be taken directly 
to obtain the following conclusion: 
when the conditioning is towards some first-encounter-time distribution 
$\gamma^*(z_a,T_a )$ with the normalization obtained from the limit $T \to +\infty$ of Eq. \ref{normatstar0}
\begin{eqnarray}
  \int_{0}^{+\infty} dT_a  \int_{-\infty}^{+\infty} dz_a\gamma^*(z_a,T_a )    =1
\label{normatstar0infinity}
\end{eqnarray}
the function $Q_{\infty}^{[\gamma^*(.,.);S^*(\infty )=0]}(x,y,t) $ is obtained from the limit $T \to +\infty$ of Eq. \ref{Qtime}
\begin{eqnarray}
Q_{\infty}^{[\gamma^*(.,.);S^*(\infty )=0]}(x,y,t) (x,y,t) = 
\int_t^{+\infty} dT_a \int_{-\infty}^{+\infty} dz_a 
 \frac{\gamma^*( z_a,T_a) }{\gamma(z_a,T_a\vert x_0,y_0,0)}\gamma(z_a,T_a \vert x,y,t) 
 \label{Qtimeinfinity}
\end{eqnarray}


\subsection{ Cases $ S^*(\infty )=1$ where the conditioning is towards full survival 
at the infinite horizon $T=+\infty$ }

When one wishes to consider the limit of the infinite horizon $T \to +\infty$ 
with full survival $ S^*(T \to \infty )=1$,
the limit $T \to +\infty$ can be taken on Eq. \ref{Qsurvivingalone} to obtain that the function
\begin{eqnarray}
Q_{\infty}^{[  S^*(\infty )=1]}(x,y,t)   = 
 \lim_{T \to +\infty} \left( \frac{S(T \vert x,y,t)  }{ S(T \vert x_0,y_0,0)} \right)  
 \label{Qspacesurvivalinfinity}
\end{eqnarray}
reduces to the asymptotic ratio of the two survival probabilities $S (T \vert x,y,t)  $ and $S (T \vert x_0,y_0,0) $ of the unconditioned process.

In practice, one thus needs to distinguish two cases :

(a) If the forever-survival probability $S(\infty \vert .)$ of the unconditioned process is finite,
the limit of Eq. \ref{Qspacesurvivalinfinity} will only involve the ratio of the two 
forever-survival probabilities $S(\infty \vert x,y)$ and $S(\infty \vert x_0,y_0)$
\begin{eqnarray}
Q_{\infty}^{[  S^*(\infty )=1]}(x,y,t)
 = \frac{ S(\infty \vert x,y)   }
 {S(\infty \vert x_0,y_0)   }
\label{Qspacesurvivalinftycasea}
\end{eqnarray}

(b) If the forever-survival probability of the unconditioned process vanishes $S(\infty \vert .)=0$,
 the limit of Eq. \ref{Qspacesurvivalinfinity} will involve the asymptotic behavior of the
 annihilation-time distributions $\gamma(T_a \vert x,y,t) $ and $\gamma(T_a \vert x_0,y_0,0) $ of the unconditioned process
\begin{eqnarray}
Q_{\infty}^{[  S^*(\infty )=1]}(x,y,t)
 =  \lim_{T \to +\infty} \frac{  \int_{T}^{+\infty} dT_a  \gamma(T_a \vert x,y,t)  }
 { \int_{T}^{+\infty} dT_a \gamma(T_a \vert x_0,y_0,0)  }
\label{Qspacesurvivalinftycaseb}
\end{eqnarray}


\subsection{ Cases $ S^*(\infty)\in ]0,1[$ where the conditioning is towards partial survival 
at the infinite horizon $T=+\infty$ }

Let us now consider the limit of the infinite horizon $T \to +\infty$
 when one wishes to impose some partial forever-survival $ S^*(\infty ) \in ]0,1[$.
 The limit $T \to +\infty$ can be taken on Eq. \ref{Qannihilationalone} to obtain the function
 \begin{eqnarray}
Q_{\infty}^{[  \gamma^*(.,.)  ; S^*(\infty) ]}(x,y,t)    
 =  \int_t^{\infty} dT_a \int_{-\infty}^{+\infty} dz_a \gamma^*( z_a,T_a)
 \frac{ \gamma(z_a,T_a \vert x,y,t)}{\gamma(z_a,T_a\vert x_0,y_0,0)}
 +S^*(\infty)  \lim_{T \to +\infty} \frac{ S (T \vert x,y,t)  }{ S (T \vert x_0,y_0,0)} 
 \label{Qinfinitypartial}
\end{eqnarray}
where the evaluation of the last limit will depend 
on whether the forever-survival probability $S(\infty \vert .)$ 
of the unconditioned process vanishes or not, as already discussed in Eqs \ref{Qspacesurvivalinftycasea}
and \ref{Qspacesurvivalinftycaseb}.


\section{ Application to the conditioning of two Brownian motions }

\label{sec_Brown}

In this section, the framework described in the previous sections
is applied to the simplest case,
where the unconditioned process $[X(t);Y(t)]$ corresponds to two independent Brownian motions that annihilate upon meeting.


\subsection{ Unconditioned process $[X(t);Y(t)]$ : two Brownian motions that annihilate upon meeting }

When the single diffusion of Eq. \ref{ito} corresponds to the vanishing drift $\mu(x)=0$ and to the diffusion coefficient $D(x)=1/2$,
the 1-particle Gaussian propagator  
\begin{eqnarray}
p(x_2,t_2 \vert x_1,t_1) = \frac{1}{\sqrt{2 \pi (t_2-t_1)}}  e^{- \frac{(x_2-x_1 )^2}{2(t_2-t_1)}} 
\label{gauss}
\end{eqnarray}
leads to the Karlin-McGregor determinant of Eq. \ref{pdet} 
\begin{eqnarray}
P(x_2,y_2,t_2 \vert x_1,y_1,t_1) 
&&  =  \frac{1}{2 \pi (t_2-t_1)} \left[ 
 e^{- \frac{(x_2-x_1 )^2+(y_2-y_1 )^2 }{2(t_2-t_1)}} 
-  e^{- \frac{(x_2-y_1 )^2+(y_2-x_1 )^2 }{2(t_2-t_1)}} 
\right] 
\nonumber \\
&&  =  \frac{1}{ \pi (t_2-t_1)} e^{ \frac{(x_2+y_2)(x_1+y_1)  }{2(t_2-t_1)}
 - \frac{x_2^2+x_1^2+y_2^2+y_1^2 }{2(t_2-t_1)}} 
\sinh \left(\frac{(x_2-y_2)(x_1-y_1)  }{2(t_2-t_1)} \right)
\label{pdetbrown}
\end{eqnarray}
The probability $\gamma(z_2,t_2 \vert x_1,y_1,t_1) $ 
of annihilation at position $z_2$ at time $t_2$ of Eq. \ref{gammazt} 
can be computed using Eq. \ref{pdetbrown}
\begin{eqnarray}
\gamma(z_2,t_2 \vert x_1,y_1,t_1) 
&& = \frac{1}{2} \left[ 
\left(    \partial_{x_2} -\partial_{y_2} \right)
P(x_2,y_2,t_2 \vert x_1,y_1,t_1) 
\right] \vert_{x_2=z_2;y_2=z_2}
\nonumber \\
&& =
 \frac{ (x_1-y_1) }{2 \pi (t_2-t_1)^2 }  
 e^{- \frac{(z_2-x_1 )^2+(z_2-y_1 )^2 }{2(t_2-t_1)}} 
\label{gammaztbrown}
\end{eqnarray}
The integration over the position $z_2$ gives the probability of annihilation at time $t_2$
of Eq. \ref{survivalderi}
\begin{eqnarray}
\gamma(t_2 \vert x_1,y_1,t_1)
&& =  \int_{-\infty}^{+\infty}  dz_2   \gamma(z_2,t_2 \vert x_1,y_1,t_1) 
 = \frac{ (x_1-y_1) }{2 \pi (t_2-t_1)^2 }    \int_{-\infty}^{+\infty}  dz_2   e^{- \frac{(z_2-x_1 )^2+(z_2-y_1 )^2 }{2(t_2-t_1)}} 
\nonumber \\
&& =  \frac{ (x_1-y_1) }{2 \sqrt{ \pi} (t_2-t_1)^{\frac{3}{2}} }      e^{- \frac{(x_1-y_1 )^2 }{4(t_2-t_1)}} 
\label{gammatbrown}
\end{eqnarray}
This means that via the change of variable from $t_2 \in [t_1,+\infty[$ to
\begin{eqnarray}
u \equiv \frac{(x_1-y_1 )^2 }{4(t_2-t_1)} \in [0,+\infty[
\label{ubrown}
\end{eqnarray}
the distribution $\gamma(t_2 \vert x_1,y_1,t_1) $ of Eq. \ref{gammatbrown} 
translates into the Gamma distribution of index $1/2$ for the appropriate rescaled variable $u$ of Eq. \ref{ubrown}
\begin{eqnarray}
\phi(u ) \equiv \frac{1}{\sqrt{\pi u} }      e^{- u} 
\label{gammatbrownnorma}
\end{eqnarray}
which is normalized on $u \in [0,+\infty[$.
The probability 
of forever-survival of Eq. \ref{gammatnorma}
vanishes 
\begin{eqnarray}
S(\infty \vert x_1,y_1) = 1- \int_{t_1}^{+\infty} dt_2 \gamma(t_2 \vert x_1,y_1,t_1) =1- \int_{0}^{+\infty} du \phi(u) =0
\label{foreversurvivalbrown}
\end{eqnarray}
while the probability $S(T \vert x_1,y_1,t_1) $ to be surviving at time $T$ reads
\begin{eqnarray}
S(T \vert x_1,y_1,t_1) && = 1- \int_{t_1}^{T} dt_2 \gamma(t_2 \vert x_1,y_1,t_1) 
=1- \int_{\frac{(x_1-y_1 )^2 }{4(T-t_1)}}^{+\infty} du \phi(u) = \erf \left( \frac{x_1 - y_1}{2 \sqrt{T-t_1}} \right)
\label{survivalbrown}
\end{eqnarray}
where $\erf(x)$ is the Error function.


\subsection{ Conditioned process $[X^*(t);Y^*(t)]$ with respect to the finite horizon $T$ }

For the two Brownian motions starting at the positions $x_0$ and $y_0$ at time $t=0$ :

 (i) the probability to be surviving at the positions $x$ and $y$ at time $T$ is given by Eq. \ref{pdetbrown}
 \begin{eqnarray}
P(x,y,T \vert x_0,y_0,0) 
&&  =  \frac{1}{2 \pi T} \left[ 
 e^{- \frac{(x-x_0 )^2+(y-y_0 )^2 }{2T}} 
-  e^{- \frac{(x-y_0 )^2+(y-x_0 )^2 }{2T}} 
\right] 
\nonumber \\
&&  =  \frac{1}{ \pi T} e^{ \frac{(x+y)(x_0+y_0)  }{2T}
 - \frac{x^2+x_0^2+y^2+y_0^2 }{2T}} 
\sinh \left(\frac{(x-y)(x_0-y_0)  }{2T} \right)
\label{pdetbrown0}
\end{eqnarray}

 (ii) the probability to have been annihilated at position $z_a$ at the time  $T_a \in ]0,T]$ 
 is given by Eq. \ref{gammaztbrown}
 \begin{eqnarray}
\gamma(z_a,T_a \vert x_0,y_0,0) 
 =
 \frac{ (x_0-y_0) }{2 \pi T_a^2 }  
 e^{- \frac{(z_a-x_0 )^2+(z_a-y_0 )^2 }{2T_a}} 
\label{gammaztbrown0}
\end{eqnarray}

As explained around Eqs \ref{survivalTstar} and \ref{deadTstar},
we now wish to impose to the conditioned process the following properties instead :

(i) another probability $P^*(x,y,T )$ to be surviving at the positions $x$ and $y$ at time $T$;

 (ii) another probability $\gamma^*(z_a, T_a ) $ to have been annihilated at position $z_a$ at the time  $T_a $.

The normalizations of Eqs \ref{survivalTstar} and \ref{deadTstar}
involve the conditioned survival probability $S^*(T ) $ at the time $T$
\begin{eqnarray}
S^*(T ) =  \int_{-\infty}^{+\infty}  dx   \int_{-\infty}^{+\infty}  dy \theta(x-y)P^*(x,y,T )
= 1-    \int_{0}^T dT_a \int_{-\infty}^{+\infty} dz_a \gamma^*(z_a,T_a ) 
\label{normatstarBrown}
\end{eqnarray}

The ratio of the annihilation distributions computed using Eq. \ref{gammaztbrown},
\begin{eqnarray}
\frac{\gamma(z_a,T_a \vert x,y,t)}{\gamma(z_a,T_a\vert x_0,y_0,0)}
 = \frac{ 
 \frac{ (x-y) }{2 \pi (T_a-t)^2 }  
 e^{- \frac{(z_a-x )^2+(z_a-y )^2 }{2(T_a-t)}} }
 { \frac{ (x_0-y_0) }{2 \pi T_a^2 }  
 e^{- \frac{(z_a-x_0 )^2+(z_a-y_0 )^2 }{2T_a}}}
 = \left( \frac{T_a}{T_a-t} \right)^2\left( \frac{x-y}{x_0-y_0} \right)
 e^{\frac{(z_a-x_0 )^2+(z_a-y_0 )^2 }{2T_a}- \frac{(z_a-x )^2+(z_a-y )^2 }{2(T_a-t)}}
\label{ratioexpli}
\end{eqnarray}
and the ratio of the propagators computed using Eq. \ref{pdetbrown}
\begin{eqnarray}
&&  \frac{ P(x_T,y_T,T \vert x,y,t) }{P(x_T,y_T,T \vert x_0,y_0,0) }
 = \frac{  \frac{1}{ \pi (T-t)} e^{ \frac{(x_T+y_T)(x+y)  }{2(T-t)}
 - \frac{x_T^2+x^2+y_T^2+y^2 }{2(T-t)}} 
\sinh \left(\frac{(x_T-y_T)(x-y)  }{2(T-t)} \right) }
{  \frac{1}{ \pi T} e^{ \frac{(x_T+y_T)(x_0+y_0)  }{2T}
 - \frac{x_T^2+x_0^2+y_T^2+y_0^2 }{2T}} 
\sinh \left(\frac{(x_T-y_T)(x_0-y_0)  }{2T} \right) }
\nonumber \\
&& 
= \frac{T}{T-t} 
e^{ - \frac{(x_T+y_T)(x_0+y_0)  }{2T}
 + \frac{x_T^2+x_0^2+y_T^2+y_0^2 }{2T} + \frac{(x_T+y_T)(x+y)  }{2(T-t)}
 - \frac{x_T^2+x^2+y_T^2+y^2 }{2(T-t)}} 
 \frac{   \sinh \left(\frac{(x_T-y_T)(x-y)  }{2(T-t)} \right) }
{  \sinh \left(\frac{(x_T-y_T)(x_0-y_0)  }{2T} \right) }
\label{ratioexpliP}
\end{eqnarray}
can be plugged 
into Eq. \ref{Qdef}
to obtain 
\begin{eqnarray}
&& Q_T(x,y,t)   = 
\left( \frac{x-y}{x_0-y_0} \right)
 \int_t^{T} dT_a \int_{-\infty}^{+\infty} dz_a \gamma^*( z_a,T_a)
 \left( \frac{T_a}{T_a-t} \right)^2
 e^{\frac{(z_a-x_0 )^2+(z_a-y_0 )^2 }{2T_a}- \frac{(z_a-x )^2+(z_a-y )^2 }{2(T_a-t)}}
  \label{Qbrown}
\\ &&
 +  \frac{T}{T-t} \int_{-\infty}^{+\infty}  dx_T   \int_{-\infty}^{x_T}  dy_T
 P^*(x_T,y_T,T )
e^{ - \frac{(x_T+y_T)(x_0+y_0)  }{2T}
 + \frac{x_T^2+x_0^2+y_T^2+y_0^2 }{2T} + \frac{(x_T+y_T)(x+y)  }{2(T-t)}
 - \frac{x_T^2+x^2+y_T^2+y^2 }{2(T-t)}} 
 \frac{   \sinh \left(\frac{(x_T-y_T)(x-y)  }{2(T-t)} \right) }
{  \sinh \left(\frac{(x_T-y_T)(x_0-y_0)  }{2T} \right) }
 \nonumber 
\end{eqnarray}
and the corresponding conditioned drift of Eq. \ref{driftdoob}
\begin{eqnarray}
\mu^*_X (x,y,t) && = \partial_x \ln Q_T(x,y,t) 
\nonumber \\
\mu^*_Y (x,y,t) && = \partial_y \ln Q_T(x,y,t) 
\label{driftdoobbrown}
\end{eqnarray}
Let us now describe some simple examples.


\subsection{  Example : conditioning the two Brownian motions towards the annihilation at position $z^*$ at time $T^*$ }

When the annihilation-time $T_a$ takes the single value $T^* $
and when the annihilation-position $z_a$ takes the single value $z^*$
\begin{eqnarray}
\gamma^*( z_a,T_a)=\delta(z_a-z^* ) \delta(T_a-T^* ) 
 \label{gammaDead1delta}
\end{eqnarray}
the function of Eq. \ref{Qbrown} for $t \in [0,T^*[$
\begin{eqnarray}
Q_T(x,y,t)   =  
\left( \frac{x-y}{x_0-y_0} \right)
 \left( \frac{T^*}{T^*-t} \right)^2
 e^{\frac{(z^*-x_0 )^2+(z^*-y_0 )^2 }{2T^*}- \frac{(z^*-x )^2+(z^*-y )^2 }{2(T^*-t)}}
 \label{Qtimebrown}
\end{eqnarray}
leads to the conditioned drift of Eq. \ref{driftdoobbrown} for $t \in [0,T^*[$ 
\begin{eqnarray}
\mu^*_X (x,y,t) && =  \partial_x \ln Q_T(x,y,t) = \frac{1}{x-y} +  \frac{z^*-x}{T^*-t} 
 \nonumber \\
\mu^*_Y (x,y,t) && =  \partial_y \ln Q_T(x,y,t) = - \frac{1}{x-y} +  \frac{z^*-y}{T^*-t} 
\label{driftdoobbrown1delta}
\end{eqnarray}


\begin{figure}[h]
\centering
\includegraphics[width=4.2in,height=3.2in]{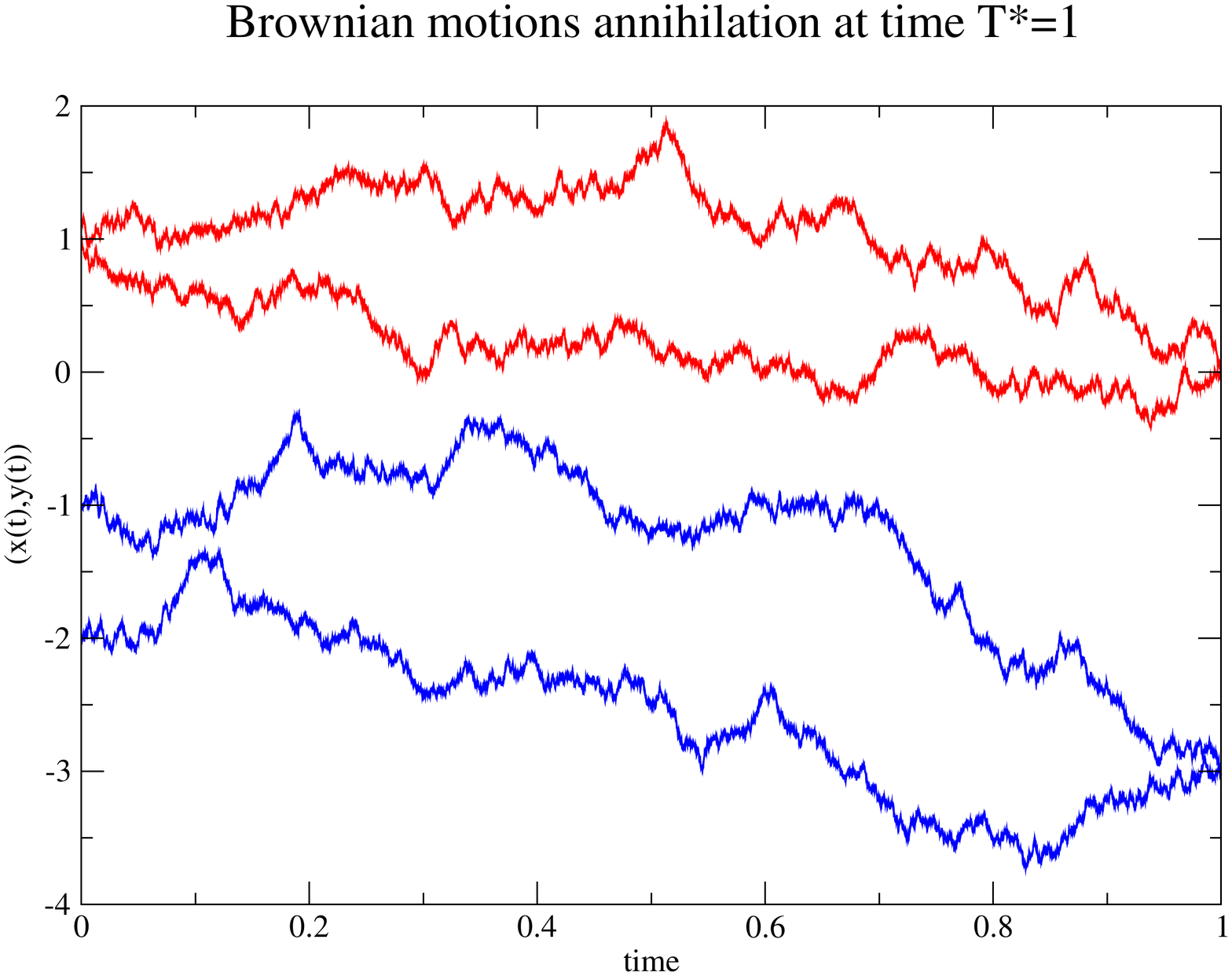}
\setlength{\abovecaptionskip}{15pt}  
\caption{A sample of diffusions for the drift given by Eq. \ref{driftdoobbrown1delta}. Each color corresponds to the realization of one  process. Red lines: the processes start at $x_0 =1.1$ and $y_0 = 1$ and both terminate at $0$ at time $T^*=1$. Blue lines: the processes start at $x_0 = -1$ and $y_0 = -2$ and terminate at $-3$ at time $T^*=1$. The time step used in the discretization is $dt = 10^{-4}$.}
\end{figure}


\subsection{ Example : conditioning towards two Brownian bridges at time $T$ without meeting }

When the two Brownian motions are constrained to be surviving at the given positions $(x_*,y_*)$ at time $T$
\begin{eqnarray}
P^*(x_T,y_T,T )=\delta(x_T-x_* ) \delta(y_T-y_* ) 
 \label{gammaDead2delta}
\end{eqnarray}
the function $Q_T(x,y,t)$ of Eq. \ref{Qbrown} 
\begin{eqnarray}
Q_T(x,y,t) 
= \frac{T}{T-t} 
e^{ - \frac{(x_*+y_*)(x_0+y_0)  }{2T}
 + \frac{x_*^2+x_0^2+y_*^2+y_0^2 }{2T} + \frac{(x_*+y_*)(x+y)  }{2(T-t)}
 - \frac{x_*^2+x^2+y_*^2+y^2 }{2(T-t)}} 
 \frac{   \sinh \left(\frac{(x_*-y_*)(x-y)  }{2(T-t)} \right) }
{  \sinh \left(\frac{(x_*-y_*)(x_0-y_0)  }{2T} \right) }
 \label{Qbrownspace2delta}
\end{eqnarray}
leads to the conditioned drift of Eq. \ref{driftdoobbrown}
\begin{eqnarray}
\mu^*_X (x,y,t) && = \partial_x \ln Q_T(x,y,t) = - \frac{x }{(T-t)}
 +  \frac{(x_*+y_*)  }{2(T-t)}
 +  \frac{(x_*-y_*)  }{2(T-t)}  \coth \left(\frac{(x_*-y_*)(x-y)  }{2(T-t)} \right) 
 \nonumber \\
\mu^*_Y (x,y,t) && = \partial_y \ln Q_T(x,y,t) =- \frac{y }{(T-t)}
+  \frac{(x_*+y_*)  }{2(T-t)}
  -  \frac{(x_*-y_*)  }{2(T-t)}  \coth \left(\frac{(x_*-y_*)(x-y)  }{2(T-t)} \right) 
\label{driftdoobbrown2deltaspace}
\end{eqnarray}

\begin{figure}[h]
\centering
\includegraphics[width=4.2in,height=3.2in]{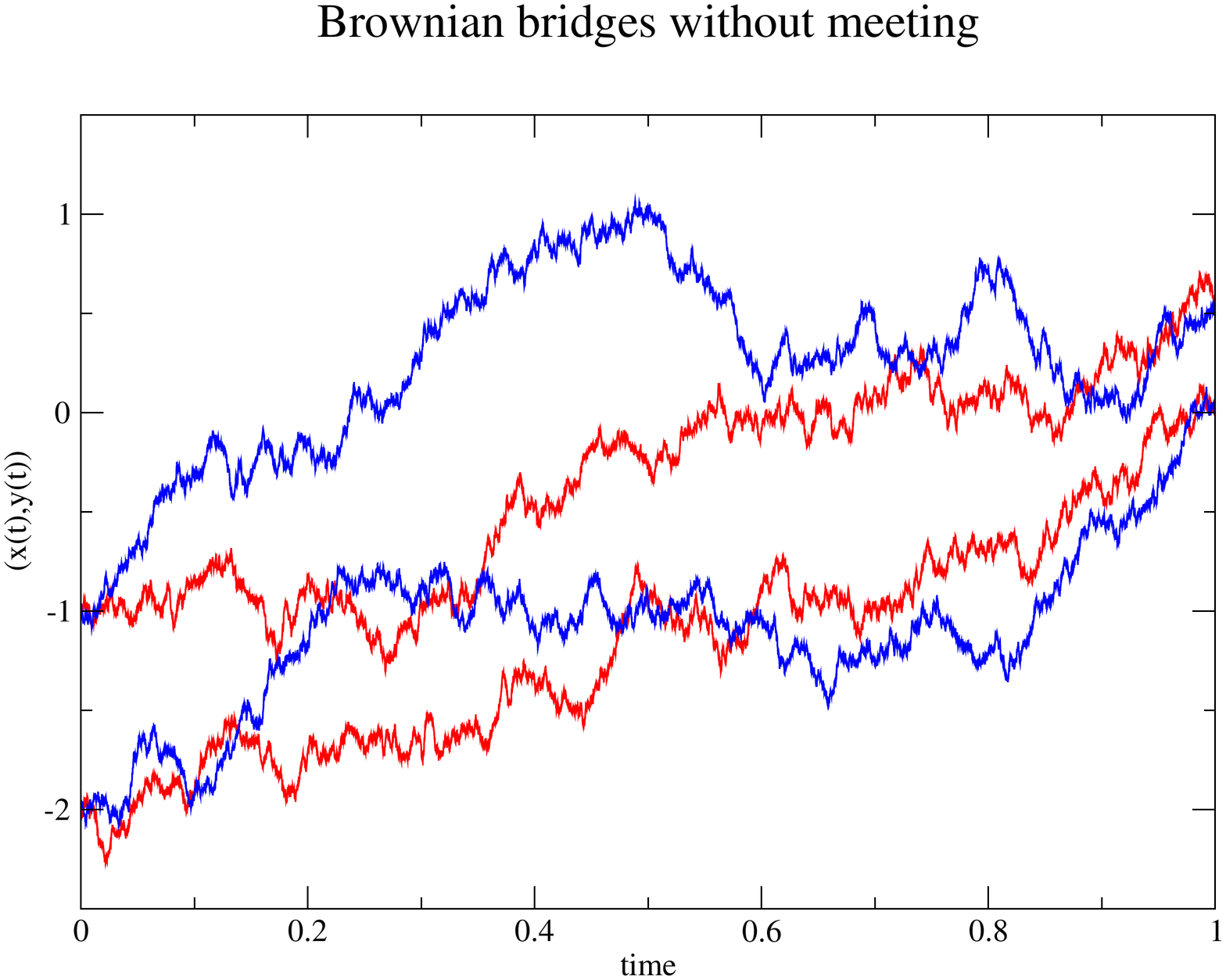}
\setlength{\abovecaptionskip}{15pt}  
\caption{A sample of diffusions for the drift given by Eq. \ref{driftdoobbrown2deltaspace}. Each color corresponds to the realization of one  process. Both processes start at $x_0 = -1$ and $y_0 = -2$ and terminate at $x_* = 1$ and $y_* = 0$ at time $T=1$ without meeting. The time step used in the discretization is $dt = 10^{-4}$.}
\end{figure}


\subsection{ Example : conditioning the two Brownian motions towards full survival without meeting }

\subsubsection{ Conditioning towards full survival $S^*(T)=1$ for the finite horizon $T$}

The conditioning towards full survival $S^*(T)=1$ for finite horizon $T$
involves the function of Eq. \ref{Qsurvivingalone}
that can be computed from the survival probability of Eq. \ref{survivalbrown}
\begin{eqnarray}
Q_T^{[  S^*(T)=1]}(x,y,t)   = 
 \frac{S(T \vert x,y,t)  }{ S(T \vert x_0,y_0,0)} 
  = \frac{\erf \left( \frac{x - y}{2 \sqrt{T-t}} \right)}{\erf \left( \frac{x_0 - y_0}{2 \sqrt{T}} \right)}
 \label{Qspacesurvivalbrown}
\end{eqnarray}
The corresponding conditioned drift of Eq. \ref{driftdoobbrown} reads
\begin{eqnarray}
\mu^*_X (x,y,t) && = \partial_x \ln Q_T^{[  S^*(T)=1]}(x,y,t)  = \frac{e^{-\frac{(x-y )^2 }{4(T-t)}} }{\sqrt{T-t} \erf \left( \frac{x - y}{2 \sqrt{T-t}} \right)}
 \nonumber \\
\mu^*_Y (x,y,t) && = \partial_y \ln Q_T^{[  S^*(T)=1]}(x,y,t)  = -\frac{e^{-\frac{(x-y )^2 }{4(T-t)}} }{\sqrt{T-t} \erf \left( \frac{x - y}{2 \sqrt{T-t}} \right)}
\label{driftdoobbrownsurviving}
\end{eqnarray}


\subsubsection{ Conditioning towards full survival $S^*(\infty)=1$ for the infinite horizon $T=+\infty$ }

Using the behavior near the origin $x \to 0$ of the Error function, $\erf (x) \simeq 2 x/\sqrt{\pi} $, the limit $T \to +\infty$ of the function $Q_T^{[  S^*(T)=1]}(x,y,t) $ of Eq. \ref{Qspacesurvivalbrown}
reads
\begin{eqnarray}
Q_{\infty}^{[  S^*(\infty)=1]}(x,y,t)
 && 
=  \lim_{T \to +\infty}  \frac{\erf \left( \frac{x - y}{2 \sqrt{T-t}} \right)}{\erf \left( \frac{x_0 - y_0}{2 \sqrt{T}} \right)} = \frac{x-y}{x_0-y_0}
 \label{Qspacesurvivalinfinitybrown}
\end{eqnarray}
So the corresponding conditioned drift is time-independent and reduces to 
\begin{eqnarray}
\mu^*_X (x,y) && = \partial_x \ln Q_{\infty}^{[  S^*(\infty)=1]}(x,y,t) = \frac{1}{x-y}
\nonumber \\
\mu^*_Y (x,y) && = \partial_y \ln Q_{\infty}^{[  S^*(\infty)=1]}(x,y,t) = - \frac{1}{x-y}
\label{driftspacesurvivalinfinitybrown}
\end{eqnarray}
This drift that prevents the meeting of the two particles
is reminiscent of the well-known Bessel drift when a single Brownian motion is conditioned to stay positive.

\begin{figure}[h]
\centering
\includegraphics[width=4.2in,height=3.2in]{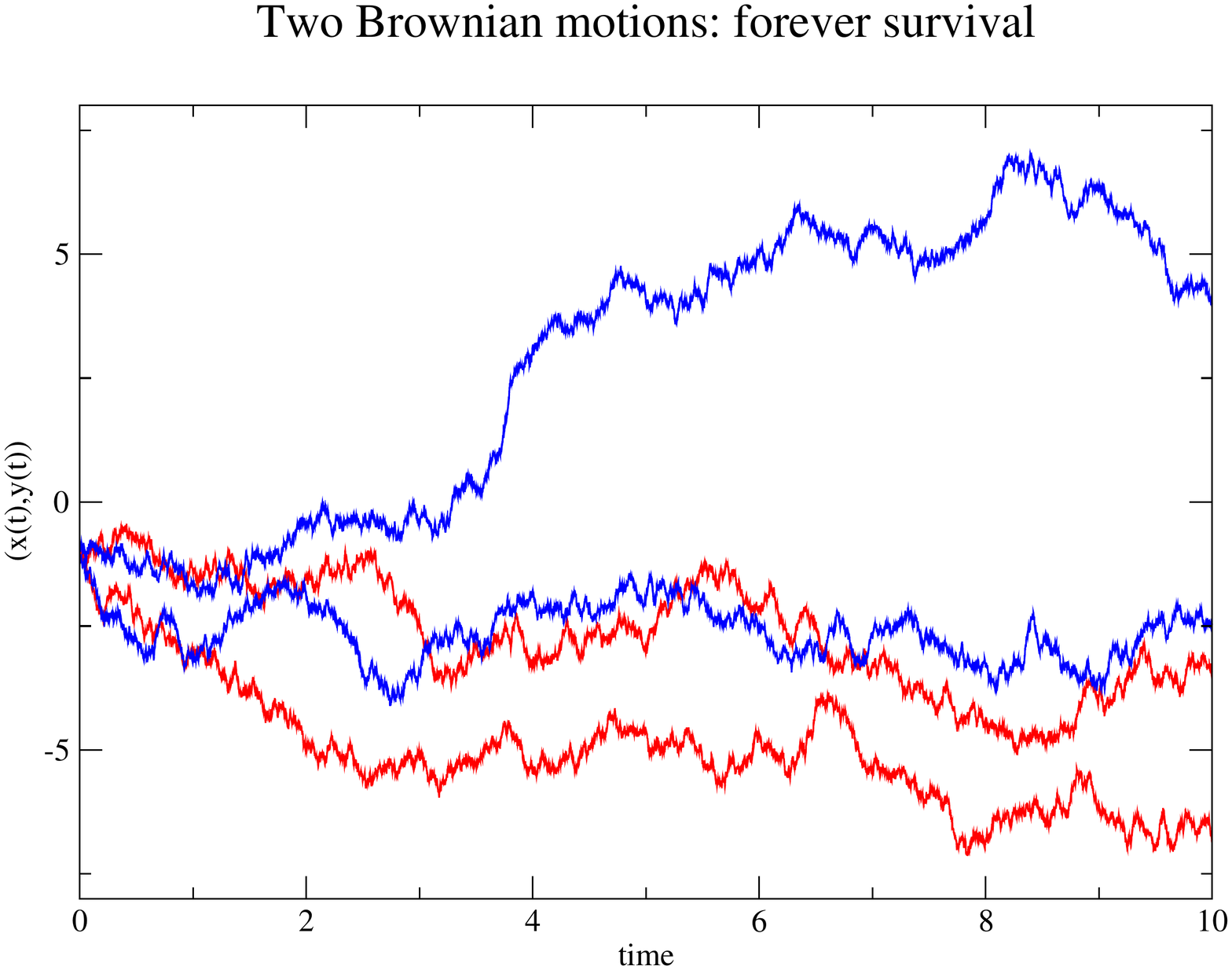}
\setlength{\abovecaptionskip}{15pt}  
\caption{A sample of diffusions for the drift given by Eq. \ref{driftspacesurvivalinfinitybrown}. Each color corresponds to the realization of one  process. Both processes start at $x_0 = -1$ and $y_0 = -1.1$ and live forever without meeting. The time step used in the discretization is $dt = 10^{-4}$.}
\end{figure}



\section{ Application to the conditioning of two Ornstein-Uhlenbeck processes }

\label{sec_OU}

In this section, we consider the case
where the unconditioned process $[X(t);Y(t)]$ corresponds to two independent
Ornstein-Uhlenbeck processes
 that annihilate upon meeting.
 

\subsection{ Unconditioned process $[X(t);Y(t)]$ : two Ornstein-Uhlenbeck processes that annihilate upon meeting }

When the single diffusion of Eq. \ref{ito} corresponds to the linear restoring drift towards the origin $x=0$
with the parameter $k>0$
\begin{eqnarray}
\mu(x)=- k x
\label{muOU}
\end{eqnarray} 
 and to the diffusion coefficient $D(x)=1/2$,
the 1-particle Ornstein-Uhlenbeck propagator  
\begin{eqnarray}
p(x_2,t_2 \vert x_1,t_1) 
=\sqrt{  \frac{k }{ \pi \left[ 1- e^{-2 k (t_2-t_1)} \right] } }  
e^{- k \frac{\left[ x_2-x_1 e^{- k (t_2-t_1)}  \right]^2}{\left[ 1- e^{-2 k (t_2-t_1)} \right]}} 
\label{gaussOU}
\end{eqnarray}
describes the convergence towards the Boltzmann equilibrium $p_{eq}(x_2) $ in a quadratic potential
\begin{eqnarray}
p(x_2,t_2 \vert x_1,t_1) 
\opsimeq_{(t_2-t_1)  \to +\infty} \sqrt{  \frac{k }{ \pi } }  e^{- k  x_2^2} \equiv p_{eq}(x_2)
\label{gaussOUeq}
\end{eqnarray}

The Karlin-McGregor determinant of Eq. \ref{pdet} 
\begin{eqnarray}
&& P(x_2,y_2,t_2 \vert x_1,y_1,t_1) 
\nonumber \\
&& = \frac{k }{ \pi \left[ 1- e^{-2 k (t_2-t_1)} \right] }  \left[ 
 e^{- k \frac{\left[ x_2-x_1 e^{- k (t_2-t_1)}  \right]^2
 +\left[ y_2-y_1 e^{- k (t_2-t_1)}  \right]^2 }{\left[ 1- e^{-2 k (t_2-t_1)} \right]}} 
-  e^{- k \frac{\left[ x_2-y_1 e^{- k (t_2-t_1)}  \right]^2
+\left[ y_2-x_1 e^{- k (t_2-t_1)}  \right]^2 }{\left[ 1- e^{-2 k (t_2-t_1)} \right]}} 
\right]  
\nonumber \\
&& = \frac{k e^{- k \frac{(x_2^2+ y_2^2)+(x_1^2 +y_1^2)  e^{- 2k (t_2-t_1)}   }{\left[ 1- e^{-2 k (t_2-t_1)} \right]}} }{ \pi \left[ 1- e^{-2 k (t_2-t_1)} \right] }  
\left[ 
 e^{ k \frac{  (x_2x_1+y_2 y_1)      }{ \sinh \left[ k (t_2-t_1) \right] } }
-   e^{ k \frac{  (x_2y_1+y_2 x_1)      }{ \sinh \left[ k (t_2-t_1) \right] } }
\right]
\nonumber \\
&& = \frac{k 
e^{k \frac{  (x_2+y_2 ) (x_1+y_1 )  -(x_2^2+ y_2^2)e^{ k (t_2-t_1)}-(x_1^2 +y_1^2)  e^{- k (t_2-t_1)}
    }{2 \sinh \left[ k (t_2-t_1) \right] }} }
{ \pi e^{- k (t_2-t_1)}  \sinh \left[ k (t_2-t_1) \right] }  
 \sinh \left[ k \frac{  (x_2-y_2 ) (x_1-y_1 )      }{2 \sinh \left[ k (t_2-t_1) \right] }  \right]
\label{pdetOU}
\end{eqnarray}
allows to compute
the probability $\gamma(z_2,t_2 \vert x_1,y_1,t_1) $ 
of annihilation at position $z_2$ at time $t_2$ of Eq. \ref{gammazt} 
\begin{eqnarray}
\gamma(z_2,t_2 \vert x_1,y_1,t_1) 
&& = \frac{1}{2} \left[ 
\left(    \partial_{x_2} -\partial_{y_2} \right)
\left[  P(x_2,y_2,t_2 \vert x_1,y_1,t_1)  \right]
\right] \vert_{x_2=z_2;y_2=z_2}
\nonumber \\
&& =
 \frac{2 k^2 (x_1-y_1) e^{- k (t_2-t_1)}}{ \pi \left[ 1- e^{-2 k (t_2-t_1)} \right]^2 }  
 e^{- k \frac{\left[ z_2-x_1 e^{- k (t_2-t_1)}  \right]^2
 +\left[ z_2-y_1 e^{- k (t_2-t_1)}  \right]^2 }{\left[ 1- e^{-2 k (t_2-t_1)} \right]}} 
\label{gammaztOU}
\end{eqnarray}

The integration over the position $z_2$ gives the probability of annihilation at time $t_2$
of Eq. \ref{survivalderi}
\begin{eqnarray}
\gamma(t_2 \vert x_1,y_1,t_1)
&& =  \int_{-\infty}^{+\infty}  dz_2   \gamma(z_2,t_2 \vert x_1,y_1,t_1) 
 =  \frac{2 k^2 (x_1-y_1) e^{- k (t_2-t_1)}}{ \pi \left[ 1- e^{-2 k (t_2-t_1)} \right]^2 }  
    \int_{-\infty}^{+\infty}  dz_2    e^{- k \frac{\left[ z_2-x_1 e^{- k (t_2-t_1)}  \right]^2
 +\left[ z_2-y_1 e^{- k (t_2-t_1)}  \right]^2 }{\left[ 1- e^{-2 k (t_2-t_1)} \right]}} 
\nonumber \\
&& =  \frac{ \sqrt{2} ( k)^{\frac{3}{2}} (x_1-y_1) e^{- k (t_2-t_1)}}
{ \sqrt{ \pi}  \left[ 1- e^{-2 k (t_2-t_1)} \right]^{\frac{3}{2}} }  
      e^{-  \frac{ k (x_1- y_1)^2 }{2 \left[  e^{2 k (t_2-t_1)} -1 \right]}} 
\label{gammatOU}
\end{eqnarray}
This means that via the change of variable from $t_2 \in [t_1,+\infty[$ to
\begin{eqnarray}
v \equiv \frac{ k (x_1- y_1)^2 }{2 \left[  e^{2 k (t_2-t_1)} -1 \right]} \in [0,+\infty[
\label{vOU}
\end{eqnarray}
the distribution $\gamma(t_2 \vert x_1,y_1,t_1) $ of Eq. \ref{gammatOU} 
translates into the Gamma distribution of index $1/2$ for the appropriate rescaled variable $v$ of Eq. \ref{vOU}
\begin{eqnarray}
\phi(v ) \equiv \frac{1}{\sqrt{\pi v} }      e^{- v} 
\label{gammatOUnorma}
\end{eqnarray}
which is normalized on $v \in [0,+\infty[$.
The probability 
of forever-survival of Eq. \ref{gammatnorma}
vanishes 
\begin{eqnarray}
S(\infty \vert x_1,y_1) = 1- \int_{t_1}^{+\infty} dt_2 \gamma(t_2 \vert x_1,y_1,t_1) =1- \int_{0}^{+\infty} dv \phi(v) =0
\label{foreversurvivalOU}
\end{eqnarray}
while the probability $S(T \vert x_1,y_1,t_1) $ to be surviving at time $T$ reads
\begin{eqnarray}
S(T \vert x_1,y_1,t_1) && = 1- \int_{t_1}^{T} dt_2 \gamma(t_2 \vert x_1,y_1,t_1) 
=1- \int_{\frac{ k (x_1- y_1)^2 }{2 \left[  e^{2 k (T-t_1)} -1 \right]}}^{+\infty} dv \phi(v) 
= \int_0^{\frac{ k (x_1- y_1)^2 }{2 \left[  e^{2 k (T-t_1)} -1 \right]}} dv \phi(v)
\nonumber \\
&&
= \erf \left(  \sqrt{\frac{k (x_1 - y_1)^2}{2(e^{2 k (T-t_1)} -1)} }\right) 
\label{survivalOU}
\end{eqnarray}


\subsection{ Conditioned process $[X^*(t);Y^*(t)]$ with respect to the finite horizon $T$ }

For the two Ornstein-Uhlenbeck processes starting at the positions $x_0$ and $y_0$ at time $t=0$ :

 (i) the probability to be surviving at positions the positions $x$ and $y$ at time $T$ is given by Eq. \ref{pdetOU}
 \begin{eqnarray}
P(x,y,T \vert x_0,y_0,0) 
&&  =  \frac{k }{ \pi \left[ 1- e^{-2 k T} \right] }  \left[ 
 e^{- k \frac{\left[ x-x_0 e^{- k T}  \right]^2
 +\left[ y-y_0 e^{- k T}  \right]^2 }{\left[ 1- e^{-2 k T} \right]}} 
-  e^{- k \frac{\left[ x-y_0 e^{- k T}  \right]^2
+\left[ y-x_0 e^{- k T}  \right]^2 }{\left[ 1- e^{-2 k T} \right]}} 
\right]  
\nonumber \\
&& = \frac{2k 
e^{- k \frac{(x^2+ y^2)+(x_0^2 +y_0^2)  e^{- 2k T}   }{\left[ 1- e^{-2 k T} \right]}
+k \frac{  (x+y ) (x_0+y_0 )      }{2 \sinh \left[ k T \right] }} }
{ \pi \left[ 1- e^{-2 k T} \right] }  
 \sinh 
\left[ k \frac{  (x-y ) (x_0-y_0 )      }{2 \sinh \left[ kT \right] } 
 \right]
\label{pdetOU0}
\end{eqnarray}

 (ii) the probability to have been annihilated at position $z_a$ at the time  $T_a \in ]0,T[$ 
 is given by Eq. \ref{gammaztOU}
 \begin{eqnarray}
\gamma(z_a,T_a \vert x_0,y_0,0) 
 = \frac{2 k^2 (x_0-y_0) e^{- k T_a}}{ \pi \left[ 1- e^{-2 k T_a} \right]^2 }  
 e^{- k \frac{\left[ z_a-x_0 e^{- k T_a}  \right]^2
 +\left[ z_a-y_0 e^{- k T_a}  \right]^2 }{\left[ 1- e^{-2 k T_a} \right]}} 
\label{gammaztOU0}
\end{eqnarray}

As explained around Eqs \ref{survivalTstar} and \ref{deadTstar},
we now wish to impose to the conditioned process the following properties instead :

(i) another probability $P^*(x,y,T )$ to be surviving at the positions $x$ and $y$ at time $T$;

 (ii) another probability $\gamma^*(z_a, T_a ) $ to have been annihilated at position $z_a$ at the time  $T_a $.

The normalizations of Eqs \ref{survivalTstar} and \ref{deadTstar}
involve the conditioned survival probability $S^*(T ) $ at the time $T$
\begin{eqnarray}
S^*(T ) =  \int_{-\infty}^{+\infty}  dx   \int_{-\infty}^{+\infty}  dy \theta(x-y)P^*(x,y,T )
= 1-    \int_{0}^T dT_a \int_{-\infty}^{+\infty} dz_a \gamma^*(z_a,T_a ) 
\label{normatstarOU}
\end{eqnarray}

The ratio of the annihilation distributions computed using Eq. \ref{gammaztbrown},
\begin{eqnarray}
\frac{\gamma(z_a,T_a \vert x,y,t)}{\gamma(z_a,T_a\vert x_0,y_0,0)}
&& = \frac{  \frac{2 k^2 (x-y) e^{- k (T_a-t)}}{ \pi \left[ 1- e^{-2 k (T_a-t)} \right]^2 }  
 e^{- k \frac{\left[ z_a-x e^{- k (T_a-t)}  \right]^2
 +\left[ z_a-y e^{- k (T_a-t)}  \right]^2 }{\left[ 1- e^{-2 k (T_a-t)} \right]}}
}
{
\frac{2 k^2 (x_0-y_0) e^{- k T_a }}{ \pi \left[ 1- e^{-2 k T_a} \right]^2 }  
 e^{- k \frac{\left[ z_a-x_0 e^{- k T_a}  \right]^2
 +\left[ z_a-y_0 e^{- k T_a}  \right]^2 }{\left[ 1- e^{-2 k T_a} \right]}}
}
\nonumber \\
&&
= \frac{ (x-y) e^{ k t}  \left[ 1- e^{-2 k T_a} \right]^2}{ (x_0-y_0) \left[ 1- e^{-2 k (T_a-t)} \right]^2 }
e^{k \frac{\left[ z_a-x_0 e^{- k T_a}  \right]^2
 +\left[ z_a-y_0 e^{- k T_a}  \right]^2 }{\left[ 1- e^{-2 k T_a} \right]}
 - k \frac{\left[ z_a-x e^{- k (T_a-t)}  \right]^2
 +\left[ z_a-y e^{- k (T_a-t)}  \right]^2 }{\left[ 1- e^{-2 k (T_a-t)} \right]}}
\label{ratioexpliOU}
\end{eqnarray}
and the ratio of the propagators computed using Eq. \ref{pdetbrown}
\begin{eqnarray}
&&  \frac{ P(x_T,y_T,T \vert x,y,t) }{P(x_T,y_T,T \vert x_0,y_0,0) }
 = \frac{ \frac{k 
e^{k \frac{  (x_T+y_T ) (x+y )  -(x_T^2+ y_T^2)e^{ k (T-t)}-(x^2 +y^2)  e^{- k (T-t)}
    }{2 \sinh \left[ k (T-t) \right] }} }
{ \pi e^{- k (T-t)}  \sinh \left[ k (T-t) \right] }  
 \sinh \left[ k \frac{  (x_T-y_T ) (x-y )      }{2 \sinh \left[ k (T-t) \right] }  \right] }
 { 
  \frac{k 
e^{k \frac{  (x_T+y_T ) (x_0+y_0 )  -(x_T^2+ y_T^2)e^{ k T}-(x_0^2 +y_0^2)  e^{- k T}
    }{2 \sinh \left[ k T \right] }} }
{ \pi e^{- k T}  \sinh \left[ k T \right] }  
 \sinh \left[ k \frac{  (x_T-y_T ) (x_0-y_0 )      }{2 \sinh \left[ k T \right] }  \right]
 }
 \nonumber \\
 &&
 =  e^{
  k \frac{ (x_T^2+ y_T^2)e^{ k T}+(x_0^2 +y_0^2)  e^{- k T} - (x_T+y_T ) (x_0+y_0 )     }{2 \sinh \left[ k T \right] }
+ k \frac{  (x_T+y_T ) (x+y )  -(x_T^2+ y_T^2)e^{ k (T-t)}-(x^2 +y^2)  e^{- k (T-t)}    }{2 \sinh \left[ k (T-t) \right] } 
    }
 \nonumber \\
 && \times 
  \frac{  e^{kt}
 \sinh \left[ k T \right]   \sinh \left[ k \frac{  (x_T-y_T ) (x-y )      }{2 \sinh \left[ k (T-t) \right] }  \right] }
 {   \sinh \left[ k (T-t) \right]   \sinh \left[ k \frac{  (x_T-y_T ) (x_0-y_0 )      }{2 \sinh \left[ k T \right] }  \right]
 }
\label{ratioexpliPOU}
\end{eqnarray}
can be plugged 
into Eq. \ref{Qdef}
to obtain 
\begin{eqnarray}
&& Q_T(x,y,t)  
\nonumber \\
&& = 
 \int_t^{T} dT_a \int_{-\infty}^{+\infty} dz_a 
\gamma^*( z_a,T_a) 
 \frac{ (x-y) e^{ k t}  \left[ 1- e^{-2 k T_a} \right]^2}{ (x_0-y_0) \left[ 1- e^{-2 k (T_a-t)} \right]^2 }
e^{k \frac{\left[ z_a-x_0 e^{- k T_a}  \right]^2
 +\left[ z_a-y_0 e^{- k T_a}  \right]^2 }{\left[ 1- e^{-2 k T_a} \right]}
 - k \frac{\left[ z_a-x e^{- k (T_a-t)}  \right]^2
 +\left[ z_a-y e^{- k (T_a-t)}  \right]^2 }{\left[ 1- e^{-2 k (T_a-t)} \right]}}
 \nonumber \\ &&
 + \int_{-\infty}^{+\infty}  dx_T   \int_{-\infty}^{x_T}  dy_T
    P^*(x_T,y_T,T ) 
 e^{
  k \frac{ (x_T^2+ y_T^2)e^{ k T}+(x_0^2 +y_0^2)  e^{- k T} - (x_T+y_T ) (x_0+y_0 )     }{2 \sinh \left[ k T \right] }
+ k \frac{  (x_T+y_T ) (x+y )  -(x_T^2+ y_T^2)e^{ k (T-t)}-(x^2 +y^2)  e^{- k (T-t)}    }{2 \sinh \left[ k (T-t) \right] } 
    }
 \nonumber \\
 && \times 
  \frac{  e^{kt}
 \sinh \left[ k T \right]   \sinh \left[ k \frac{  (x_T-y_T ) (x-y )      }{2 \sinh \left[ k (T-t) \right] }  \right] }
 {   \sinh \left[ k (T-t) \right]   \sinh \left[ k \frac{  (x_T-y_T ) (x_0-y_0 )      }{2 \sinh \left[ k T \right] }  \right]
 }    
 \label{QOU}
\end{eqnarray}
and the corresponding conditioned drift of Eq. \ref{driftdoob}
\begin{eqnarray}
\mu^*_X (x,y,t) && = - k x + \partial_x \ln Q_T(x,y,t) 
\nonumber \\
\mu^*_Y (x,y,t) && = - k y + \partial_y \ln Q_T(x,y,t) 
\label{driftdoobOU}
\end{eqnarray}
Let us now describe some simple examples.


\subsection{ Example :  conditioning the two Ornstein-Uhlenbeck processes towards annihilation at position $z^*$ at $T^*$ }

When the annihilation-time $T_a$ takes the single value $T^* $
and when the annihilation-position $z_a$ takes the single value $z^*$
\begin{eqnarray}
\gamma^*( z_a,T_a)=\delta(z_a-z^* ) \delta(T_a-T^* ) 
 \label{gammaDead1deltaOU}
\end{eqnarray}
the function of Eq. \ref{QOU} for $t \in [0,T^*[$
\begin{eqnarray}
Q_T(x,y,t)   = 
\frac{ (x-y) e^{ k t} }{(x_0-y_0)}
 \frac{  \left[ 1- e^{-2 k T^*} \right]^2}{  \left[ 1- e^{-2 k (T^*-t)} \right]^2 }
e^{k \frac{\left[ z^*-x_0 e^{- k T^*}  \right]^2
 +\left[ z^*-y_0 e^{- k T^*}  \right]^2 }{\left[ 1- e^{-2 k T^*} \right]}
 - k \frac{\left[ z^*-x e^{- k (T^*-t)}  \right]^2
 +\left[ z^*-y e^{- k (T^*-t)}  \right]^2 }{\left[ 1- e^{-2 k (T^*-t)} \right]}}
 \label{QtimeOU1delta}
\end{eqnarray}
leads to the conditioned drift of Eq. \ref{driftdoobOU} for $t \in [0,T^*[$ 
\begin{eqnarray}
\mu^*_X (x,y,t) && = - k x + \frac{1}{x-y}  + \frac{ k \left[ z^* -x e^{- k (T^*-t)}\right]}{\sinh \left[ k (T^*-t)\right]} 
\nonumber \\
\mu^*_Y (x,y,t) && = - k y - \frac{1}{x-y}  + \frac{ k \left[ z^* -y e^{- k (T^*-t)}\right]}{\sinh \left[ k (T^*-t)\right]}
\label{driftdoobOU1delta}
\end{eqnarray}

\begin{figure}[h]
\centering
\includegraphics[width=4.2in,height=3.2in]{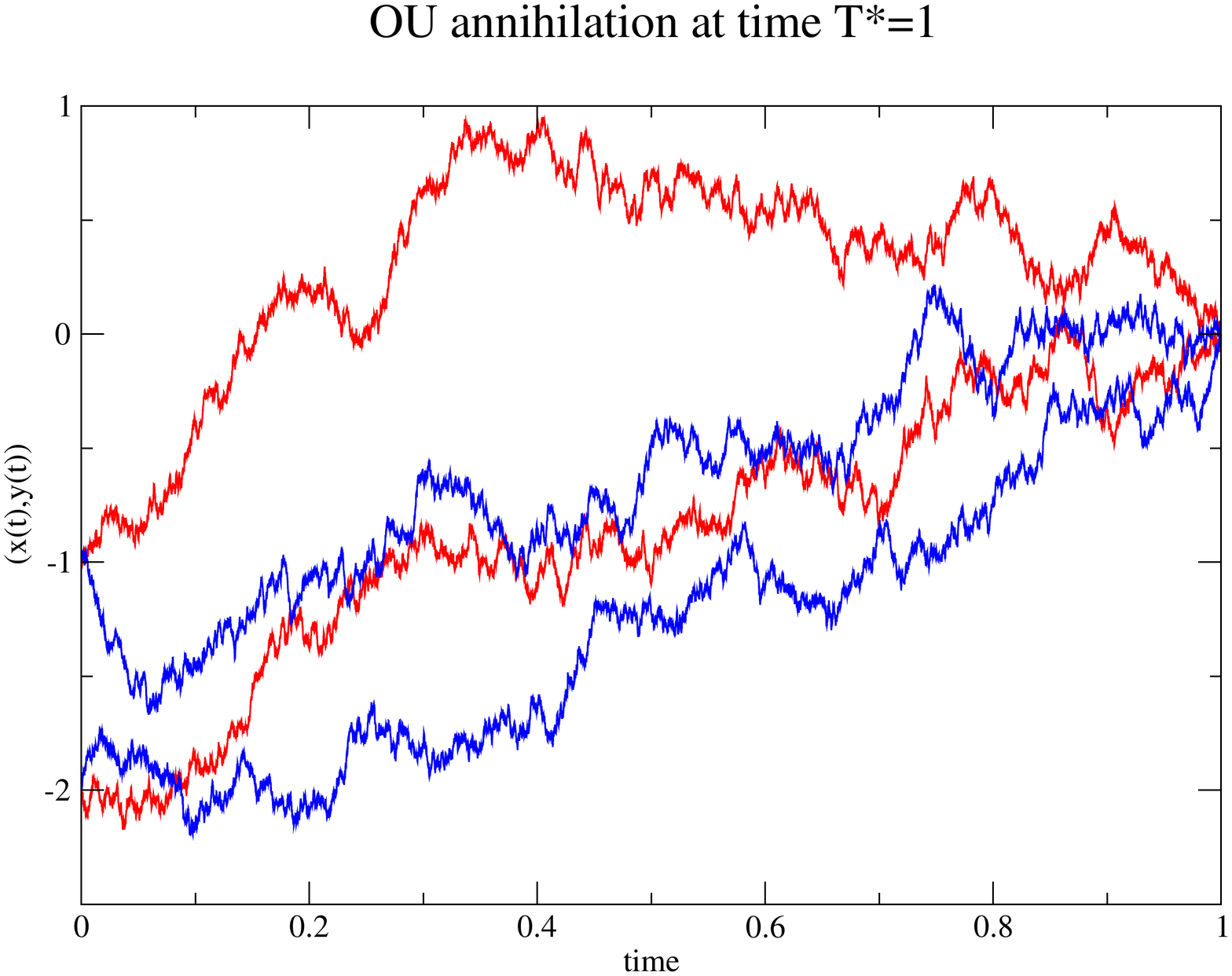}
\setlength{\abovecaptionskip}{15pt}  
\caption{A sample of diffusions for the drift given by Eq. \ref{driftdoobOU1delta} with $k = 1$. Each color corresponds to the realization of one  process. Both processes start at $x_0 = -1$ and $y_0 = -2$ and terminate at $0$ at time $T^*=1$. The time step used in the discretization is $dt = 10^{-4}$.}
\end{figure}


\subsection{ Example : conditioning towards two Ornstein-Uhlenbeck bridges at time $T$ without meeting }

When the two Ornstein-Uhlenbeck processes are constrained to be surviving at the positions $(x_*,y_*)$ at time $T$
\begin{eqnarray}
P^*(x_T,y_T,T )=\delta(x_T-x_* ) \delta(y_T-y_* ) 
 \label{gammaDead2deltaOU}
\end{eqnarray}
the function $Q_T(x,y,t)$ of Eq. \ref{QOU}  
\begin{eqnarray}
Q_T^{[space]}(x,y,t) 
&& =  \frac{  e^{kt} \sinh \left[ k T \right]    }
 {   \sinh \left[ k (T-t) \right]    }    
  e^{
  k \frac{ (x_0^2 +y_0^2)  e^{- k T}   }{2 \sinh \left[ k T \right] }
- k \frac{ (x^2 +y^2)  e^{- k (T-t)}    }{2 \sinh \left[ k (T-t) \right] } 
    }
\nonumber \\
 && \times  e^{
  k \frac{ (x_*^2+ y_*^2)e^{ k T} - (x_*+y_* ) (x_0+y_0 )     }{2 \sinh \left[ k T \right] }
+ k \frac{  (x_*+y_* ) (x+y )  -(x_*^2+ y_*^2)e^{ k (T-t)}    }{2 \sinh \left[ k (T-t) \right] } 
    }
  \frac{    \sinh \left[ k \frac{  (x_*-y_* ) (x-y )      }{2 \sinh \left[ k (T-t) \right] }  \right] }
 {     \sinh \left[ k \frac{  (x_*-y_* ) (x_0-y_0 )      }{2 \sinh \left[ k T \right] }  \right]
 }    
\label{QOUspace2delta}
\end{eqnarray}
leads to the conditioned drift of Eq. \ref{driftdoobOU} 
\begin{eqnarray}
\mu^*_X (x,y,t) && = - k x 
- k \frac{ x   e^{- k (T-t)}    }{ \sinh \left[ k (T-t) \right] } 
+ k \frac{  (x_*+y_* )    }{2 \sinh \left[ k (T-t) \right] }
 + k \frac{  (x_*-y_* )       }{2 \sinh \left[ k (T-t) \right] }
 \coth \left[ k \frac{  (x_*-y_* ) (x-y )      }{2 \sinh \left[ k (T-t) \right] }  \right]
\nonumber \\
\mu^*_Y (x,y,t) && = - k y
- k \frac{y  e^{- k (T-t)}    }{ \sinh \left[ k (T-t) \right] } 
+ k \frac{  (x_*+y_* )    }{2 \sinh \left[ k (T-t) \right] }
 - k \frac{  (x_*-y_* )       }{2 \sinh \left[ k (T-t) \right] }
 \coth \left[ k \frac{  (x_*-y_* ) (x-y )      }{2 \sinh \left[ k (T-t) \right] }  \right]
\label{driftdoobOU2deltaspace}
\end{eqnarray}

\begin{figure}[h]
\centering
\includegraphics[width=4.2in,height=3.2in]{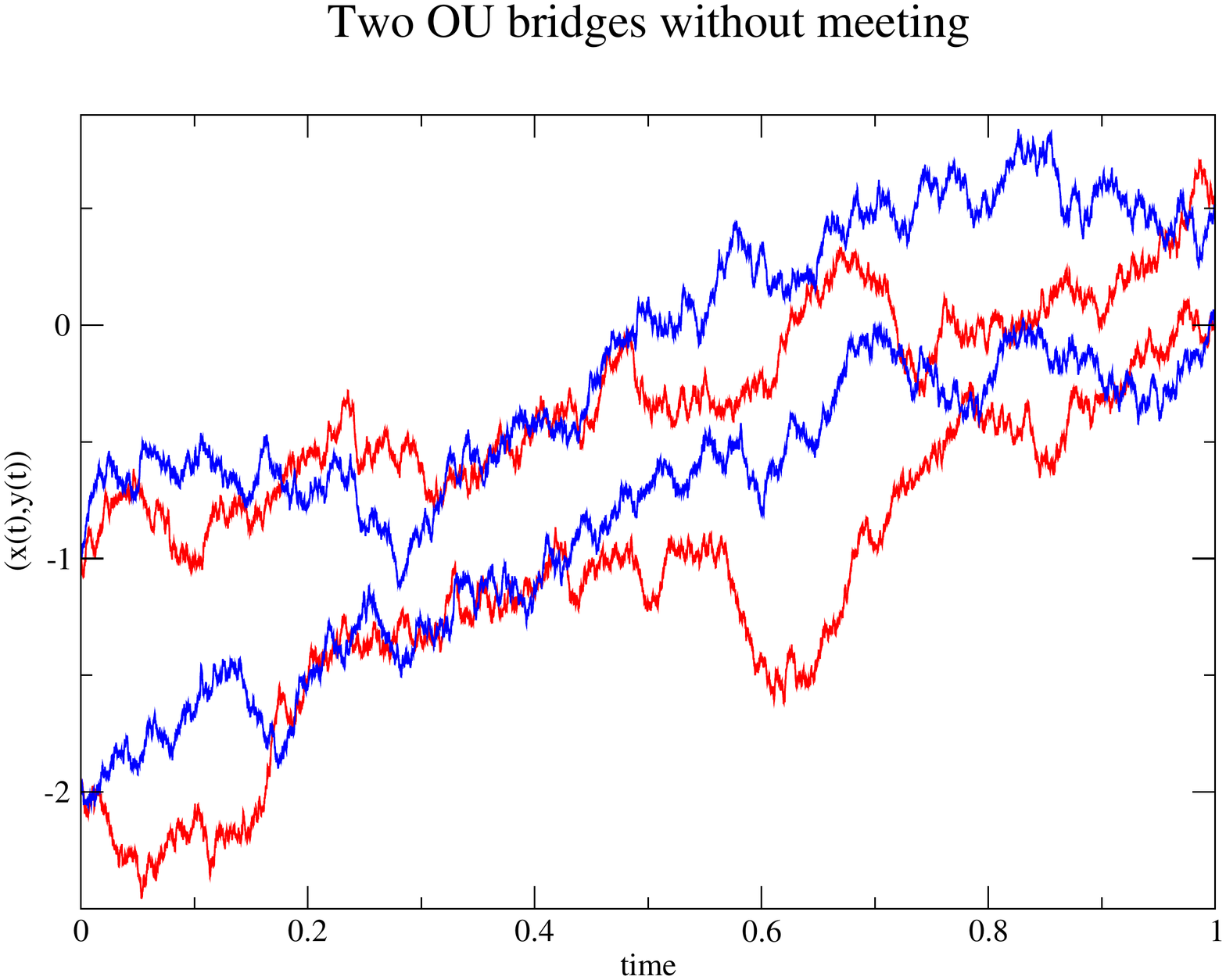}
\setlength{\abovecaptionskip}{15pt}  
\caption{A sample of diffusions for the drift given by Eq. \ref{driftdoobOU2deltaspace} with $k = 1$. Each color corresponds to the realization of one process. Both processes start at $x_0 = -1$ and $y_0 = -2$ and terminate at $x_* = 1$ and $y_* = 0$ at time $T=1$ without meeting. The time step used in the discretization is $dt = 10^{-4}$.}
\end{figure}

\subsection{ Example : conditioning the two Ornstein-Uhlenbeck processes towards full survival without meeting }

\subsubsection{ Conditioning towards full survival $S^*(T)=1$ for the finite horizon $T$}

The conditioning towards full survival $S^*(T)=1$ for finite horizon $T$
involves the function of Eq. \ref{Qsurvivingalone}
that can be computed from the survival probability of Eq. \ref{survivalOU}
\begin{eqnarray}
Q_T^{[  S^*(T)=1]}(x,y,t) 
 = \frac{ S (T \vert x,y,t)  }{ S (T \vert x_0,y_0,0)}
 = \frac{\erf \left(  \sqrt{\frac{k (x - y)^2}{2(e^{2 k (T-t)} -1)} }\right)}
             {\erf \left(  \sqrt{\frac{k (x_0 - y_0)^2}{2(e^{2 k T} -1)} }\right)}
 \label{QspacesurvivalOU}
\end{eqnarray}
The corresponding conditioned drift of Eq. \ref{driftdoobbrown} reads
\begin{eqnarray}
\mu^*_X (x,y,t) && = -kx + \partial_x \ln Q_T(x,y,t) = -kx +  \frac{ \sqrt{\frac{2k}{e^{2 k (T-t)} -1} } e^{- \frac{ k (x- y)^2 }{2 \left[  e^{2 k (T-t)} -1 \right]}} }
{\erf \left(  \sqrt{\frac{k (x - y)^2}{2(e^{2 k (T-t)} -1)} }\right)  }
 \nonumber \\
\mu^*_Y (x,y,t) && = -ky + \partial_y \ln Q_T(x,y,t) =-ky -  \frac{ \sqrt{\frac{2k}{e^{2 k (T-t)} -1} } e^{- \frac{ k (x- y)^2 }{2 \left[  e^{2 k (T-t)} -1 \right]}} }
{\erf \left(  \sqrt{\frac{k (x - y)^2}{2(e^{2 k (T-t)} -1)} }\right)  }
\label{driftdoobOUsurviving}
\end{eqnarray}


\subsubsection{ Conditioning towards full survival $S^*(\infty)=1$ for the infinite horizon $T=+\infty$ }

Since $\erf(x) \simeq_{x \to 0} 2 x/\sqrt{\pi}$, the limit $T \to +\infty$ of the function $Q_T^{[  S^*(T)=1]}(x,y,t)  $ of Eq. \ref{QspacesurvivalOU} reads
\begin{eqnarray}
Q_{\infty}^{[S^*(\infty)=1]}(x,y,t)  
 = \frac{x-y}{x_0-y_0} \ e^{kt} 
  \label{QspacesurvivalinfinityOU}
\end{eqnarray}
So the corresponding conditioned drift is time-independent and reduces to
\begin{eqnarray}
\mu^*_X (x,y) && = -k x + \partial_x \ln Q_{\infty}^{[surviving]}(x,y,t) = -kx  + \frac{1}{x-y} 
\nonumber \\
\mu^*_Y (x,y) && = -k y + \partial_y \ln Q_{\infty}^{[surviving]}(x,y,t) = -ky -  \frac{1}{x-y} 
\label{driftspacesurvivalinfinityOU}
\end{eqnarray}
The supplementary contributions with respect to the initial linear restoring drift are exactly the same as in Eq. \ref{driftspacesurvivalinfinitybrown}: they prevent the meeting of the two particles and are the analog of the Bessel drift.

\begin{figure}[h]
\centering
\includegraphics[width=4.2in,height=3.2in]{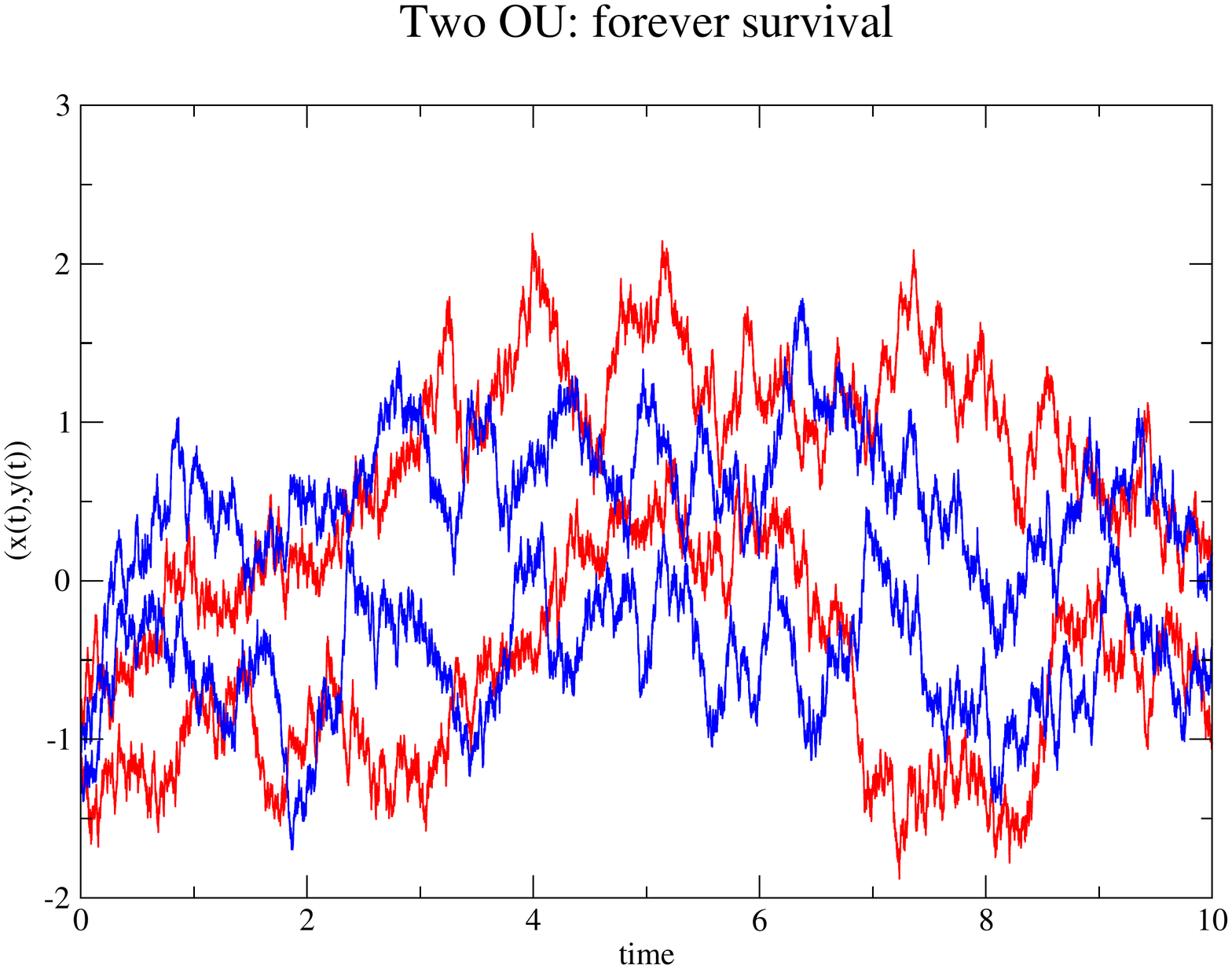}
\setlength{\abovecaptionskip}{15pt}  
\caption{A sample of diffusions for the drift given by Eq. \ref{driftspacesurvivalinfinityOU} with $k = 1$. Each color corresponds to the realization of one  process. Both processes start at $x_0 = -1$ and $y_0 = -1.1$ and live forever without meeting. The time step used in the discretization is $dt = 10^{-4}$.}
\end{figure}

The conditioned dynamics of Eq. \ref{conditionaldyn}
\begin{eqnarray}
  \partial_t P^*(x,y,t)  && =
  -  \partial_x \left[   \mu^*_X (x,y) P^*(x,y,t) -\frac{1}{2}\partial_x  P^*(x,y,t)   \right]
-  \partial_y \left[  \mu^*_Y (x,y)  P^*(x,y,t) -\frac{1}{2}\partial_x  P^*(x,y,t) \right]
\label{conditionaldynOU}
\end{eqnarray}
will converge towards the conditioned equilibrium $P_{eq}^*(x,y) $ without steady currents
\begin{eqnarray}
\partial_x  \ln P_{eq} ^*(x,y)&& = 2  \mu^*_X (x,y) =  - 2 kx  + \frac{2}{x-y} \equiv  - \partial_x V^*(x,y)
\nonumber \\
\partial_y  \ln P_{eq} ^*(x,y)&& = 2  \mu^*_X (x,y) =  - 2 ky  - \frac{2}{x-y} \equiv - \partial_y V^*(x,y)
\label{conditionaldynOUeqzerocurrent}
\end{eqnarray}
that corresponds to the conditioned potential
\begin{eqnarray}
 V^*(x,y) = k (x^2+y^2) -2 \ln (x-y)
\label{conditionaldynOUeqpotential}
\end{eqnarray}
So the conditioning of forever-survival has introduced the logarithmic repulsive interaction between the two particles that are confined in their initial quadratic potential $V(x,y) = k (x^2+y^2) $.
The corresponding equilibrium distribution
\begin{eqnarray}
P_{eq} ^*(x,y)  = \frac{ e^{ - V^*(x,y) } }{ Z } =  \frac{ (x-y)^2 e^{ - k (x^2+y^2) } }{ Z }
\label{conditionaldynOUeq}
\end{eqnarray}
normalized by the partition function
\begin{eqnarray}
Z  = \int_{-\infty}^{+\infty}  dx   \int_{-\infty}^{+\infty}  dy \theta(x-y)  e^{ - V^*(x,y) }
= \int_{-\infty}^{+\infty}  dx   \int_{-\infty}^{x}  dy (x-y)^2 e^{ - k (x^2+y^2) }
\label{ZconditionaldynOUeq}
\end{eqnarray}
corresponds to the simple case $N=2$ of the famous spectral statistics
for the $N$ eigenvalues of random matrices drawn from the
Gaussian Unitary Ensemble (see the books \cite{handbook,vivo,potters}
and references therein).
Accordingly, the conditioned dynamics of Eq. \ref{conditionaldynOU} with the conditioned drift of Eq. \ref{driftspacesurvivalinfinityOU}
corresponds to the Dyson Brownian motion \cite{dyson} for only $N=2$ eigenvalues.


\section{Application to the conditioning of two tanh-drift diffusion processes}

\label{sec_tanh}

\subsection{ Unconditioned process $[X(t);Y(t)]$ : two tanh-drift diffusions that annihilate upon meeting }

When the single diffusion of Eq. \ref{ito} corresponds to the diffusion coefficient $D(x)=1/2$
and to the drift pushing away from the origin $x=0$ \footnote{In the mathematical literature this process is sometimes called hyperbolic Ornstein-Uhlenbeck process \cite{BorodinHyperbolic}, or Bene\v{s} process \cite{Sarkka}. To avoid confusion, we will simply call it the tanh-drift process.}
\begin{eqnarray}
\mu(x)=\alpha \tanh (\alpha x) 
\label{mutanh}
\end{eqnarray} 
of parameter $\alpha>0$ ,
the 1-particle propagator \cite{Borodin,Sarkka} 
\begin{eqnarray}
p(x_2,t_2 \vert x_1,t_1)  && =  \frac{ e^{  - \frac{(x_2-x_1)^2 }{2(t_2-t_1)} - \frac{\alpha^2}{2} (t_2-t_1)} }{  \sqrt{2 \pi (t_2-t_1)}}  \left(\frac{ \cosh( \alpha x_2 ) }{ \cosh( \alpha x_1 ) } \right)
= \frac{ e^{  - \frac{x_2^2+x_1^2 }{2(t_2-t_1)} - \frac{\alpha^2}{2} (t_2-t_1)} }{  \sqrt{2 \pi (t_2-t_1)}}  \left(\frac{ \cosh( \alpha x_2 ) }{ \cosh( \alpha x_1 ) } \right) e^{ \frac{x_2 x_1 }{t_2-t_1} }
\label{tanh}
\end{eqnarray}
leads to the Karlin-McGregor determinant of Eq. \ref{pdet} 
\begin{eqnarray}
&& P(x_2,y_2,t_2 \vert x_1,y_1,t_1) 
  =   \frac{ e^{  - \frac{x_2^2+x_1^2 + y_2^2+y_1^2}{2(t_2-t_1)} - \alpha^2 (t_2-t_1)} }
  {  2 \pi (t_2-t_1)}  \left(\frac{ \cosh( \alpha x_2 ) \cosh( \alpha y_2 )}{ \cosh( \alpha x_1 ) \cosh( \alpha y_1 )} \right) \left[e^{ \frac{x_2 x_1 +y_2 y_1  }{t_2-t_1} } - e^{ \frac{x_2 y_1 +y_2 x_1  }{t_2-t_1} }\right]
  \nonumber \\
&&   =   \frac{ e^{ \frac{(x_2+y_2)(x_1+y_1)  }{2(t_2-t_1)} - \frac{x_2^2+x_1^2 + y_2^2+y_1^2}{2(t_2-t_1)} - \alpha^2 (t_2-t_1)} }
  {   \pi (t_2-t_1)}  \left(\frac{ \cosh( \alpha x_2 ) \cosh( \alpha y_2 )}{ \cosh( \alpha x_1 ) \cosh( \alpha y_1 )} \right) \sinh \left(\frac{(x_2-y_2)(x_1-y_1)  }{2(t_2-t_1)} \right)
  \label{pdettanh}
\end{eqnarray}

The probability $\gamma(z_2,t_2 \vert x_1,y_1,t_1) $ 
of annihilation at position $z_2$ at time $t_2$ of Eq. \ref{gammazt} 
can be computed using Eq. \ref{pdettanh}
\begin{eqnarray}
\gamma(z_2,t_2 \vert x_1,y_1,t_1) 
&& = \frac{1}{2} \left[ 
\left(    \partial_{x_2} -\partial_{y_2} \right)
P(x_2,y_2,t_2 \vert x_1,y_1,t_1) 
\right] \vert_{x_2=z_2;y_2=z_2}
\nonumber \\
&& =  \frac{ (x_1-y_1) e^{ z_2 \frac{(x_1+y_1)  }{(t_2-t_1)} 
- \frac{z_2^2}{(t_2-t_1)}
- \frac{x_1^2 + y_1^2}{2(t_2-t_1)} - \alpha^2 (t_2-t_1)} }
  {  2 \pi (t_2-t_1)^2}  \left(\frac{ \cosh^2( \alpha z_2 ) }{ \cosh( \alpha x_1 ) \cosh( \alpha y_1 )} \right) 
\label{gammazttanh}
\end{eqnarray}

The integration over the position $z_2$ gives the probability of annihilation at time $t_2$
of Eq. \ref{survivalderi}
\begin{eqnarray}
 \gamma(t_2 \vert x_1,y_1,t_1) && =  \int_{-\infty}^{+\infty}  dz_2   \gamma(z_2,t_2 \vert x_1,y_1,t_1)     
\nonumber \\
&& =   
 \frac{ (x_1-y_1) e^{
- \frac{x_1^2 + y_1^2}{2(t_2-t_1)} - \alpha^2 (t_2-t_1)} }
  {  2 \pi (t_2-t_1)^2 \cosh( \alpha x_1 ) \cosh( \alpha y_1 )}
\int_{-\infty}^{+\infty}  dz_2  e^{ z_2 \frac{(x_1+y_1)  }{(t_2-t_1)} 
- \frac{z_2^2}{(t_2-t_1)} }\cosh^2( \alpha z_2 )
  \nonumber \\
&& =   \frac{ (x_1-y_1) e^{- \frac{ (x_1-y_1)^2}{4(t_2-t_1)} } }
  {  4 \sqrt{\pi}  (t_2-t_1)^{\frac{3}{2}} \cosh( \alpha x_1 ) \cosh( \alpha y_1 )}
  \left[ \cosh[ \alpha (x_1+y_1) ] +  e^{- \alpha^2 (t_2-t_1)}\right]
\label{gammattanh}
\end{eqnarray}

The normalization over finite times $t_2 \in [t_1,+\infty[$ 
can be computed using 
the integral for $\alpha \geq 0$ and for $\beta >0$
\begin{eqnarray} 
\int_0^{+\infty} \frac{d\tau }{  \tau^{\frac{3}{2}} }  e^{- \frac{\beta^2}{\tau}-  \alpha^2 \tau}
 = \frac{ \sqrt{ \pi }}{  \beta } e^{ - 2  \alpha \beta   }
\label{integral}
\end{eqnarray}
to obtain 
\begin{eqnarray}
 \int_{t_1}^{+\infty} dt_2 \gamma(t_2 \vert x_1,y_1,t_1) 
&&  =  
 \frac{ (x_1-y_1)  }
  {  4 \sqrt{\pi}   \cosh( \alpha x_1 ) \cosh( \alpha y_1 )}
  \left[ \cosh[ \alpha (x_1+y_1) ] \int_0^{+\infty} \frac{d\tau }{  \tau^{\frac{3}{2}} } e^{- \frac{ (x_1-y_1)^2}{4 \tau} }+  \int_0^{+\infty} \frac{d\tau }{  \tau^{\frac{3}{2}} } e^{- \frac{ (x_1-y_1)^2}{4 \tau}- \alpha^2 \tau}\right]
  \nonumber \\
  && =     
 \frac{ 1  }  {  2  \cosh( \alpha x_1 ) \cosh( \alpha y_1 )}
  \left[ \cosh[ \alpha (x_1+y_1) ] +  e^{ - \alpha (x_1-y_1) } \right]
    \nonumber \\
  && =     
 \frac{ 1  }  {  2  \cosh( \alpha x_1 ) \cosh( \alpha y_1 )}
  \left[ \cosh[ \alpha (x_1+y_1) ] + \cosh[ \alpha (x_1-y_1) ] - \sinh[ \alpha (x_1-y_1) ] \right]
   \nonumber \\
  &&  = 1 - \frac{ \sinh[ \alpha (x_1-y_1) ]   }  {  2  \cosh( \alpha x_1 ) \cosh( \alpha y_1 )}
\label{gammattanhnorma}
\end{eqnarray}
i.e. the probability 
of forever-survival of Eq. \ref{gammatnorma} is finite 
\begin{eqnarray}
S(\infty \vert x_1,y_1) && = 1- \int_{t_1}^{+\infty} dt_2 \gamma(t_2 \vert x_1,y_1,t_1) 
=\frac{ \sinh[ \alpha (x_1-y_1) ]   }  {  2  \cosh( \alpha x_1 ) \cosh( \alpha y_1 )}
\nonumber \\
&& = \frac{ \sinh( \alpha x_1 ) \cosh( \alpha y_1) -  \cosh( \alpha x_1 ) \sinh( \alpha y_1)   }  {  2  \cosh( \alpha x_1 ) \cosh( \alpha y_1 )}
= \frac{\tanh( \alpha x_1 ) - \tanh( \alpha y_1)}{2} 
 \label{foreversurvivaltanh}
\end{eqnarray}

Observe that when $x_1 \gg 0$ and $y_1 \ll 0$, then $S(\infty \vert x_1,y_1)$ is close to $1$, meaning that the two tanh-drift process is almost certain to survive.


\subsection{ Conditioned process $[X^*(t);Y^*(t)]$ with respect to the finite horizon $T$ }

For the two tanh-drift processes starting at the positions $x_0$ and $y_0$ at time $t=0$ :

 (i) the probability to be surviving at the positions $x$ and $y$ at time $T$ is given by Eq. \ref{pdettanh}
 \begin{eqnarray}
P(x,y,T \vert x_0,y_0,0) 
  =  \frac{ e^{ \frac{(x+y)(x_0+y_0)  }{2T} - \frac{x^2+x_0^2 + y^2+y_0^2}{2T} - \alpha^2 T} }
  {   \pi T}  \left(\frac{ \cosh( \alpha x ) \cosh( \alpha y )}{ \cosh( \alpha x_0 ) \cosh( \alpha y_0 )} \right) \sinh \left(\frac{(x-y)(x_0-y_0)  }{2T} \right)
\label{pdettanh0}
\end{eqnarray}

 (ii) the probability to have been annihilated at position $z_a$ at the time  $T_a \in ]0,T]$ 
 is given by Eq. \ref{gammazttanh}
 \begin{eqnarray}
\gamma(z_a,T_a \vert x_0,y_0,0) 
 = \frac{ (x_0-y_0) e^{ z_a \frac{(x_0+y_0)  }{T_a} 
- \frac{z_a^2}{T_a}
- \frac{x_0^2 + y_0^2}{2T_a} - \alpha^2 T_a} }
  {  2 \pi T_a^2}  \left(\frac{ \cosh^2( \alpha z_a ) }{ \cosh( \alpha x_0 ) \cosh( \alpha y_0 )} \right) 
\label{gammazttanh0}
\end{eqnarray}

As explained around Eqs \ref{survivalTstar} and \ref{deadTstar},
we now wish to impose to the conditioned process the following properties instead :

(i) another probability $P^*(x,y,T )$ to be surviving at the positions $x$ and $y$ at time $T$;

 (ii) another probability $\gamma^*(z_a, T_a ) $ to have been annihilated at position $z_a$ at the time  $T_a $.

The normalizations of Eqs \ref{survivalTstar} and \ref{deadTstar}
involve the conditioned survival probability $S^*(T ) $ at the time $T$
\begin{eqnarray}
S^*(T ) =  \int_{-\infty}^{+\infty}  dx   \int_{-\infty}^{+\infty}  dy \theta(x-y)P^*(x,y,T )
= 1-    \int_{0}^T dT_a \int_{-\infty}^{+\infty} dz_a \gamma^*(z_a,T_a ) 
\label{normatstartanh}
\end{eqnarray}

The ratio of the annihilation distributions computed using Eq. \ref{gammazttanh},
\begin{eqnarray}
&& \frac{\gamma(z_a,T_a \vert x,y,t)}{\gamma(z_a,T_a\vert x_0,y_0,0)}
 =
 \frac{\frac{ (x-y) e^{ z_a \frac{(x+y)  }{(T_a-t)} - \frac{z_a^2}{(T_a-t)}- \frac{x^2 + y^2}{2(T_a-t)} - \alpha^2 (T_a-t)} }
  {  2 \pi (T_a-t)^2}  \left(\frac{ \cosh^2( \alpha z_a ) }{ \cosh( \alpha x ) \cosh( \alpha y )} \right)}
 {\frac{ (x_0-y_0) e^{ z_a \frac{(x_0+y_0)  }{T_a} - \frac{z_a^2}{T_a}- \frac{x_0^2 + y_0^2}{2T_a} - \alpha^2 T_a} }
  {  2 \pi T_a^2}  \left(\frac{ \cosh^2( \alpha z_a ) }{ \cosh( \alpha x_0 ) \cosh( \alpha y_0 )} \right)}
  \nonumber \\
  && = \left( \frac{T_a}{T_a-t} \right)^2
   \frac{(x-y)\cosh( \alpha x_0 ) \cosh( \alpha y_0 )}{(x_0-y_0) \cosh( \alpha x ) \cosh( \alpha y )}
 e^{ z_a \frac{(x+y)  }{(T_a-t)}  -z_a \frac{(x_0+y_0)  }{T_a}
+ \frac{z_a^2}{T_a} - \frac{z_a^2}{(T_a-t)}+\frac{x_0^2 + y_0^2}{2T_a}- \frac{x^2 + y^2}{2(T_a-t)} + \alpha^2 t}
 \label{ratioexplitanh}
\end{eqnarray}
and the ratio of the propagators computed using Eq. \ref{pdettanh}
\begin{eqnarray}
&&  \frac{ P(x_T,y_T,T \vert x,y,t) }{P(x_T,y_T,T \vert x_0,y_0,0) }
 = \frac{ \frac{ e^{ \frac{(x_T+y_T)(x+y)  }{2(T-t)} - \frac{x_T^2+x^2 + y_T^2+y^2}{2(T-t)} - \alpha^2 (T-t)} }
  {   \pi (T-t)}  \left(\frac{ \cosh( \alpha x_T ) \cosh( \alpha y_T )}{ \cosh( \alpha x ) \cosh( \alpha y )} \right) \sinh \left(\frac{(x_T-y_T)(x-y)  }{2(T-t)} \right) }
{\frac{ e^{ \frac{(x_T+y_T)(x_0+y_0)  }{2T} - \frac{x_T^2+x_0^2 + y_T^2+y_0^2}{2T} - \alpha^2 T} }
  {   \pi T}  \left(\frac{ \cosh( \alpha x_T ) \cosh( \alpha y_T )}{ \cosh( \alpha x_0 ) \cosh( \alpha y_0 )} \right) \sinh \left(\frac{(x_T-y_T)(x_0-y_0)  }{2T} \right)}
\nonumber \\
&& 
= \left( \frac{T}{T-t} \right) 
\frac{\cosh( \alpha x_0 ) \cosh( \alpha y_0 )\sinh \left(\frac{(x_T-y_T)(x-y)  }{2(T-t)} \right) }
{ \cosh( \alpha x ) \cosh( \alpha y )\sinh \left(\frac{(x_T-y_T)(x_0-y_0)  }{2T} \right)}
e^{ (x_T+y_T) \left[\frac{x+y  }{2(T-t)} - \frac{x_0+y_0  }{2T} \right] 
+ \frac{x_T^2+x_0^2 + y_T^2+y_0^2}{2T}- \frac{x_T^2+x^2 + y_T^2+y^2}{2(T-t)} + \alpha^2 t}
\ \ \ \ 
\label{ratioexpliPtanh}
\end{eqnarray}
can be plugged 
into Eq. \ref{Qdef}
to obtain 
\begin{eqnarray}
&& Q_T(x,y,t)   = e^{ \alpha^2 t} \frac{(x-y)\cosh( \alpha x_0 ) 
\cosh( \alpha y_0 )}{(x_0-y_0) \cosh( \alpha x ) \cosh( \alpha y )}
\nonumber \\
&& \times  \int_t^{T} dT_a \int_{-\infty}^{+\infty} dz_a 
\gamma^*( z_a,T_a) 
 \left( \frac{T_a}{T_a-t} \right)^2
 e^{ z_a \frac{(x+y)  }{(T_a-t)}  -z_a \frac{(x_0+y_0)  }{T_a}
+ \frac{z_a^2}{T_a} - \frac{z_a^2}{(T_a-t)}+\frac{x_0^2 + y_0^2}{2T_a}- \frac{x^2 + y^2}{2(T_a-t)} }
 \nonumber \\ &&
 +e^{ \alpha^2 t} \left( \frac{T}{T-t} \right) 
\frac{\cosh( \alpha x_0 ) \cosh( \alpha y_0 ) }
{ \cosh( \alpha x ) \cosh( \alpha y )} 
e^{\frac{x_0^2 +y_0^2}{2T}- \frac{x^2 +y^2}{2(T-t)}}
 \int_{-\infty}^{+\infty}  dx_T   \int_{-\infty}^{+\infty}  dy_T
\theta(x_T-y_T)    P^*(x_T,y_T,T ) 
\nonumber \\
&& \times  
\frac{\sinh \left(\frac{(x_T-y_T)(x-y)  }{2(T-t)} \right) }
{ \sinh \left(\frac{(x_T-y_T)(x_0-y_0)  }{2T} \right)}
e^{ (x_T+y_T) \left[\frac{x+y  }{2(T-t)} - \frac{x_0+y_0  }{2T} \right] 
+ \frac{x_T^2 + y_T^2}{2T}- \frac{x_T^2 + y_T^2}{2(T-t)} }
  \label{Qtanh}
\end{eqnarray}
and the corresponding conditioned drift of Eq. \ref{driftdoob}
\begin{eqnarray}
\mu^*_X (x,y,t) && =\alpha \tanh (\alpha x)  +  \partial_x \ln Q_T(x,y,t) 
\nonumber \\
\mu^*_Y (x,y,t) && =\alpha \tanh (\alpha y)  +  \partial_y \ln Q_T(x,y,t) 
\label{driftdoobtanh}
\end{eqnarray}
Let us now describe some simple examples.


\subsection{  Example : conditioning the two tanh-drift processes
 towards the annihilation at position $z^*$ at time $T^*$ }

When the annihilation-time $T_a$ takes the single value $T^* $
and when the annihilation-position $z_a$ takes the single value $z^*$
\begin{eqnarray}
\gamma^*( z_a,T_a)=\delta(z_a-z^* ) \delta(T_a-T^* ) 
 \label{gammaDead1deltatanh}
\end{eqnarray}
the function of Eq. \ref{Qtanh} for $t \in [0,T^*[$
\begin{eqnarray}
Q_T(x,y,t)   =  e^{ \alpha^2 t} \frac{(x-y)\cosh( \alpha x_0 ) 
\cosh( \alpha y_0 )}{(x_0-y_0) \cosh( \alpha x ) \cosh( \alpha y )} 
 \left( \frac{T^*}{T^*-t} \right)^2
 e^{ z^* \frac{(x+y)  }{(T^*-t)}  -z^* \frac{(x_0+y_0)  }{T^*}
+ \frac{[z^*]^2}{T^*} - \frac{[z^*]^2}{(T^*-t)}+\frac{x_0^2 + y_0^2}{2T^*}- \frac{x^2 + y^2}{2(T^*-t)} }
 \label{Qtimetanh}
\end{eqnarray}
leads to the conditioned drift of Eq. \ref{driftdoobbrown} for $t \in [0,T^*[$ 
\begin{eqnarray}
\mu^*_X (x,y,t) && = \alpha \tanh (\alpha x)  + \partial_x \ln Q_T(x,y,t) = 
 \frac{1}{x-y} + \frac{z^*-x}{T^*-t}
 \nonumber \\
\mu^*_Y (x,y,t) && = \alpha \tanh (\alpha y)  + \partial_y \ln Q_T(x,y,t) 
= -\frac{1}{x-y}+  \frac{z^* -y}{T^*-t} 
\label{driftdoobtanh1delta}
\end{eqnarray}
that actually coincide with the drift of Eq. \ref{driftdoobbrown1delta}.


\subsection{ Example : conditioning towards two tanh-drift bridges at time $T$ without meeting }

When the two tanh-drift processes are constrained to be surviving at the given positions $(x_*,y_*)$ at time $T$
\begin{eqnarray}
P^*(x_T,y_T,T )=\delta(x_T-x_* ) \delta(y_T-y_* ) 
 \label{gammaDead2deltatanh}
\end{eqnarray}
the function $Q_T(x,y,t)$ of Eq. \ref{Qtanh} 
\begin{eqnarray}
Q_T(x,y,t) 
=  \left( \frac{T}{T-t} \right) 
\frac{\cosh( \alpha x_0 ) \cosh( \alpha y_0 ) \sinh \left(\frac{(x_*-y_*)(x-y)  }{2(T-t)} \right)}
{ \cosh( \alpha x ) \cosh( \alpha y )\sinh \left(\frac{(x_*-y_*)(x_0-y_0)  }{2T} \right)} 
e^{\frac{x_0^2 +y_0^2}{2T}- \frac{x^2 +y^2}{2(T-t)}
+ (x_*+y_*) \left[\frac{x+y  }{2(T-t)} - \frac{x_0+y_0  }{2T} \right] 
+ \frac{x_*^2 + y_*^2}{2T}- \frac{x_*^2 + y_*^2}{2(T-t)} +\alpha^2 t } 
\ \ \ 
 \label{Qtanhspace2delta}
\end{eqnarray}
leads to the conditioned drift of Eq. \ref{driftdoobtanh}
\begin{eqnarray}
\mu^*_X (x,y,t) && = \alpha \tanh (\alpha x)  + \partial_x \ln Q_T(x,y,t) 
= 
- \frac{x}{(T-t)}
+  \frac{ x_*+y_* }{2(T-t)} 
+  \frac{(x_*-y_*)  }{2(T-t)}  \coth \left(\frac{(x_*-y_*)(x-y)  }{2(T-t)} \right) 
 \nonumber \\
\mu^*_Y (x,y,t) && =\alpha \tanh (\alpha y)  + \partial_y \ln Q_T(x,y,t) 
=
- \frac{y}{T-t}
+  \frac{x_*+y _* }{2(T-t)} 
-  \frac{(x_*-y_*)  }{2(T-t)}  \coth \left(\frac{(x_*-y_*)(x-y)  }{2(T-t)} \right) 
\label{driftdoobtanh2deltaspace}
\end{eqnarray}
that actually coincide with the drift of Eq. \ref{driftdoobbrown2deltaspace}.

As previously mentioned, Eq. \ref{driftdoobtanh1delta} and Eq. \ref{driftdoobbrown1delta} are identical as are Eq. \ref{driftdoobtanh2deltaspace} and Eq. \ref{driftdoobbrown2deltaspace}, and it is not a coincidence. Indeed, in this article we have developed a method that allows to impose several constraints on processes at once. However, an equivalent way to proceed (but probably less elegant) consists in imposing the constraints one after the other \cite{refMazzoloJstat}. Now, if we condition the tanh-drift process on its final state, according to  \cite{Benjamini,Fitzsimmons,BorodinBridge} we know that the resulting process is a Brownian bridge \footnote{More precisely, Benjamini and Lee \cite{Benjamini} show that processes with constant drifts or of the form $\mu(x) = \alpha \tanh(\alpha x + c)$ with $\alpha, c \in \mathcal{R}$ have their bridges that coincide with the Brownian bridge.}. Therefore, imposing other constraints (whatever they might be) on such a process is the same as imposing these constraints on the Brownian bridge. This result explains why Eq. \ref{driftdoobtanh1delta} and Eq. \ref{driftdoobbrown1delta} as well as Eq. \ref{driftdoobtanh2deltaspace} and Eq. \ref{driftdoobbrown2deltaspace} are identical. However, this reasoning does not apply to the next case when one wishes that two tanh-drift processes survive forever.


\subsection{ Example : conditioning the two tanh-drift processes towards full survival $S^*(\infty)=1 $ at $T=+\infty$ }

\begin{figure}[h]
\centering
\includegraphics[width=4.2in,height=3.2in]{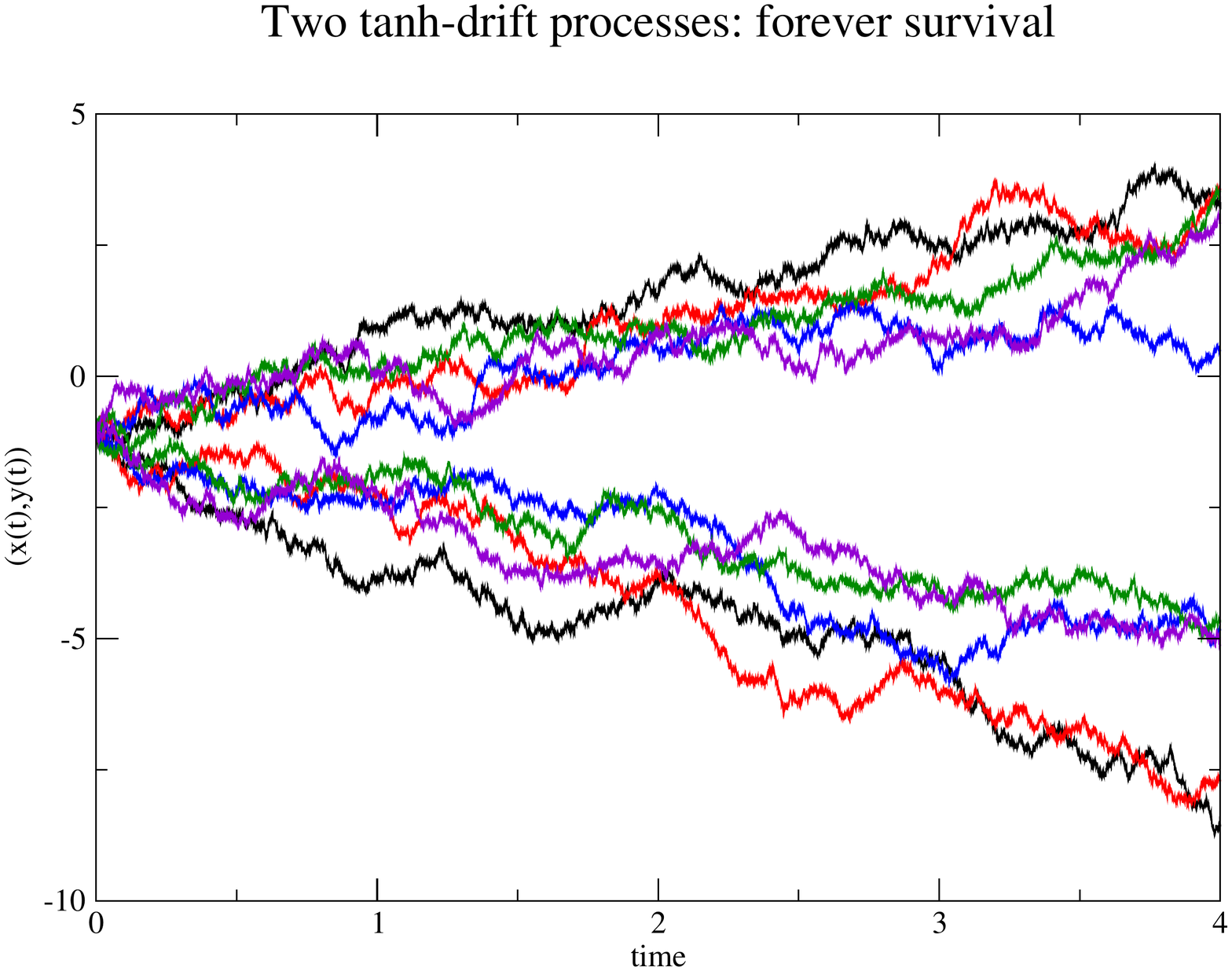}
\setlength{\abovecaptionskip}{15pt}  
\caption{A sample of diffusions for the drift given by Eq. \ref{driftspacesurvivalinfinitytanh} with $\alpha = 1$. Each color corresponds to the realization of one  process. Both processes start at $x_0 = -1$ and $y_0 = -1.1$ and live forever without meeting. At large times the behavior of the conditioned two tanh-drift process corresponds to a Brownian motion with positive drift $\alpha$ for $x>0$ and negative drift -$\alpha$ for $y < 0$. The time step used in the discretization is $dt = 10^{-4}$.}
\end{figure}

Since the unconditioned probability 
of forever-survival of Eq. \ref{foreversurvivaltanh} is finite,
the function $Q_{\infty}^{[  S^*(\infty )=1]}(x,y,t) $
is given by Eq. \ref{Qspacesurvivalinftycasea}
\begin{eqnarray}
Q_{\infty}^{[  S^*(\infty )=1]}(x,y,t)
 = \frac{ S(\infty \vert x,y)   }
 {S(\infty \vert x_0,y_0)   }
  = \frac{ \tanh( \alpha x ) - \tanh( \alpha y)  }
 { \tanh( \alpha x_0 ) - \tanh( \alpha y_0)  }
\label{Qspacesurvivalinftycaseatanh}
\end{eqnarray}

So the corresponding conditioned drift is time-independent and reduces to
\begin{eqnarray}
\mu^*_X (x,y) && =\alpha \tanh (\alpha x)  + \partial_x \ln Q_{\infty}^{[  S^*(\infty)=1]}(x,y,t) 
= \alpha \left[ \tanh (\alpha x)  + \frac{1-\tanh^2( \alpha x )}{ \tanh( \alpha x ) - \tanh( \alpha y)  } \right]
\nonumber \\
&& = \alpha  \frac{1-\tanh( \alpha x )\tanh( \alpha y )}{ \tanh( \alpha x ) - \tanh( \alpha y)  } 
= \frac{\alpha}{ \tanh [\alpha(x-y) ]}
\nonumber \\
\mu^*_Y (x,y) && = \alpha \tanh (\alpha y)  +\partial_y \ln Q_{\infty}^{[  S^*(\infty)=1]}(x,y,t) 
= \alpha \left[ \tanh (\alpha y)  - \frac{1-\tanh^2( \alpha y )}{ \tanh( \alpha x ) - \tanh( \alpha y)  } \right]
\nonumber \\
&& = - \alpha  \frac{1-\tanh( \alpha x )\tanh( \alpha y )}{ \tanh( \alpha x ) - \tanh( \alpha y)  } 
= - \frac{\alpha}{ \tanh [\alpha(x-y) ]}
\label{driftspacesurvivalinfinitytanh}
\end{eqnarray}


\section{ Conclusions }

\label{sec_conclusion}

In this paper, we have considered two independent identical diffusion processes that annihilate upon meeting in order to analyze various types of conditioning with respect to their first-encounter properties.
We have first described the case of the finite horizon $T<+\infty$, where one imposes the probability $P^*(x,y,T ) $ that the two particles are surviving at positions $x$ and $y$ at time $T$, and the probability $\gamma^*(z,t) $ of annihilation at position $z$ at the intermediate times $t \in [0,T]$. We have then focused on various conditioning constraints that are less-detailed than these full distributions, via the optimization of the appropriate relative entropy with respect to the unconditioned processes. We have also studied the limit of infinite horizon $T =+\infty$, where the maximum conditioning consists in imposing the first-encounter probability $\gamma^*(z,t) $ at position $z$ at all finite times $t \in [0,+\infty[$, while its normalization $[1- S^*(\infty )]$ determines the conditioned probability $S^*(\infty ) \in [0,1]$ of forever-survival. We have then applied this general framework to the cases where the unconditioned processes are two Brownian motions, two Ornstein-Uhlenbeck processes, or two tanh-drift processes, in order to generate stochastic trajectories satisfying various types of conditioning constraints. Finally, the link with the stochastic control theory is described in the Appendix, where one optimizes the dynamical large deviations at Level 2.5 in the presence of the conditioning constraints that one wishes to impose.

We thank an anonymous referee for suggesting three directions
to extend the present work in the future :

(i) instead of being identical, the two unconditioned diffusion processes could have different properties,
i.e. different drifts and/or different diffusion coefficients.

(ii) instead of the annihilation at first encounter described by the Dirichlet boundary condition,
one could consider the case of partial reactivity described by other boundary conditions.

(iii) one could consider higher dimensions $d>1$ with an appropriate notion of "encounter".


\appendix

\section{Links with the dynamical large deviations at Level 2.5 and the stochastic control theory}

\label{app_control}

In this Appendix, as in the subsection \ref{subsec_sanov} of the main text, 
one considers a large number $N$ of independent realizations  $[X_n(t),Y_n(t)]$
 of the unconditioned process
starting all at the same initial condition $[X_n(0)=x_0;Y_n(0)=y_0]$.
  The analysis of their dynamical large deviations properties 
gives another perspective on the conditioned process $[X^*(t),Y^*(t)] $ described in 
the main text.

\subsection{ Empirical ensemble-averaged observables for $N$ independent unconditioned processes $[X_n(t),Y_n(t)]$ }

The basic empirical observable is the ensemble-averaged density ${\hat P}(x,y,t) $ of the positions $x$ and $y$
when the two processes are still surviving at time $t$
 \begin{eqnarray}
 {\hat P}(x,y,t) \equiv  \frac{1}{N} \sum_{n=1}^N \delta( X_n(t) -x) \delta( Y_n(t) -x) 
\label{empiP}
\end{eqnarray}
that vanishes at coinciding points $x=y$
\begin{eqnarray}
 {\hat P}(z,z,t)   =0
\label{empiAbs}
\end{eqnarray}
The empirical dynamics can be written as the continuity equation
 \begin{eqnarray}
 \partial_t {\hat P}(x,y,t) = - \partial_x {\hat J}_X(x,y,t) - \partial_y {\hat J}_Y(x,y,t)
\label{empidyn}
\end{eqnarray}
where the empirical current $[ {\hat J}_X(x,y,t); {\hat J}_Y(x,y,t) ]$ can be parametrized in terms some empirical drift 
$[{\hat \mu}_X(x,y,t); {\hat \mu}_Y(x,y,t)]$, while the diffusion coefficient $D(x)  $ is fixed
\begin{eqnarray}
{\hat J}_X(x,y,t) &&= {\hat \mu }_X(x,y,t)  {\hat P}(x,y,t) -  \partial_x  \left[ D(x) {\hat P}(x,y,t) \right]
\nonumber \\
{\hat J}_Y(x,y,t) && = {\hat \mu }_Y(x,y,t)  {\hat P}(x,y,t) -  \partial_y  \left[ D(y) {\hat P}(x,y,t) \right]
\label{empiJ}
\end{eqnarray}
The normalization of the empirical density $ {\hat P}(x,y,t) $ 
over the positions $x >y$ gives the empirical survival probability 
${\hat S}(t )$ at time $t$
\begin{eqnarray}
{\hat S}(t ) \equiv \int_{-\infty}^{+\infty}  dy   \int_{-\infty}^{+\infty}  dx \theta(x-y)  {\hat P}(x,y,t)
\label{empiS}
\end{eqnarray}
Its time-decay corresponds to the empirical distribution ${\hat \gamma}(t) $
of the annihilation-time $t$ 
\begin{eqnarray}
{\hat \gamma}(t)  = - \frac{  d {\hat S}(t ) }{dt}  
=- \int_{-\infty}^{+\infty}  dy   \int_{-\infty}^{+\infty}  dx \theta(x-y) \partial_t {\hat P}(x,y,t)  
\label{empigammat}
\end{eqnarray}
Using the continuity Eq. \ref{empidyn}, one obtains
that ${\hat \gamma}(t)  $ can be decomposed spatially into into 
\begin{eqnarray}
{\hat \gamma}(t) && =  \int_{-\infty}^{+\infty}  dy   \int_{-\infty}^{+\infty}  dx \theta(x-y) 
\left(  \partial_x {\hat J}_X(x,y,t) + \partial_y {\hat J}_Y(x,y,t) \right) 
\nonumber \\
&& =  \int_{-\infty}^{+\infty}  dy   
  \left[ {\hat J}_X(x,y,t) \right]_{x=y}^{x=+\infty}
+ \int_{-\infty}^{+\infty}  dx   
 \left[ {\hat J}_Y(x,y,t)  \right]_{y=-\infty}^{y=x}
 \nonumber \\
&& = - \int_{-\infty}^{+\infty}  dy    {\hat J}_X(y,y,t) 
+ \int_{-\infty}^{+\infty}  dx    {\hat J}_Y(x,x,t)
\nonumber \\
&& \equiv  \int_{-\infty}^{+\infty} dz {\hat \gamma}(z,t) 
 \label{empigammatz}
\end{eqnarray}
where the empirical probability ${\hat \gamma}(z,t)  $ 
of annihilation at position $z$ at time $t$ involves the empirical current 
entering the absorbing diagonal $x=y$ at position $(x=z,y=z)$
\begin{eqnarray}
{\hat \gamma}(z,t) = -  {\hat J}_X(z,z,t) +  {\hat J}_Y(z,z,t)
 = D(z) \left[ 
\left(    \partial_{x} -\partial_{y} \right)
\left[  {\hat P}(x,y,t)  \right]
\right] \vert_{x=z;y=z}
\label{empigammatzcurrent}
\end{eqnarray}

In the thermodynamic limit $N \to +\infty$, all these empirical observables concentrate 
on their typical values given by the corresponding observables without hats described 
in section \ref{sec_unconditioned}.
However for large finite $N$, dynamical fluctuations around these typical values are possible
and can be analyzed via the theory of large deviations discussed
 in the next subsection.


\subsection{ Large deviations at Level 2.5 for the empirical dynamics during the time-window 
$t \in [0,T]$}

In the field of dynamical large deviations for Markov processes 
(see the reviews \cite{oono,ellis,review_touchette} and references therein),
the initial standard classification into Levels 1,2,3
has turned out to be inappropriate :
indeed, the Level 2 concerning the empirical density alone cannot be written explicitly in most cases,
while the Level 3 concerning the whole empirical process is actually far too general for many purposes.
As a consequence, a new Level has been introduced between the Level 2 and the Level 3
and has been called "Level 2.5", even if it is actually much closer in spirit to the Level 2,
since the "Level 2.5" describes the large deviations properties
of the joint distribution of the empirical density and of the empirical flows.
In contrast to the Level 2, the Level 2.5 can be written explicitly for general Markov processes,
including discrete-time Markov chains
 \cite{fortelle_thesis,fortelle_chain,review_touchette,c_largedevdisorder,c_reset,c_inference},
continuous-time Markov jump processes
\cite{fortelle_thesis,fortelle_jump,maes_canonical,maes_onandbeyond,wynants_thesis,chetrite_formal,BFG1,BFG2,chetrite_HDR,c_ring,c_interactions,c_open,c_detailed,barato_periodic,chetrite_periodic,c_reset,c_inference,c_runandtumble,c_jumpdiff,c_skew,c_metastable,c_east,c_exclusion}
and Diffusion processes 
\cite{wynants_thesis,maes_diffusion,chetrite_formal,engel,chetrite_HDR,c_reset,c_lyapunov,c_inference,c_metastable}.
In summary, the Level 2.5 plays an essential role because it is the smallest Level that is explicit
in full generality. Let us now describe the particular application to our present setting.

\subsubsection{ Large deviations at Level 2.5 for the empirical density $ {\hat P }(.,.,.) $
and the empirical current ${\hat J}_.(.,.,.)$ for $t \in [0,T]$}

For the $N$ independent unconditioned processes $[X_n(t);Y_n(t)]$ introduced in the previous subsection,
the application of Level 2.5 yields that
the joint probability ${\cal P}^{[2.5]}_{[0,T]} \left[ {\hat P }(.,.,.) ; {\hat J}_.(.,.,.)  \right] $
 to see the empirical density 
$ {\hat P}(x,y,t) $ of Eq. \ref{empiP}
and the empirical current $[{\hat J}_X(x,y,t);{\hat J}_Y(x,y,t) ] $ 
of Eq. \ref{empiJ} over the time-window $t \in [0,T]$
follows the large deviation form for large $N$
\begin{eqnarray}
{\cal P}^{[2.5]}_{[0,T]} \left[ {\hat P }(.,.,.) ; {\hat J}_.(.,.,.)  \right]  \opsimeq_{N \to +\infty} 
{\cal C}^{[2.5]}_{[0,T]} \left[ {\hat P }(.,.,.) ; {\hat J}_.(.,.,.)\right]
e^{- N {\cal I}^{[2.5]}_{[0,T]} \left[{\hat P }(.,.,.) ; {\hat J}_.(.,.,.)  \right] }
\label{Level2.5}
\end{eqnarray}
with the following notations :

(i)  the rate function ${\cal I}^{[2.5]}_{[0,T]} \left[{\hat P }(.,.,.) ; {\hat J}_.(.,.,.)  \right] $ at Level 2.5 
is given by the usual explicit form for diffusion processes 
\cite{wynants_thesis,maes_diffusion,chetrite_formal,engel,chetrite_HDR,c_reset,c_lyapunov,c_inference,c_metastable}
applied to our present case
\begin{eqnarray}
&& {\cal I}^{[2.5]}_{[0,T]} \left[{\hat P }(.,.,.) ; {\hat J}_.(.,.,.)  \right]
 = \int_0^T dt
\int_{-\infty}^{+\infty}  dy   \int_{-\infty}^{+\infty}  dx \theta(x-y) 
\label{rate2.5diff}
 \\ &&
\left(  \frac{  \left[  {\hat J}_X(x,y,t) - \mu (x) {\hat P}(x,y,t) +\partial_x \left( D(x) {\hat P}(x,y,t) \right) \right]^2}{ 4 D(x) {\hat P}(x,y,t) }
 + \frac{  \left[  {\hat J}_Y(x,y,t) - \mu (y) {\hat P}(x,y,t) +\partial_y \left( D(y) {\hat P}(x,y,t) \right) \right]^2}{ 4 D(y) {\hat P}(x,y,t) }
\right)
\nonumber
\end{eqnarray}

(ii) the constitutive constraints ${\cal C}^{[2.5]}_{[0,T]} \left[ {\hat P }(.,.,.) ; {\hat J}_.(.,.,.) \right] $ 
at Level 2.5 can be decomposed 
\begin{eqnarray}
{\cal C}^{[2.5]}_{[0,T]} \left[ {\hat P }(.,.,.) ; {\hat J}_.(.,.,.)  \right]
&& = \delta( {\hat P}(x,y,0) - \delta(x-x_0)\delta(y-y_0) ) \ 
\nonumber \\
&&  {\cal C}^{Bulk}_{[0,T]} \left[ {\hat P }(.,.,.) ; {\hat J}_.(.,.,.)  \right]
 {\cal C}^{Boundary}_{[0,T]} \left[ {\hat P }(.,.,.) ; {\hat J}_.(.,.,.) \right]
  \label{C2.5}
\end{eqnarray}
into the initial condition $ {\hat P}(x,y,t=0) = \delta(x-x_0)\delta(y-y_0)$, 
the empirical dynamics of Eq. \ref{empidyn}
\begin{eqnarray}
{\cal C}^{Bulk}_{[0,T]} \left[ {\hat P }(.,.,.) ; {\hat J}_.(.,.,.)  \right]
= \prod_{t \in [0,T]} \prod_{x \in ]-\infty,+\infty[}\prod_{y \in ]-\infty,x[} \left( 
 \partial_t {\hat P}(x,y,t) + \partial_x {\hat J}_X(x,y,t) + \partial_y {\hat J}_Y(x,y,t)
\right)
  \label{C2.5diff}
\end{eqnarray}
and the boundary conditions of Eqs \ref{empiAbs} and \ref{empigammatzcurrent} at coinciding points $x=y$
during the time-window $t \in [0,T] $
\begin{eqnarray}
&& {\cal C}^{Boundary}_{[0,T]} \left[ {\hat P }(.,.,.) ;  {\hat J}_.(.,.,.) \right]
\nonumber \\
&&  = \prod_{t \in [0,T]}  \prod_{z \in ]-\infty,+\infty[} 
\delta \left( {\hat P}(z,z,t) \right)
 \delta \left( 
   {\hat J}_X(z,z,t) -  {\hat J}_Y(z,z,t)
 + D(z) \left[ 
\left(    \partial_{x} -\partial_{y} \right)
\left[  {\hat P}(x,y,t)  \right]
\right] \vert_{x=z;y=z}
  \right)
  \label{C2.5abs}
\end{eqnarray}


\subsubsection{ Level 2.5 for the empirical density ${\hat P}(.,.,.) $, 
the empirical drift $ {\hat \mu }_.(.,.,.) $, and the empirical annihilation distribution ${\hat \gamma}(.,.) $ }

In the bulk $x >y$ where the empirical density does not vanish ${\hat P}(x,y,t) >0$,
the parametrization of Eq. \ref{empiJ} allows to replace the empirical current ${\hat J}_.(.,.,.) $
 by the empirical drift  
\begin{eqnarray}
 {\hat \mu }_X(x,y,t) && = \frac{ {\hat J}_X(x,y,t) + \partial_x  \left[ D(x) {\hat P}(x,y,t) \right] }{ {\hat P}(x,y,t) }
 \nonumber \\
  {\hat \mu }_Y(x,y,t) && = \frac{ {\hat J}_Y(x,y,t) + \partial_y  \left[ D(y) {\hat P}(x,y,t) \right] }{ {\hat P}(x,y,t) }
\label{empiJtomu}
\end{eqnarray}
On the diagonal $x=y$ where the empirical density vanishes ${\hat P}(z,z,t) =0$, 
the current component that appears in the boundary condition of Eq. \ref{C2.5abs}
can be replaced by the empirical annihilation distribution ${\hat \gamma}(z,t) $ of Eq. \ref{empigammatzcurrent}
\begin{eqnarray}
{\hat \gamma}(z,t) = -  {\hat J}_X(z,z,t) +  {\hat J}_Y(z,z,t)
\label{empigammatzcurrentbis}
\end{eqnarray}

As a consequence, the large deviations at Level 2.5 of Eq. \ref{Level2.5}
can be directly translated into the joint probability ${\cal P}^{[2.5]}_{[0,T]} \left[ {\hat P }(.,.,.) ; {\hat \mu}_.(.,.,.) ; {\hat \gamma}(.,.) \right] $
to see the empirical density 
$ {\hat P}(x,y,t) $, the empirical drift $[{\hat \mu}_X(x,y,t);{\hat \mu}_Y(x,y,t) ] $, and the empirical annihilation distribution ${\hat \gamma}(z,t) $
\begin{eqnarray}
{\cal P}^{[2.5]}_{[0,T]} \left[ {\hat P }(.,.,.) ; {\hat \mu}_.(.,.,.) ; {\hat \gamma}(.,.) \right] \opsimeq_{N \to +\infty} 
{\cal C}^{[2.5]}_{[0,T]} \left[ {\hat P }(.,.,.) ; {\hat \mu}_.(.,.,.) ; {\hat \gamma}(.,.) \right]
e^{- N {\cal I}^{[2.5]}_{[0,T]} \left[{\hat P }(.,.,.) ; {\hat \mu}_.(.,.,.)  \right] }
\label{Level2.5mu}
\end{eqnarray}
The rate function translated from Eq. \ref{rate2.5diff} 
reduces to the simpler Gaussian form
for the empirical drift 
\begin{eqnarray}
{\cal I}^{[2.5]}_{[0,T]} \left[{\hat P }(.,.,.) ; {\hat \mu}_.(.,.,.)   \right]
&& =\int_0^T dt
\int_{-\infty}^{+\infty}  dy   \int_{-\infty}^{+\infty}  dx \theta(x-y) {\hat P}(x,y,t)
\bigg(  \frac{  \left[  {\hat \mu}_X(x,y,t) - \mu (x)  \right]^2}{ 4 D(x)  }
 +  \frac{  \left[  {\hat \mu}_Y(x,y,t) - \mu (y)  \right]^2}{ 4 D(y)  }
\bigg)
\nonumber \\
\label{rate2.5diffmu}
\end{eqnarray} 
The constitutive constraints translated from Eq. \ref{C2.5}
\begin{eqnarray}
{\cal C}^{[2.5]}_{[0,T]} \left[ {\hat P }(.,.,.) ; {\hat \mu}_.(.,.,.); {\hat \gamma}(.,.)  \right]
&& = \delta( {\hat P}(x,y,0) - \delta(x-x_0)\delta(y-y_0) ) \ 
 \nonumber \\ &&  
{\cal C}^{Bulk}_{[0,T]} \left[ {\hat P }(.,.,.) ; {\hat \mu}_.(.,.,.)  \right]
 {\cal C}^{Boundary}_{[0,T]} \left[ {\hat P }(.,.,.) ; {\hat \gamma}(.,.) \right]
  \label{C2.5mu}
\end{eqnarray}
involve the contribution of the empirical dynamics
translated from Eq. \ref{C2.5diff}
\begin{eqnarray}
&& {\cal C}^{Bulk}_{[0,T]} \left[ {\hat P }(.,.,.) ; {\hat \mu}_.(.,.,.)  \right]
= \prod_{t \in [0,T]} 
\prod_{x \in ]-\infty,+\infty[}\prod_{y \in ]-\infty,x[}
  \label{C2.5diffmu}
\nonumber \\
&& \left( \partial_t {\hat P}(x,y,t) 
+ \partial_x \left[ {\hat \mu }_X(x,y,t)  {\hat P}(x,y,t) \right] 
+ \partial_y \left[ {\hat \mu }_Y(x,y,t)  {\hat P}(x,y,t) \right] 
-  \partial_x^2  \left[ D(x) {\hat P}(x,y,t) \right]
-  \partial_y^2  \left[ D(y) {\hat P}(x,y,t) \right]
\right)
\nonumber
\end{eqnarray}
and the contribution of the boundary conditions on the diagonal  $x=y$
 translated from Eq. \ref{C2.5abs}
 \begin{eqnarray}
 {\cal C}^{Boundary}_{[0,T]} \left[ {\hat P }(.,.,.) ; {\hat \gamma}(.,.) \right]
  = \prod_{t \in [0,T]}  \prod_{z \in ]-\infty,+\infty[} \delta \left[ {\hat P}(z,z,t) \right]  
 \delta \left( 
  {\hat \gamma}(z,t)  -  D(z) \left[ 
\left(    \partial_{x} -\partial_{y} \right)
\left[  {\hat P}(x,y,t)  \right]
\right] \vert_{x=z;y=z}
  \right) \ \ \ 
  \label{C2.5absmu}
\end{eqnarray}


\subsection{ Link with the stochastic control theory }

In this subsection, one assumes that the empirical density $ {\hat P}(x,y,T) $ at time $T$ is given
and that the empirical annihilation distribution ${\hat \gamma}(z,t) $ is given for $0 \leq t \leq T$
\begin{eqnarray}
  {\hat P}(x,y,T) && = P^*(x,y,T) \ \ \ {\rm for } \ \ \ \ \ -\infty<y<x<+\infty
  \nonumber \\
  {\hat \gamma}(z,t) && = \gamma^*(z,t)  \ \ \ \ \ \ \ {\rm for } \ \ \ \ \ \ z \in ]-\infty,+\infty[ \ \ \ \ {\rm and} \ \ \ \ \ \ t \in [0,T]
  \label{empiT}
\end{eqnarray}
The goal is then to optimize the rate function at Level 2.5 of Eq. \ref{rate2.5diffmu}
over the empirical density ${\hat P}(x,y,t) $ and over the empirical drift ${\hat \mu }_.(x,y,t)  $
at all the intermediate times $t \in ]0,T[$, 
in the presence of the constitutive constraints of Eq. \ref{C2.5mu}
and the supplementary constraints of Eq. \ref{empiT}.

\subsubsection{ Lagrangian for the optimization problem }

It is convenient to separate the constraints of Eqs \ref{C2.5mu} and \ref{empiT}
into :

(i) the time-boundary-conditions for the empirical density $ {\hat P }(.,.,.) $
at the initial time $t=0$ and at the final time $t=T$ 
\begin{eqnarray}
 {\hat P}(x,y,t=0) && = \delta(x-x_0)  \delta(y-y_0) 
 \nonumber \\
   {\hat P}(x,t=T) &&  = P^*(x,y,T)  
 \label{timeboundaries}
\end{eqnarray} 

(ii) the space-boundary-conditions on the diagonal $x=y$ for $t \in ]0,T[$
\begin{eqnarray}
  {\hat P}(z,z,t) && =0
  \nonumber \\
  D(z) \left[ 
\left(    \partial_{x} -\partial_{y} \right)
\left[  {\hat P}(x,y,t)  \right]
\right] \vert_{x=z;y=z} && =     \gamma^*(z,t)  
 \label{spaceboundaries}
\end{eqnarray} 

(iii) the bulk constraint ${\cal C}^{Bulk}_{[0,T]} \left[ {\hat P }(.,.,.) ; {\hat \mu}_.(.,.,.)  \right] $ of Eq. \ref{C2.5diffmu} concerning the
empirical dynamics 
\begin{eqnarray}
 \partial_t {\hat P}(x,y,t) 
&& = - \partial_x \left[ {\hat \mu }_X(x,y,t)  {\hat P}(x,y,t) \right] 
- \partial_y \left[ {\hat \mu }_Y(x,y,t)  {\hat P}(x,y,t) \right] 
\nonumber \\
&& +  \partial_x^2  \left[ D(x) {\hat P}(x,y,t) \right]
+  \partial_y^2  \left[ D(y) {\hat P}(x,y,t) \right]
\label{bulkempi}
\end{eqnarray} 

As a consequence, in the space-time-bulk region $\left( y<x ; t \in ]0,T[\right) $,
one only needs to optimize the rate function at Level 2.5 of Eq. \ref{rate2.5diffmu}
in the presence of the bulk constraint (iii). 
This optimization can be done via the introduction of the Lagrangian
\begin{eqnarray}
{\cal L}^{Bulk}\left[ {\hat P }(.,.,.) ; {\hat \mu}_.(.,.,.)  \right] 
 = {\cal I}^{[2.5]}_{[0,T]} \left[ {\hat P }(.,.,.) ; {\hat \mu}_.(.,.,.)  \right]
+ {\cal L}^{Empi}\left[ {\hat P }(.,.,.) ; {\hat \mu}_.(.,.,.)  \right] 
\label{lagrangiantot}
\end{eqnarray} 
where the contribution 
\begin{eqnarray}
&& {\cal L}^{Empi}\left[ {\hat P }(.,.,.) ; {\hat \mu}_.(.,.,.)  \right] 
 \equiv  \int_0^T dt
\int_{-\infty}^{+\infty}  dy   \int_{-\infty}^{+\infty}  dx \theta(x-y)
 \ \psi(x,y,t) 
 \label{lagrangianpsi}
 \\
 && \left( 
\partial_t {\hat P}(x,y,t) 
+ \partial_x \left[ {\hat \mu }_X(x,y,t)  {\hat P}(x,y,t) \right] 
+ \partial_y \left[ {\hat \mu }_Y(x,y,t)  {\hat P}(x,y,t) \right] 
-  \partial_x^2  \left[ D(x) {\hat P}(x,y,t) \right]
-  \partial_y^2  \left[ D(y) {\hat P}(x,y,t) \right] 
\right)
 \nonumber
\end{eqnarray} 
involves the Lagrange multiplier $\psi(x,y,t)$ introduced in order to impose
the constraint of Eq. \ref{bulkempi} concerning the empirical dynamics.


\subsubsection{ The adjoint-equation method to analyze the optimization problem }

As usual in stochastic control theory,
it is useful to make some transformation of the Lagrangian of Eq. \ref{lagrangiantot} before its optimization. 
The goal is to eliminate the derivatives of the empirical density via integrations by parts.
In our present case, this amounts to rewrite the three terms of Eq. \ref{lagrangianpsi}
via integrations by parts,
 either over time $t \in ]0,T[$ using the time-boundary-conditions of Eq. \ref{timeboundaries}
\begin{eqnarray}
&&\int_0^T dt \ \psi(x,y,t) \partial_t {\hat P}(x,y,t) 
 = \left[ \psi(x,y,t)  {\hat P}(x,y,t)\right]_{t=0}^{t=T} - \int_0^T dt {\hat P}(x,y,t) \partial_t  \psi(x,y,t)
\nonumber \\
&& =  \psi(x,y,T)  P^*(x,y,T) -  \psi(x,y,0) \delta(x-x_0)\delta(y-y_0)
- \int_0^T dt {\hat P}(x,y,t) \partial_t  \psi(x,y,t)
\label{integtime}
\end{eqnarray} 
or over space using the space-boundary-conditions of Eq. \ref{spaceboundaries}, namely
\begin{eqnarray}
 \int_y^{+\infty} dx \ \psi(x,y,t)  \partial_x \left( {\hat \mu }_X(x,y,t)  {\hat P}(x,y,t)  \right)
&&   =  -  \int_y^{+\infty} dx     {\hat P}(x,y,t)  {\hat \mu }_X(x,y,t)  \partial_x\psi(x,y,t) 
\nonumber \\
 \int_{-\infty}^x dy \ \psi(x,y,t)  \partial_y \left( {\hat \mu }_Y(x,y,t)  {\hat P}(x,y,t)  \right)
&&   =  -  \int_{-\infty}^x dy   {\hat P}(x,y,t)  {\hat \mu }_Y(x,y,t)  \partial_y\psi(x,y,t) 
\label{integspace}
\end{eqnarray} 
together with
\begin{eqnarray}
 && -  \int_y^{+\infty} dx \ \psi(x,y,t)  \partial^2_x  \left[ D(x) {\hat P}(x,y,t) \right] 
 = -  \left[ \psi(x,y,t)   \partial_x  \left( D(x) {\hat P}(x,y,t) \right)  \right]_{x=y}^{x=+\infty}
 +  \int_y^{+\infty} dx
  \left(  \partial_x  \left[ D(x) {\hat P}(x,y,t) \right] \right) \partial_x\psi(x,y,t) 
  \nonumber \\
  && =  \psi(y,y,t) D(y)  \left(  \partial_x    {\hat P}(x,y,t)   \right)_{x=y}
  - \int_y^{+\infty} dx
D(x) {\hat P}(x,y,t)  \partial^2_x\psi(x,y,t) 
\label{integspace2}
\end{eqnarray} 
and
\begin{eqnarray}
 && -   \int_{-\infty}^x dy \ \psi(x,y,t)  \partial^2_y  \left[ D(y) {\hat P}(x,y,t) \right] 
 = -  \left[ \psi(x,y,t)   \partial_y  \left( D(y) {\hat P}(x,y,t) \right)  \right]_{y=-\infty}^{y=x}
 +  \int_{-\infty}^x dy 
  \left(  \partial_y  \left[ D(y) {\hat P}(x,y,t) \right] \right) \partial_y \psi(x,y,t) 
  \nonumber \\
  && =  - \psi(x,x,t)  D(x)  \left(  \partial_y   {\hat P}(x,y,t)   \right)_{y=x}
  -  \int_{-\infty}^x dy 
D(y) {\hat P}(x,y,t)  \partial^2_y\psi(x,y,t) 
\label{integspace2y}
\end{eqnarray}

Putting everything together, the contribution of Eq. \ref{lagrangianpsi} reads
\begin{eqnarray}
&& {\cal L}^{Empi}\left[ {\hat P }(.,.,.) ; {\hat \mu}_.(.,.,.)  \right] 
\nonumber \\
&& = 
\int_{-\infty}^{+\infty}  dy   \int_{-\infty}^{+\infty}  dx \theta(x-y)
\left(
 \psi(x,y,T)  P^*(x,y,T) -  \psi(x,y,0) \delta(x-x_0)\delta(y-y_0)
- \int_0^T dt {\hat P}(x,y,t) \partial_t  \psi(x,y,t)
 \right)
 \nonumber \\
 &&+ 
  \int_0^T dt
\int_{-\infty}^{+\infty}  dy  
\left(  -  \int_y^{+\infty} dx     {\hat P}(x,y,t)  {\hat \mu }_X(x,y,t)  \partial_x\psi(x,y,t) 
\right)
 \nonumber \\
 &&+ \int_0^T dt
   \int_{-\infty}^{+\infty}  dx 
\left(
-  \int_{-\infty}^x dy   {\hat P}(x,y,t)  {\hat \mu }_Y(x,y,t)  \partial_y\psi(x,y,t) 
\right)
 \nonumber \\
 &&+  \int_0^T dt
\int_{-\infty}^{+\infty}  dy 
\left(
 \psi(y,y,t) D(y)  \left(  \partial_x    {\hat P}(x,y,t)   \right)_{x=y}
  - \int_y^{+\infty} dx
D(x) {\hat P}(x,y,t)  \partial^2_x\psi(x,y,t) 
\right)
 \nonumber \\
 &&+  \int_0^T dt
   \int_{-\infty}^{+\infty}  dx 
   \left(
- \psi(x,x,t)  D(x)  \left(  \partial_y   {\hat P}(x,y,t)   \right)_{y=x}
  -  \int_{-\infty}^x dy 
D(y) {\hat P}(x,y,t)  \partial^2_y\psi(x,y,t)    
   \right)
   \nonumber \\
   && =\int_{-\infty}^{+\infty}  dy   \int_{-\infty}^{+\infty}  dx \theta(x-y)
 \psi(x,y,T)  P^*(x,y,T)-  \psi(x_0,y_0,0)
+  \int_0^T dt \int_{-\infty}^{+\infty}  dz  \psi(z,z,t) \gamma^*(z,t)
   \nonumber \\
   &&  -  \int_0^T dt \int_{-\infty}^{+\infty}  dy   \int_{-\infty}^{+\infty}  dx \theta(x-y)
 {\hat P}(x,y,t) 
 \nonumber \\
 && \left[ \partial_t  \psi(x,y,t)
  + {\hat \mu }_X(x,y,t)  \partial_x\psi(x,y,t) 
  +{\hat \mu }_Y(x,y,t)  \partial_y\psi(x,y,t) 
  + D(x)   \partial^2_x\psi(x,y,t) 
  + D(y)   \partial^2_y\psi(x,y,t)    
\right]
\label{lagrangianpsifinal}
\end{eqnarray} 

As a consequence, the bulk lagrangian of Eq. \ref{lagrangiantot} becomes 
using the explicit rate function at Level 2.5 of Eq. \ref{rate2.5diffmu} 
\begin{eqnarray}
&& {\cal L}^{Bulk}\left[ {\hat P }(.,.,.) ; {\hat \mu}_.(.,.,.)  \right] 
 = \int_{-\infty}^{+\infty}  dy   \int_{-\infty}^{+\infty}  dx \theta(x-y)
 \psi(x,y,T)  P^*(x,y,T)-  \psi(x_0,y_0,0)
+  \int_0^T dt \int_{-\infty}^{+\infty}  dz  \psi(z,z,t) \gamma^*(z,t)
  \nonumber \\
   &&
+ \int_0^T dt
\int_{-\infty}^{+\infty}  dy   \int_{-\infty}^{+\infty}  dx \theta(x-y) {\hat P}(x,y,t)
\bigg(  \frac{  \left[  {\hat \mu}_X(x,y,t) - \mu (x)  \right]^2}{ 4 D(x)  }
 +  \frac{  \left[  {\hat \mu}_Y(x,y,t) - \mu (y)  \right]^2}{ 4 D(y)  }
  \nonumber \\
   && -  \left[ \partial_t  \psi(x,y,t)
  + {\hat \mu }_X(x,y,t)  \partial_x\psi(x,y,t) 
  +{\hat \mu }_Y(x,y,t)  \partial_y\psi(x,y,t) 
  + D(x)   \partial^2_x\psi(x,y,t) 
  + D(y)   \partial^2_y\psi(x,y,t)    
\right]
\bigg)
\label{lagrangianbulk}
\end{eqnarray} 
The optimization of Eq. \ref{lagrangianbulk}
over the empirical drift $[{\hat \mu }_X(x,y,t); {\hat \mu }_Y(x,y,t)] $
\begin{eqnarray}
0 && = \frac{ {\cal L}^{Bulk}\left[ {\hat P }(.,.,.) ; {\hat \mu}_.(.,.,.)  \right] }{ \partial {\hat \mu }_X(x,y,t) } = 
 {\hat P}(x,y,t)  \left( \frac{   {\hat \mu }_X(x,y,t)  - \mu (x)  }{ 2 D(x)   }
 -   \partial_x\psi(x,y,t) \right)
 \nonumber \\
0 && = \frac{ {\cal L}^{Bulk}\left[ {\hat P }(.,.,.) ; {\hat \mu}_.(.,.,.)  \right] }{ \partial {\hat \mu }_Y(x,y,t) } = 
 {\hat P}(x,y,t)  \left( \frac{   {\hat \mu }_Y(x,y,t)  - \mu (y)  }{ 2 D(y)   }
 -   \partial_y\psi(x,y,t) \right) 
 \label{lagrangianbulkderimu}
\end{eqnarray} 
allows to 
evaluate the optimal empirical drift ${\hat \mu }^{opt}_.(x,y,t) $ in terms of the Lagrange multiplier $\psi(x,y,t)  $
\begin{eqnarray}
  {\hat \mu }^{opt}_X(x,y,t)  && = \mu (x) + 2 D(x)   \partial_x\psi(x,y,t)
  \nonumber \\
  {\hat \mu }^{opt}_Y(x,y,t)  && = \mu (y) + 2 D(y)   \partial_y\psi(x,y,t)  
 \label{mupsi}
\end{eqnarray} 
The further optimization of Eq. \ref{lagrangianbulk}
over the empirical density ${\hat P}(x,y,t) $ reads using the optimal drift of Eq. \ref{mupsi}
\begin{eqnarray}
0 && = - \frac{ {\cal L}^{Bulk}\left[ {\hat P }(.,.,.) ; {\hat \mu}(.,.,.)  \right] }{ \partial {\hat P }(x,y,t) }
 = 
- \frac{  \left[  {\hat \mu}^{opt}_X(x,y,t) - \mu (x)  \right]^2}{ 4 D(x)  }
 -  \frac{  \left[  {\hat \mu}^{opt}_Y(x,y,t) - \mu (y)  \right]^2}{ 4 D(y)  }
 + \partial_t  \psi(x,y,t)
  \nonumber \\
   &&  + {\hat \mu}^{opt}_X(x,y,t)  \partial_x\psi(x,y,t) 
  +{\hat \mu}^{opt}_Y(x,y,t)  \partial_y\psi(x,y,t) 
  + D(x)   \partial^2_x\psi(x,y,t) 
  + D(y)   \partial^2_y\psi(x,y,t)     
 \nonumber \\
&& o =  - D(x)  \left[  \partial_x\psi(x,y,t)  \right]^2  - D(y)  \left[  \partial_y\psi(x,y,t)  \right]^2
 + \partial_t  \psi(x,y,t)
  \nonumber \\
   &&  + \left(  \mu (x) + 2 D(x)   \partial_x\psi(x,y,t)\right)  \partial_x\psi(x,y,t) 
  + \left( \mu (y) + 2 D(y)   \partial_y\psi(x,y,t)   \right)  \partial_y\psi(x,y,t) 
  + D(x)   \partial^2_x\psi(x,y,t) 
  + D(y)   \partial^2_y\psi(x,y,t)     
      \nonumber \\
&& =   \partial_t  \psi(x,y,t)  + \mu (x)       \partial_x\psi(x,y,t) + \mu (y)       \partial_y\psi(x,y,t)
   \nonumber \\
&& + D(x) \left( \partial^2_x\psi(x,y,t) + \left[    \partial_x\psi(x,y,t) \right]^2 \right)
 +  D(y) \left( \partial^2_y\psi(x,y,t) + \left[    \partial_y\psi(x,y,t) \right]^2 \right)
\label{lagrangianbulkderiP}
\end{eqnarray} 
This Hamilton-Jacobi-Bellman equation for $\psi(x,y,t)$ can be transformed via
the change of variables
\begin{eqnarray}
 \psi(x,y,t)  = \ln q(x,y,t)
 \label{psiQ}
\end{eqnarray} 
into the linear backward unconditioned dynamics for the function $q(x,y,t)$
\begin{eqnarray}
- \partial_t  q(x,y,t)    =   \mu (x)   \partial_x  q(x,y,t) + \mu (y)   \partial_y  q(x,y,t)
 + D(x)   \partial^2_x  q(x,y,t) + D(y)   \partial^2_y  q(x,y,t) 
\label{backwardsmallq}
\end{eqnarray} 

Using Eq. \ref{psiQ},
the optimal empirical drift of Eq. \ref{mupsi} becomes
\begin{eqnarray}
  {\hat \mu }^{opt}_X(x,y,t)  && = \mu (x) + 2 D(x)   \partial_x  \ln q(x,y,t)
  \nonumber \\
   {\hat \mu }^{opt}_Y(x,y,t)  && = \mu (y) + 2 D(y)   \partial_y  \ln q(x,y,t) 
   \label{mupsiQ}
\end{eqnarray} 
while the optimal empirical density ${\hat P }^{opt}(x,y,t) $ should be the solution of the corresponding empirical forward dynamics of Eq. \ref{bulkempi}
\begin{eqnarray}
 \partial_t {\hat P}^{opt}(x,y,t) && =
- \partial_x \left[ {\hat \mu }^{opt}_X(x,y,t)  {\hat P}^{opt}(x,y,t) \right] 
- \partial_y \left[ {\hat \mu }^{opt}_Y(x,y,t)  {\hat P}^{opt}(x,y,t) \right] 
+  \partial_x^2  \left[ D(x) {\hat P}^{opt}(x,y,t) \right]
+  \partial_y^2  \left[ D(y) {\hat P}^{opt}(x,y,t) \right]
  \nonumber \\
 && =
- \partial_x \left[ \left(\mu (x) + 2 D(x)   \partial_x  \ln q(x,y,t) \right) {\hat P}^{opt}(x,y,t) \right] 
- \partial_y \left[ \left( \mu (y) + 2 D(y)   \partial_y  \ln q(x,y,t)  \right)  {\hat P}^{opt}(x,y,t) \right] 
\nonumber \\
 && +  \partial_x^2  \left[ D(x) {\hat P}^{opt}(x,y,t) \right]
+  \partial_y^2  \left[ D(y) {\hat P}^{opt}(x,y,t) \right]  
\label{forwardhat}
\end{eqnarray} 

Using the backward unconditioned dynamics of Eq. \ref{backwardsmallq} for the function $q(x,y,t)$
and the forward optimal dynamics of Eq. \ref{forwardhat} for ${\hat P}^{opt}(x,y,t) $,
one obtains that the ratio
\begin{eqnarray}
  p(x,y,t)  \equiv \frac{{\hat P}^{opt}(x,y,t)}{ q(x,y,t)}
   \label{defsmallp}
\end{eqnarray} 
satisfies the forward unconditioned dynamics 
\begin{eqnarray}
  \partial_t p(x,y,t)  && = \frac{ 1}{ q(x,y,t)} \partial_t {\hat P}^{opt}(x,y,t) - \frac{{\hat P}^{opt}(x,y,t)}{ q^2(x,y,t)} \partial_t q(x,y,t)
  \nonumber \\
  && =   -  \partial_x \left[ \mu(x) p(x,y,t)  \right] -  \partial_y \left[ \mu(y) p(x,y,t)  \right] 
  + \partial^2_{x} \left[ D(x) p(x,y,t)  \right]  + \partial^2_y \left[ D(y) p(x,y,t)  \right] 
   \label{forwardsmallp}
\end{eqnarray} 


\subsubsection{ Taking into account the space-time boundary conditions to obtain the final optimal solution }

In summary, the optimal solution ${\hat P}^{opt}(x,y,t) $ is given by the product of Eq. \ref{defsmallp}
\begin{eqnarray}
 {\hat P}^{opt}(x,y,t) =  q(x,y,t) p(x,y,t)
   \label{optimalprod}
\end{eqnarray} 
where $q(x,y,t)$ satisfies the backward unconditioned dynamics of Eq. \ref{backwardsmallq},
while $p(x,y,t)$ satisfies the forward unconditioned dynamics of Eq. \ref{forwardsmallp}.
In addition, we have to take into account
the time-boundary-conditions of Eq. \ref{timeboundaries}
 at the initial time $t=0$ and at the final time $t=T$
\begin{eqnarray}
 \delta(x-x_0)  \delta(y-y_0) && = {\hat P}^{opt}(x,y,t=0) =  q(x,y,0) p(x,y,0)
 \nonumber \\
 P^*(x,y,T) &&  = {\hat P}^{opt}(x,y,t=T) =  q(x,y,T) p(x,y,T)
 \label{timeboundariessmall}
\end{eqnarray} 
as well as the space-boundary-conditions of Eq. \ref{spaceboundaries}
concerning the diagonal $x=y$
 for $t \in ]0,T[$ 
\begin{eqnarray}
0 && =   {\hat P}^{opt} (z,z,t) = q(z,z,t) p(z,z,t)
  \nonumber \\
\gamma^*(z,t)  && =  D(z) \left[ 
\left(    \partial_{x} -\partial_{y} \right) 
\left[  {\hat P}^{opt}(x,y,t)  \right] \right] \vert_{x=z;y=z}      
 \nonumber \\
 && = D(z) q(z,z,t) \left[ \left(    \partial_{x} -\partial_{y} \right)\left[  p(x,y,t)  \right] \right] \vert_{x=z;y=z}
+ D(z) p(z,z,t) \left[ \left(    \partial_{x} -\partial_{y} \right)\left[  q(x,y,t)  \right] \right] \vert_{x=z;y=z}
 \label{spaceboundariessmall}
\end{eqnarray} 

For the function $p(x,y,t)$, it is natural to choose the unconditioned propagator $P(x, y,t \vert x_0,y_0,0)$
that would be the solution if one were not imposing atypical constraints
\begin{eqnarray}
p(x,y,t) = P(x,y,t \vert x_0,y_0,0)
 \label{smallpBigP}
\end{eqnarray} 
Plugging this choice into Eq. \ref{timeboundariessmall}, one obtains that the function $q(x,y,t)$ should satisfy the
time-boundary-conditions of Eq. \ref{timeboundariessmall}
  at the initial time $t=0$ and at the final time $t=T$
\begin{eqnarray}
 q(x,y,t=0) && =1
 \nonumber \\
 q(x,y,t=T) && = \frac{P^*(x,y,T) }{ P(x,y,T \vert x_0,y_0,0)} 
 \label{timeboundariessmallq}
\end{eqnarray} 
as well as the space-boundary-condition of Eq. \ref{spaceboundaries} on the diagonal $x=y$
 for $t \in ]0,T[$ 
\begin{eqnarray}
 q(z,z,t) = \frac{  {\hat \gamma}^*(z,t) }
 { D(z)  \left[ \left(    \partial_{x} -\partial_{y} \right)\left[  p(x,y,t)  \right] \right] \vert_{x=z;y=z}
} 
 = \frac{  {\hat \gamma}^*(z,t) } { \gamma (z,t \vert x_0,y_0,0) }
 \label{spaceboundariessmallq}
\end{eqnarray} 
The solution $q(x,y,t)$
of the backward unconditioned dynamics of Eq. \ref{backwardsmallq}
that satisfies the boundary conditions of Eqs \ref{timeboundariessmallq}
and \ref{spaceboundariessmallq}
thus coincides with the function $Q_T(x,y,t)$ introduced in Eq. \ref{Qdef} of the main text.


\subsubsection{ Corresponding optimal value of the Lagrangian }

The corresponding optimal value of the Lagrangian of Eq. \ref{lagrangianbulk} 
reduces to the boundary terms, since the bulk contribution vanishes as a consequence of the optimization
Eq. \ref{lagrangianbulkderiP}
\begin{eqnarray}
 {\cal L}^{Bulk}\left[ {\hat P }^{opt}(.,.,.) ; {\hat \mu}^{opt}_.(.,.,.)  \right] 
&&  = \int_{-\infty}^{+\infty}  dy   \int_{-\infty}^{+\infty}  dx \theta(x-y) \psi(x,y,T)  P^*(x,y,T)-  \psi(x_0,y_0,0)
\nonumber \\
&& +  \int_0^T dt \int_{-\infty}^{+\infty}  dz  \psi(z,z,t) \gamma^*(z,t) 
\label{lagrangianbulkopt}
\end{eqnarray} 
Using Eq. \ref{psiQ} for $q(x,y,t)= Q_T(x,y,t) $ , the Lagrange multiplier $\psi(x,y,t)$ 
\begin{eqnarray}
\psi(x,y,t) = \ln  q(x,y,t)= \ln Q_T(x,y,t) = \ln \left( \frac{P^*(x,y,t) }{P(x,y,t \vert x_0,y_0,0)} \right)
\label{smallqbigQ}
\end{eqnarray} 
and its particular values
\begin{eqnarray}
\psi(x_0,y_0,0) && = \ln q(x_0,y_0,0) = \ln \left( \frac{P^*(x_0,y_0,0) }{P(x_0,y_0,0 \vert x_0,y_00)} \right) = \ln ( 1 )=0
   \nonumber \\
\psi(z,z,t) && = \ln q(z,z,t)      = \ln \left( \frac{\gamma^*(z,t) }{\gamma(z,t\vert x_0,y_0,0)} \right)
\label{lagrangianoptimalpsi}
\end{eqnarray} 
can be plugged into Eq. \ref{lagrangianbulkopt} to obtain that the optimal value of the Lagrangian
\begin{eqnarray}
 {\cal L}^{Bulk}\left[ P^*(.,.,.) ; \mu^*_.(.,.,.)  \right] 
&& =
 \int_{-\infty}^{+\infty}  dy   \int_{-\infty}^{+\infty}  dx \theta(x-y)  P^*(x,y,T)
 \ln \left( \frac{P^*(x,y,T) }{P(x,y,T \vert x_0,y_0,0)} \right)
\nonumber \\
&& +  \int_0^T dt \int_{-\infty}^{+\infty}  dz   \gamma^*(z,t) \ln \left( \frac{\gamma^*(z,t) }{\gamma(z,t\vert x_0,y_0,0)} \right)
 \label{lagrangianbulkoptqfin}
\end{eqnarray} 
coincides with the Sanov rate function ${\cal I}^{Sanov}_T \left[ P^*(.,.,T) ; \gamma^*(.,.)  \right] $
 of Eq. \ref{RateSanovstar} as it should for consistency.



\end{document}